\documentclass[twocolumn,tighten]{aastex63}

\usepackage{graphicx}
\usepackage{amsmath}
\usepackage{comment}
\usepackage{array,multirow,color,tablefootnote}
\usepackage{xcolor}
\shorttitle{Year 2 TESS phase curves}
\shortauthors{Wong et~al.}

\begin{document}
\title{Visible-light Phase Curves from the Second Year of the TESS Primary Mission} 
\correspondingauthor{Ian Wong}
\email{iwong@mit.edu}

\author[0000-0001-9665-8429]{Ian~Wong}
\altaffiliation{51 Pegasi b Fellow}
\affil{Department of Earth, Atmospheric and Planetary Sciences, Massachusetts Institute of Technology,
Cambridge, MA 02139, USA}

\author[0000-0003-4269-3311]{Daniel~Kitzmann}
\affiliation{University of Bern, Center for Space and Habitability, Bern, Switzerland}

\author[0000-0002-1836-3120]{Avi~Shporer}
\affil{Department of Physics and Kavli Institute for Astrophysics and Space Research, Massachusetts Institute of Technology, Cambridge, MA 02139, USA}

\author[0000-0003-1907-5910]{Kevin~Heng}
\affiliation{University of Bern, Center for Space and Habitability, Bern, Switzerland}
\affiliation{University of Warwick, Department of Physics, Astronomy and Astrophysics Group, Coventry CV4 7AL, UK}

\author[0000-0002-3551-279X]{Tara~Fetherolf}
\affiliation{Department of Earth and Planetary Sciences, University of California, Riverside, CA 92521, USA}

\author[0000-0001-5578-1498]{Bj{\" o}rn~Benneke}
\affiliation{Department of Physics and Institute for Research on Exoplanets, Universit{\' e} de Montr{\' e}al, Montr{\' e}al, QC, Canada}

\author[0000-0002-6939-9211]{Tansu~Daylan}
\altaffiliation{Kavli Fellow}
\affiliation{Department of Physics and Kavli Institute for Astrophysics and Space Research, Massachusetts Institute of Technology, Cambridge, MA 02139, USA}

\author[0000-0002-7084-0529]{Stephen~R.~Kane}
\affiliation{Department of Earth and Planetary Sciences, University of California, Riverside, CA 92521, USA}

% \author[0000-0003-2058-6662]{George~R.~Ricker}
% \affiliation{Department of Physics and Kavli Institute for Astrophysics and Space Research, Massachusetts Institute of Technology, Cambridge, MA 02139, USA}

\author[0000-0001-6763-6562]{Roland~Vanderspek}
\affiliation{Department of Physics and Kavli Institute for Astrophysics and Space Research, Massachusetts Institute of Technology, Cambridge, MA 02139, USA}

% \author[0000-0001-9911-7388]{David~W.~Latham}
% \affiliation{Center for Astrophysics ${\rm \mid}$ Harvard {\rm \&} Smithsonian, 60 Garden Street, Cambridge, MA 02138, USA}

\author[0000-0002-6892-6948]{Sara~Seager}
\affiliation{Department of Earth, Atmospheric and Planetary Sciences, Massachusetts Institute of Technology, Cambridge, MA 02139, USA}
\affiliation{Department of Physics and Kavli Institute for Astrophysics and Space Research, Massachusetts Institute of Technology, Cambridge, MA 02139, USA}
\affiliation{Department of Aeronautics and Astronautics, MIT, 77 Massachusetts Avenue, Cambridge, MA 02139, USA}

\author[0000-0002-4265-047X]{Joshua~N.~Winn}
\affiliation{Department of Astrophysical Sciences, Princeton University, Princeton, NJ 08544, USA}

\author[0000-0002-4715-9460]{Jon~M.~Jenkins}
\affiliation{NASA Ames Research Center, Moffett Field, CA 94035, USA}

%\author[0000-0003-0442-4284]{Patricia~T.~Boyd}
%\affiliation{Astrophysics Science Division, NASA Goddard Space Flight Center, Greenbelt, MD 20771, USA}

%\author[0000-0002-5322-2315]{Ana~Glidden}
%\affiliation{Department of Earth, Atmospheric and Planetary Sciences, Massachusetts Institute of Technology, Cambridge, MA 02139, USA}
%\affiliation{Department of Physics and Kavli Institute for Astrophysics and Space Research, Massachusetts Institute of Technology, Cambridge, MA 02139, USA}

%\author[0000-0003-1748-5975]{Robert~F.~Goeke}
%\affiliation{Department of Physics and Kavli Institute for Astrophysics and Space Research, Massachusetts Institute of Technology, Cambridge, MA 02139, USA}

%\author[0000-0001-5401-8079]{Lizhou~Sha}
%\affiliation{Department of Physics and Kavli Institute for Astrophysics and Space Research, Massachusetts Institute of Technology, Cambridge, MA 02139, USA}

%\author[0000-0002-6148-7903]{Jeffrey~C.~Smith}
%\affiliation{NASA Ames Research Center, Moffett Field, CA 94035, USA}
%\affiliation{SETI Institute, Mountain View, CA 94043, USA}

%\author[0000-0002-8964-8377]{Samuel~N.~Quinn}
%\affiliation{Center for Astrophysics ${\rm \mid}$ Harvard {\rm \&} Smithsonian, 60 Garden Street, Cambridge, MA 02138, USA}

%\author[0000-0002-1949-4720]{Peter~Tenenbaum}
%\affiliation{NASA Ames Research Center, Moffett Field, CA 94035, USA}
%\affiliation{SETI Institute, Mountain View, CA 94043, USA}

\author[0000-0002-8219-9505]{Eric~B.~Ting}
\affiliation{NASA Ames Research Center, Moffett Field, CA 94035, USA}

%\author[0000-0003-4755-584X]{Daniel~Yahalomi}
%\affiliation{Center for Astrophysics ${\rm \mid}$ Harvard {\rm \&} Smithsonian, 60 Garden Street, Cambridge, MA 02138, USA}

\begin{abstract}
We carried out a systematic study of full-orbit phase curves for known transiting systems in the northern ecliptic sky that were observed during Year 2 of the TESS primary mission. We applied the same methodology for target selection, data processing, and light-curve fitting as we did in our Year 1 study. Out of the 15 transiting systems selected for analysis, seven --- HAT-P-7, KELT-1, KELT-9, KELT-16, KELT-20, Kepler-13A, and WASP-12 --- show statistically significant secondary eclipses and day--night atmospheric brightness modulations. Small eastward dayside hotspot offsets were measured for KELT-9b and WASP-12b. KELT-1, Kepler-13A, and WASP-12 show additional phase-curve variability attributed to the tidal distortion of the host star; the amplitudes of these signals are consistent with theoretical predictions. We combined occultation measurements from TESS and Spitzer to compute dayside brightness temperatures, TESS-band geometric albedos, Bond albedos, and phase integrals for several systems. The new albedo values solidify the previously reported trend between dayside temperature and geometric albedo for planets with $1500<T_{\mathrm{day}}<3000$~K. For Kepler-13Ab, we carried out an atmospheric retrieval of the full secondary eclipse spectrum, which revealed a non-inverted temperature--pressure profile, significant H$_{2}$O and K absorption in the near-infrared, evidence for strong optical atmospheric opacity due to sodium, and a confirmation of the high geometric albedo inferred from our simpler analysis. We explore the implications of the phase integrals (ratios of Bond to geometric albedos) for understanding exoplanet clouds. We also report updated transit ephemerides for all of the systems studied in this work.
\end{abstract}

\section{Introduction}\label{sec:intro}

In July 2020, the Transiting Exoplanet Survey Satellite (TESS) completed its two-year primary mission to discover new exoplanets around bright stars in the solar neighborhood. With the goal of achieving almost full sky coverage, the survey has provided high-cadence visible-wavelength photometry for hundreds of thousands of stars. Among these observed targets are hundreds of previously-discovered transiting exoplanet systems. For these systems, TESS light curves enable a wide range of scientific investigations \citep{kane2021}, from refining orbital ephemerides \citep[e.g.,][]{cortes-zuleta2020,ikwut-ukwa2020,szabo2020} and detecting additional transiting planets \citep[e.g.,][]{huang2018,teske2020} to probing for orbital decay and transit-timing variations \citep[e.g.,][]{bouma2019}.

The study of exoplanet phase curves in particular has benefited immensely from the nearly continuous long-baseline observations by TESS. The full-orbit light curve of a transiting system at optical wavelengths can reveal the secondary eclipse, when the light from the planet's star-facing hemisphere is occulted by the host star, as well as synchronous flux modulations attributed to longitudinal brightness variations across the planet's surface \citep[e.g.,][]{hs15,parmentier2017}, the tidal distortion of the surfaces of both bodies \citep[e.g.,][]{morris1985,morris1993}, and periodic Doppler shifting of the stellar spectrum through the mutual star--planet gravitational interaction \citep[e.g.,][]{shakura1987,loeb2003,zucker2007,shporer2010}. Detecting and measuring these phase-curve signals can provide crucial insights into the system, including the global temperature distribution, efficiency of day--night heat transport, and reflectivity of the planet, as well as the stellar tidal response (see the review by \citealt{shporer2017}).

To date, dedicated TESS phase-curve analyses have been published for a wide range of individual exoplanet systems, including KELT-1 \citep{beatty2020,vonessen2020kelt1}, KELT-9 \citep{wong2020kelt9}, KELT-16 \citep{mancini2021}, WASP-18 \citep{shporer2019}, WASP-19 \citep{wong2020wasp19}, WASP-33 \citep{vonessen2020wasp33}, WASP-100 \citep{jansen2020}, and WASP-121 \citep{bourrier2019,daylan2019}. These are some of the brightest and most amenable targets for detailed study, yielding high signal-to-noise secondary eclipse measurements and exquisite constraints on the day--night brightness contrast. Looking beyond these benchmark targets, we have set out to compile a comprehensive body of phase-curve analyses based on TESS photometry. This effort is guided by previous systematic investigations of Kepler light curves \citep[e.g.,][]{esteves2013,esteves2015,angerhausen2015} and facilitates ensemble studies of visible-light secondary eclipses and atmospheric properties.

In \citet{wong2020year1}, hereafter referred to as Paper 1, we presented a summary of phase-curve measurements from the first year of the TESS primary mission, when TESS's four cameras surveyed the southern ecliptic sky. Ten systems displayed statistically significant secondary eclipse and/or phase-curve signals. One of the most notable results from this study emerged when combining the newly-obtained TESS-band secondary eclipses with previously-published Spitzer measurements, which allowed us to break the degeneracy between atmospheric reflectivity and dayside brightness temperature and calculate self-consistent TESS-band geometric albedos. We uncovered a tentative positive correlation between geometric albedo and dayside temperature among hot Jupiters, suggesting a steady increase in reflective cloud cover and/or systematic deviations from blackbody-like emission spectra with increasing temperature. This surprising and consequential finding necessitates further study, and the inclusion of additional data points into the body of geometric albedo measurements promises to shed more light on this emergent trend.

In this paper, we extend our previous systematic phase-curve study of southern targets into the northern ecliptic sky, which was observed by TESS during the second year of the primary mission. A comparable number of targets are considered, among which seven show robust phase-curve signals. We employ a consistent light-curve processing and fitting methodology, thereby ensuring that the analyses carried out in this paper and Paper 1 constitute a uniform set of results. 

The target-selection criteria, TESS light curves, and data-analysis techniques are described in Sections \ref{subsec:targets}--\ref{subsec:model}, respectively. Section \ref{sec:res} presents the results of our phase curve fits. In Section \ref{sec:dis}, we use published Spitzer secondary eclipse measurements to expand the list of self-consistently derived geometric albedo and dayside brightness temperatures (Section \ref{subsec:temps}) and revisit the emergent albedo vs. temperature trend for highly-irradiated planets (Section \ref{subsec:trends}). This discussion is supplemented by detailed emission spectrum modeling of high-albedo hot Jupiter Kepler-13Ab (Section \ref{subsec:modeling}), as well as an exploration of the predictive power of albedo measurements in characterizing exoplanet cloud properties (Section \ref{subsec:clouds}). Lastly, we present updated transit ephemerides in Section \ref{subsec:ephem}. A broad summary of the results of this work is given in Section \ref{sec:con}.

\section{Light-curve Analysis}\label{sec:ana}
To ensure maximum consistency with our systematic phase-curve study from the first year of the TESS mission (Paper 1), we implemented an identical methodology for target selection, data processing, phase-curve modeling, and error analysis. All steps in the light-curve analysis were carried out using the ExoTEP pipeline \citep[e.g.,][]{benneke2019,wong2020hatp12}. We briefly discuss these techniques in the following; see Paper 1 for a more detailed description of the methods.

\subsection{Target Selection}\label{subsec:targets}
Targets for phase-curve study were selected from the population of all transiting planet and brown dwarf systems published in the literature as of 2021 January 1. In order to adequately resolve the transit and secondary eclipse shapes, we limited our scope to systems that were preselected by the TESS mission to have photometry extracted at 2-minute cadence, as opposed to the 30-minute cadence of the stacked full-frame images.

The signal-to-noise and predicted signal-strength thresholds that we considered are unchanged from those defined in Paper 1. We first excluded systems with TESS-band magnitudes greater than $T=12.5$~mag. For the remaining systems, we used the predicted secondary eclipse depth as the determining factor in the selection. To compute this depth $D'_{d}$, we assumed maximally inefficient day--night heat recirculation and a dayside geometric albedo of $A_{g}=0.1$ \citep{esteves2013,esteves2015,hengalbedo,shporer2017}:
\begin{align}
\label{depth} D'_{d} & = \left(\frac{R_{p}}{R_{*}}\right)^{2}\frac{\int B_{\lambda}(T_{p})\tau(\lambda) d\lambda}{\int B_{\lambda}(T_{*})\tau(\lambda) d\lambda}+A_{g}\left(\frac{R_{p}}{a}\right)^{2}, \\
\label{temp} T_{p} &= T_{*}\sqrt{\frac{R_{*}}{a}}\left\lbrack\frac{2}{3}\left(1-\frac{3}{2}A_{g}\right)\right\rbrack^{1/4},
\end{align}
where we assumed a conversion factor of 3/2 between the Bond and geometric albedos, as is appropriate for Lambertian scattering. Here, the photon-weighted planetary and stellar emission spectra, which are approximated by blackbodies with temperatures $T_{p}$ and $T_{*}$, respectively, are integrated over the TESS transmission function $\tau(\lambda)$. We note that the TESS transmission function provided online\footnote{https://heasarc.gsfc.nasa.gov/docs/tess/data/tess-response-function-v1.0.csv} is given in energy units, so an additional factor of $\lambda/hc$ is not needed. The remaining variables are system parameters related to the shape of the transit light curve: the planet--star radius ratio $R_{p}/R_{*}$ and the scaled orbital semimajor axis $a/R_{*}$. As a preliminary cut, we calculated the predicted secondary eclipse depths using parameter values from the respective discovery papers and selected all systems with $D'_{d}>100~\mathrm{ppm}/\sqrt{s}$, where $s$ is the number of sectors that a system was observed by TESS, and the scaling reflects the approximate increase in combined signal-to-noise with additional sectors of observation. 

Next, we inspected the raw light curves of the systems that passed this threshold (see Section \ref{subsec:tess}). In order to limit our analysis to cases where the astrophysical signal can be reliably detected, we removed systems that show severe systematics and/or significant short-period stellar variability. Both of these features present difficulties for systematics detrending methodologies, particularly when the timescale of the variations is shorter than the orbital period (i.e., the characteristic timescale of the phase-curve modulation).  Several targets that are otherwise promising for phase-curve study were excluded due to excessive variability, including KELT-7, WASP-33\footnote{A dedicated analysis of the WASP-33 TESS phase curve was published in \citet{vonessen2020wasp33}, where a detailed treatment of the complex stellar pulsation frequency spectrum was applied.}, and XO-3. An exception to this exclusion condition is KELT-9, which displays a stellar pulsation signal with a period of roughly 7.6~hr \citep{wong2020kelt9}. Having previously analyzed the TESS light curve for this system, we included it in the present work to derive updated phase-curve results using a consistent methodology with the other targets on the list.

The final set of 15 targets selected for our systematic phase-curve analysis is as follows: HAT-P-7 \citep{hatp7}, HAT-P-36 \citep{hatp36}, KELT-1 \citep{kelt1}, KELT-9 \citep{kelt9}, KELT-16 \citep{kelt16}, KELT-20 \citep{kelt20}, KELT-23A \citep{kelt23}, Kepler-13A \citep{shporer2011,szabo2011}, Qatar-1 \citep{qatar1}, TrES-3 \citep{tres3}, WASP-3 \citep{wasp3}, WASP-12 \citep{wasp12}, WASP-92 \citep{wasp9293}, WASP-93 \citep{wasp9293}, and WASP-135 \citep{wasp135}.

As in Paper 1, we established an additional selection criterion based on the predicted amplitudes of the ellipsoidal distortion and Doppler boosting phase-curve signals. Ellipsoidal distortion of the host star yields a photometric modulation with a leading-order term at the first harmonic of the cosine of the orbital period, while Doppler boosting produces a contribution at the fundamental of the sine. The corresponding semiamplitudes are related to the planet--star mass ratio $q\equiv M_{p}/M_{*}$ via the following expressions \citep[e.g.,][]{shporer2017}:
\begin{align}\label{ellip}
A'_{\mathrm{ellip}} & = \alpha_{\mathrm{ellip}}q\left(\frac{R_{*}}{a}\right)^{3}\sin^2 i, \\
\label{dopp}A'_{\mathrm{Dopp}} & =\left\lbrack\frac{2\pi G} {Pc^{3}}\frac{q^2M_p\sin^3 i}{(1+q)^2}\right\rbrack^{\frac{1}{3}}  \left\langle \frac{xe^{x}}{e^{x}-1}\right\rangle_{\mathrm{TESS}}.
\end{align}
Here, $i$ is the orbital inclination, $P$ represents the orbital period, and $x\equiv hc/k\lambda T_{*}$. The expression for the ellipsoidal distortion semiamplitude includes a prefactor $\alpha_{\mathrm{ellip}}$, which depends on the linear limb- and gravity-darkening coefficients for the host star (see, for example, \citealt{morris1985} and \citealt{shporer2017}). Tabulated values of the TESS-band limb- and gravity-darkening coefficients from \cite{claret2017} were interpolated to provide appropriate coefficients for a given set of stellar parameters. In the case of Doppler boosting, the term inside the angled brackets is the logarithmic derivative of the host star's spectrum (approximated as a blackbody) and is integrated over the TESS bandpass. Using a theoretical stellar spectrum instead (e.g., a PHOENIX model; \citealt{husser2013}) results in a negligible change to the resultant Doppler boosting amplitude at the level of a few percent.

We calculated the predicted $A'_{\mathrm{ellip}}$ and $A'_{\mathrm{Dopp}}$ values for all known systems brighter than $T=12.5$~mag and set a minimum threshold of 25~ppm. We found that all systems for which $A'_{\mathrm{ellip}}$ and/or $A'_{\mathrm{Dopp}}$ exceed 25~ppm were already added to our target list by satisfying the aforementioned secondary eclipse depth benchmark, and hence no additional targets were included based on this threshold.

\subsection{TESS Light Curves}\label{subsec:tess}
During the second year of the primary mission, TESS observed the northern ecliptic hemisphere, which was divided into 13 sectors. Each sector has a combined field of view of $24^{\circ}\times96^{\circ}$ and was observed for 27.4 days, during which the spacecraft completed two eccentric orbits around the Earth, with a gap in science observations near perigee for data downlink. 

We obtained the light-curve files from the Mikulski Archive for Space Telescopes (MAST). The photometry and associated data products were produced using the official Science Processing Operations Center (SPOC) pipeline, based at NASA Ames Research Center \citep{jenkins2016}. The files contain both the raw simple aperture photometry (SAP) and the pre-search data conditioning (PDC) light curves, which were corrected for instrumental systematics using co-trending basis vectors empirically derived on a sector-by-sector basis for each camera and detector on the instrument \citep{smith2012,stumpe2012,stumpe2014}. Just as in Paper 1, we carried out analogous analyses of the SAP and PDC light curves and found that the systematics corrections by the SPOC pipeline typically result in significantly reduced long-term flux variations and reduced red noise, while crucially preserving the astrophysical phase-curve signal of interest. For most targets, we utilized the PDC light curves in the final fits presented in this paper.

However, for systems exhibiting significant stellar variability, the PDC detrending process is often unable to properly discriminate between instrumental systematics trends and flux variations from the star, resulting in poorer light-curve quality. This was previously seen in the light curves of several active targets, including WASP-19 \citep{wong2020wasp19} and WASP-121 \citep{daylan2019}. Among the Year 2 targets selected for phase-curve analysis, only TrES-3 displays notable photometric variability from stellar activity. For that system, we used the SAP light curve instead and detrended the instrumental systematics by using the publicly available co-trending basis vectors.

We note that TESS data from sectors 14--19 were reprocessed by the official SPOC pipeline after their initial release to rectify issues with the time stamps and alter treatment of scattered light, among other improvements\footnote{See the notes for Data Release 30 for full details: \texttt{archive.stsci.edu/tess/tess\_drn.html} (dated 2020 August 5).}. The previously published phase-curve analyses of KELT-1 \citep{beatty2020,vonessen2020kelt1} and KELT-9 \citep{wong2020kelt9} were based off of the original versions of the light curves. In this analysis, we used the newer versions of the photometry for all targets observed in sectors 14--19. As discussed in Sections \ref{kelt1} and \ref{kelt9}, the updated astrophysical parameter values for KELT-1 and KELT-9 in this paper do not differ significantly from the previously published results.

Following our previous work \citep{wong2020wasp19,wong2020year1,wong2020kelt9}, we split each sector's worth of photometry into smaller segments that are separated by the scheduled momentum dumps. During the second year of the primary mission, these occurred once or twice during each spacecraft orbit and were typically associated with discernible discontinuities in the photometry, with some instances showing additional flux ramps before and/or after. In cases with severe flux ramps on short timescales (i.e., shorter than the orbital period of the system), we trimmed the ramps prior to fitting, with the trimming interval selected among multiples of 0.25~d. After removing all points assigned a nonzero data-quality flag by the SPOC pipeline, we applied a 16-point-wide moving median filter to trim $3\sigma$ outliers. Lastly, we inspected each light curve and disregarded all segments shorter than one day, as well as those that show systematically larger time-correlated noise or contain large gaps due to periods of significant scattered light on the detector.

In Appendix~\ref{sec:segmentlist}, we provide a full description of the data segments used in our analysis. The raw and trimmed light curves for each target are plotted in Appendix~\ref{sec:rawplots}, with the locations of momentum dumps indicated by vertical blue lines.

\subsection{Phase-curve Model Fitting}\label{subsec:model}

The combined phase-curve and systematics model used in our fits was defined exactly as in Paper 1:
\begin{equation}\label{fullfit}
    f(t) = \psi(t)\times S_N^{\lbrace k\rbrace}(t).
\end{equation}
The first term is the astrophysical model that describes the photometric modulation of the host star and orbiting companion separately with respect to orbital phase $\phi\equiv 2\pi(t-T_{0})/P$, as well as the geometrical loss-of-light functions due to transits $\lambda_t(t)$ and secondary eclipses $\lambda_e(t)$\footnote{The formulation presented here contains a few simplifying assumptions regarding the shape of the atmospheric brightness modulation and higher-order terms in the ellipsoidal distortion modulation. See Paper 1 for a full description of the caveats and validations for our approach.}:
\begin{align}
\label{astro}\psi(t) &= \frac{\psi_{*}(t)\lambda_t(t)+\psi_{p}(t)\lambda_e(t)}{1+\bar{f_p}},\\
\label{planet}\psi_{p}(t) &= \bar{f_{p}} - A_{\mathrm{atm}} \cos(\phi+\delta),\\
\label{star}\psi_{*}(t) &= 1-A_{\mathrm{ellip}}\cos(2\phi)+A_{\mathrm{Dopp}}\sin(\phi).
\end{align}
Both transits and secondary eclipses were modeled using \texttt{batman} \citep{kreidberg2015}. The variables $\bar{f_{p}}$, $A_{\mathrm{atm}}$, and $\delta$ signify the average relative brightness of the companion, the semiamplitude of the atmospheric brightness modulation, and the corresponding phase shift, respectively. From these parameters, the dayside flux (i.e., secondary eclipse depth) and nightside flux are given by $D_{d}=\bar{f_{p}}-A_{\mathrm{atm}}\cos(\pi+\delta)$ and $D_{n}=\bar{f_{p}}-A_{\mathrm{atm}}\cos(\delta)$. 

The host star's variability includes contributions from ellipsoidal distortion and Doppler boosting. In cases where significant ellipsoidal distortion amplitudes were measured, we experimented with fitting for additional higher-order harmonics, but did not retrieve any statistically significant signals. For targets where no significant $A_{\mathrm{ellip}}$ or $A_{\mathrm{Dopp}}$ values were retrieved in unconstrained fits, we followed the methodology of Paper 1 and applied Gaussian priors on the amplitudes instead. These priors were derived using Equations~\eqref{ellip} and \eqref{dopp}, the stellar parameters from the corresponding discovery papers, and the tabulated limb- and gravity-darkening coefficients from \citet{claret2017}.

The second term in Equation~\eqref{fullfit} is the systematics detrending model, which consists of generalized polynomial functions in time that were applied separately to each data segment $k$ in the light curve:
\begin{equation}\label{systematics}
    S_N^{\lbrace k\rbrace}(t) = \sum\limits_{j=0}^{N}c_j^{\lbrace k\rbrace}(t-t_0)^j.
\end{equation}
Here, $t_0$ is the time of the first data point of the segment, and $N$ is the order of the detrending polynomial. To choose the optimal polynomial order for a given segment, we fit the segment's light curve individually, selecting the order that minimized the Bayesian information criterion (BIC). Table~\ref{tab:segments} in Appendix~\ref{sec:segmentlist} lists the optimal polynomial orders for every segment; typical values range from 0 to 2. The systematics-detrended light curves are plotted in Appendix~\ref{sec:rawplots}.

In the first step of our light-curve analysis, the astrophysical and systematics models were fit simultaneously using the affine-invariant Markov chain Monte Carlo (MCMC) routine \texttt{emcee} \citep{emcee}. All transit shape and orbital ephemeris parameters were allowed to vary freely, except in the case of KELT-9, where the transits were trimmed from the light curve and Gaussian priors were used instead (Section \ref{kelt9}). For all of the targets in our analysis, the orbit of the companion is consistent with circular, and we set the orbital eccentricity to zero. The time of secondary eclipse was adjusted for the light-travel time between inferior and superior conjunction, which is less than a minute for all targets in our study. The fitted parameters are the mid-transit time $T_{0}$, orbital period $P$, impact parameter $b$, scaled orbital semimajor axis $a/R_{*}$, planet--star radius ratio $R_{p}/R_{*}$, and modified quadratic limb-darkening coefficients, which are defined by \citet{holman2006}: $\gamma_{1}\equiv 2u_{1}+u_{2}$ and $\gamma_{2}\equiv u_{1}-2u_{2}$, where $u_1$ and $u_2$ are the standard quadratic coefficients. We also introduced a uniform per-point scatter parameter $\sigma$, which was allowed to float freely to ensure that the chains converged to models with a reduced $\chi^2$ value near unity. 

In the next step, we employed two methods to account for the additional contribution of red noise at timescales longer than the 2-minute cadence of the time series. First, following the technique first described by \citet{pont2006}, we computed the scatter in the residual series, binned at various intervals $n$, and calculated the average fractional deviation $\xi$ from the $1/\sqrt{n}$ scaling expected for pure white noise across bin sizes corresponding to time intervals between 20~min and 8~hr. These timescales are relevant to the primary features of the astrophysical model, i.e., transit ingress/egress and phase-curve inflection timescales. To incorporate this long-timescale red noise contribution into the final MCMC fits, we inflated the previously calculated per-point uncertainty values $\sigma$ by $\xi$ and reran the fitting procedure, now with the flux uncertainties fixed to the new values. The second technique was ``prayer-bead'' (PB) residual permutation \citep[e.g.,][]{gillon2009}: after dividing out the best-fit systematics detrending model from the initial MCMC fit, we cyclically shifted and readded the residual array 5000 times, each time computing the best-fit astrophysical parameters using a standard Levenberg--Marquardt optimization routine. The uncertainties on the fit parameters were derived from the resulting 5000-point posteriors of best-fit values.

For all parameters except the mid-transit time, the uncertainty-inflated MCMC analysis yielded larger uncertainties, and we present those values in the results tables below. For $T_{0}$, the PB analysis produced uncertainties that are consistently larger than those from the MCMC fits (by up to 150\%). We list both the MCMC- and PB-derived transit timings in the tables and utilize the larger PB uncertainties when calculating updated transit ephemerides (Section \ref{subsec:ephem}).

\section{Results}\label{sec:res}

For each of the 15 targets, we determined which phase-curve signals were robustly detected in the TESS light curves by running an ensemble of joint MCMC fits that included different combinations of phase-curve parameters. In cases where no significant ellipsoidal distortion and/or Doppler boosting signals were measured from an unconstrained fit, we instead applied Gaussian priors on the semiamplitudes based on the predicted values derived using Equations~\eqref{ellip} and \eqref{dopp}. For the final fit results, we selected the combination of free parameters that minimized the Akaike information criterion (AIC). The AIC penalizes the addition of free parameters less severely than the BIC, so considering the AIC allowed us to explore some comparatively marginal phase-curve signals. These weak detections will benefit the most from additional light curves obtained during the extended mission and will help orient strategies for follow-up study.

In the following subsections, we present the results for the 7 systems that yielded significant phase-curve signals. Due to the presence of stellar pulsations and additional light-curve variability, the KELT-9 light-curve was treated differently than the other six targets; that system is discussed after the other nominal cases. Finally, for the remaining 8 systems without statistically significant phase-curve signals or secondary eclipses, we present the results from transit-only light-curve fits; these targets are discussed together in the last subsection.

\begin{figure*}[t]
\includegraphics[width=\linewidth]{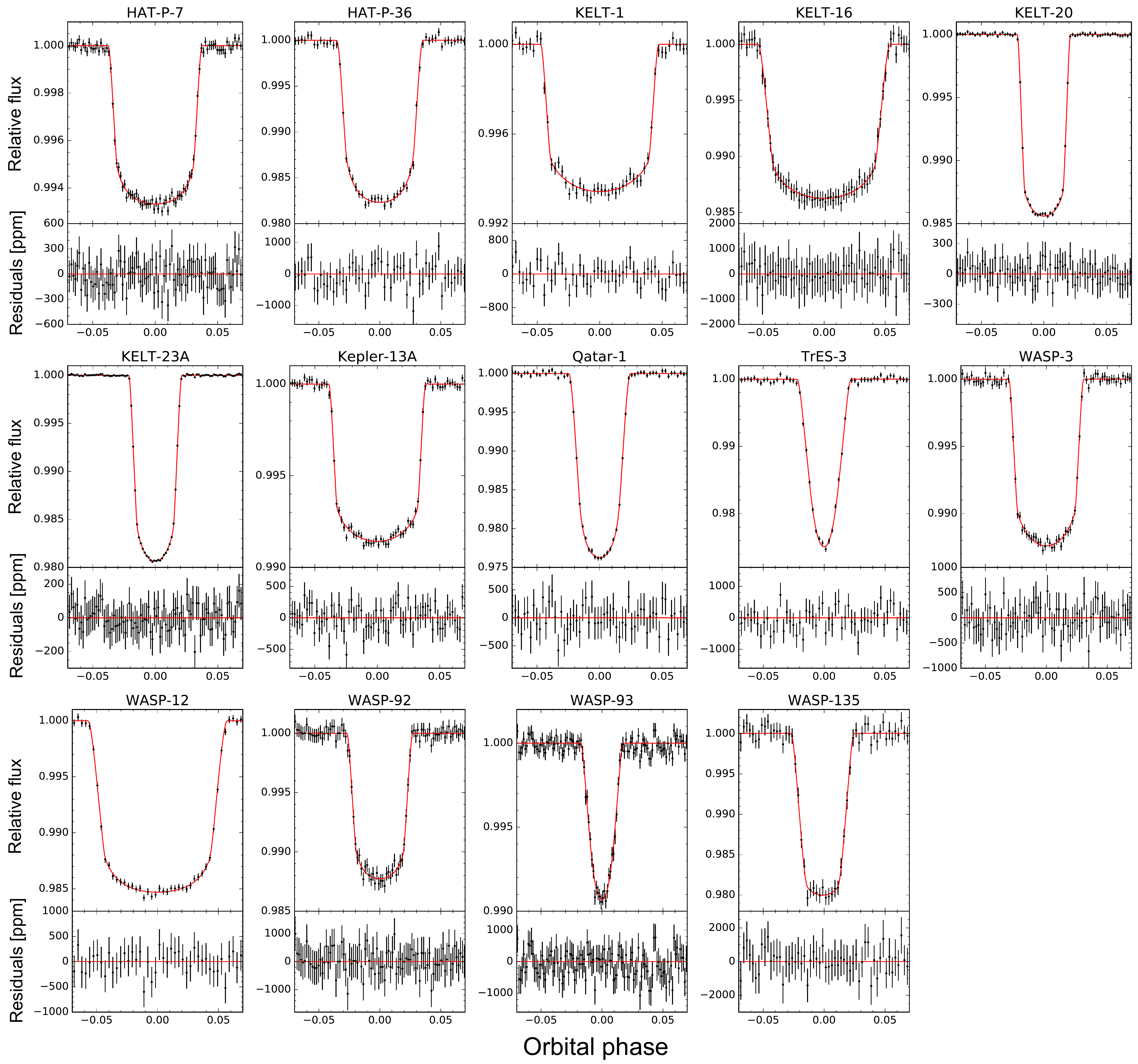}
\caption{A compilation of the systematics-corrected and phase-folded TESS light curves in the vicinity of the primary transit for 14 of the 15 targets analyzed in this work. KELT-9 is excluded, because the transits were removed from the time series prior to analysis (Section \ref{kelt9}). The best-fit phase-curve models have been removed from the data for the systems where significant signals were detected. The corresponding residuals from the best-fit model are shown in the bottom panels. The data binning interval was set to 3, 5, and 10~minutes for systems with orbital periods in the ranges $P<1$~d, $1\le P\le 3$~d, and $P>3$~d, respectively.}
\label{fig:transit}
\end{figure*}

% Model fit parameters table
%-------------------------------------------------------------------------------
\begin{splitdeluxetable*}{lllllllBlllllll}
\tablewidth{0pc}
\tabletypesize{\scriptsize}
\tablecaption{
    Results from Phase-Curve Fits without Stellar Pulsations
    \label{tab:fit}
}
\vspace{-0.3cm}
\tablehead{& \multicolumn{2}{c}{\underline{HAT-P-7}} \vspace{-0.2cm}& \multicolumn{2}{c}{\underline{KELT-1}} &
\multicolumn{2}{c}{\underline{KELT-16}} & & \multicolumn{2}{c}{\underline{KELT-20}} &  \multicolumn{2}{c}{\underline{Kepler-13A}} &
\multicolumn{2}{c}{\underline{WASP-12}} \\
    \colhead{Parameter} \vspace{-0.1cm}&
    \colhead{Value}                     &
    \colhead{Error}  &
    \colhead{Value}                     &
    \colhead{Error}  & 
    \colhead{Value}                     &
    \colhead{Error}  &  
    \colhead{Parameter} &
    \colhead{Value}                     &
    \colhead{Error}  &
    \colhead{Value}                     &
    \colhead{Error}  & 
    \colhead{Value}                     &
    \colhead{Error}  
}
\startdata
\multicolumn{2}{l}{\textit{Fitted parameters}} & & & & & & \multicolumn{2}{l}{\textit{Fitted parameters}} & & & & &\\
$R_p/R_*$     & 0.0770 & $_{-0.0011}^{+0.0009}$ & 0.07612 & $_{-0.00076}^{+0.00095}$ & 0.1099 & $_{-0.0018}^{+0.0021}$ & $R_{p}/R_{*}$ & 0.11562 & $_{-0.00064}^{+0.00056}$ & 0.08739 & $_{-0.00039}^{+0.00046}$ & 0.1169 & $_{-0.0012}^{+0.0010}$ \\
$T_{0,\mathrm{MCMC}}$ ($\mathrm{BJD}_{\mathrm{TDB}}-2458000$)\tablenotemark{\scriptsize a}  &  $709.02447$ & $_{-0.00019}^{+0.00018}$ & 778.92668 & $_{-0.00024}^{+0.00022}$ & 719.14833 & 0.00025 & $T_{0,\mathrm{MCMC}}$ ($\mathrm{BJD}_{\mathrm{TDB}}-2458000$) & 698.21073 & 0.00011 & 718.82552 & 0.00015 & 853.91923 & $_{-0.00011}^{+0.00013}$ \\
$T_{0,\mathrm{PB}}$ ($\mathrm{BJD}_{\mathrm{TDB}}-2458000$)\tablenotemark{\scriptsize a}  & 709.02466 & $^{+0.00034}_{-0.00033}$ & 778.92707 & $^{+0.00048}_{-0.00050}$ & 719.14831 & $^{+0.00045}_{-0.00042}$ & $T_{0,\mathrm{PB}}$ ($\mathrm{BJD}_{\mathrm{TDB}}-2458000$) & 698.21073 & 0.00014 & 718.82551 & $^{+0.00030}_{-0.00034}$ & 853.91918 & $^{+0.00020}_{-0.00021}$\\
$P$ (days)    & 2.204753 & $_{-0.000025}^{+0.000027}$ &  1.217537 & $_{-0.000034}^{+0.000036}$ & 0.968995 & $_{-0.000029}^{+0.000033}$ & $P$ (days) & 3.474074 & $_{-0.000045}^{+0.000042}$ & 1.7635869 & 0.0000016 & 1.091414 & $_{-0.000017}^{+0.000015}$ \\
$b$           & 0.41 & $_{-0.14}^{+0.08}$ & 0.27 & $_{-0.17}^{+0.15}$ & 0.29 & 0.16 & $b$ & 0.499 & $_{-0.024}^{+0.021}$ & 0.12 & $_{-0.09}^{+0.10}$ & 0.338 & $_{-0.084}^{+0.065}$\\
$a/R_*$       & 4.31 & $_{-0.17}^{+0.19}$ & 3.59 & $_{-0.18}^{+0.10}$ & 3.21 & $_{-0.16}^{+0.10}$ & $a/R_*$ & 7.546 & $_{-0.090}^{+0.095}$ & 4.508 & $_{-0.070}^{+0.033}$ & 3.062 & $_{-0.066}^{+0.063}$ \\
$\bar{f_p}$ (ppm)    & 71 & 30 & 213 & 61 & 240 & $_{-110}^{+120}$ & $\bar{f_p}$ (ppm)  & 64 & $_{-32}^{+34}$ & 151 & $_{-39}^{+45}$ & 184 & $_{-79}^{+82}$ \\
$A_{\mathrm{atm}}$ (ppm)   & 56 & $_{-13}^{+14}$ & 176 & $_{-30}^{+29}$ & 175 & $_{-62}^{+64}$ & $A_{\mathrm{atm}}$ (ppm)   & 43 & $_{-11}^{+13}$ & 151 & $_{-16}^{+15}$ & 264 & $_{-30}^{+33}$ \\
$\delta$ ($^{\circ}$) & $\left(5\right)$\tablenotemark{\scriptsize b} & $\left(12\right)$\tablenotemark{\scriptsize b} & $\left(5.2\right)$ & $\left(_{-7.4}^{+8.0}\right)$ & $\left(6\right)$ & $\left(_{-19}^{+18}\right)$ & $\delta$ ($^{\circ}$) & $\left(-9\right)$ & $\left(_{-15}^{+16}\right)$  & $\left(8.9\right)$ & $\left(_{-4.6}^{+5.0}\right)$ & 13.2 & 5.7 \\
$A_{\mathrm{ellip}}$ (ppm)  & $\left\lbrack 16\right\rbrack$\tablenotemark{\scriptsize b} & $\left\lbrack 4\right\rbrack$\tablenotemark{\scriptsize b} & 416 & $_{-26}^{+25}$ & $\left\lbrack 72\right\rbrack$ & $\left\lbrack 10\right\rbrack$ & $A_{\mathrm{ellip}}$ (ppm)  & \dots & \dots & 49 & $_{-16}^{+17}$ & 80 & $_{-35}^{+33}$ \\
$A_{\mathrm{Dopp}}$ (ppm)  & $\left\lbrack 2.2\right\rbrack$ & $\left\lbrack 0.1\right\rbrack$ & $\left\lbrack 43\right\rbrack$ & $\left\lbrack 2\right\rbrack$ &  $\left\lbrack 5.2\right\rbrack$ & $\left\lbrack 0.3\right\rbrack$ & $A_{\mathrm{Dopp}}$ (ppm)  & \dots & \dots & $\left\lbrack 6.8\right\rbrack$ & $\left\lbrack 1.7\right\rbrack$ & $\left\lbrack 2.3\right\rbrack$ & $\left\lbrack 0.2\right\rbrack$ \\
$\gamma_{1}$\tablenotemark{\scriptsize c}  &  0.795 & $_{-0.086}^{+0.080}$ & 0.74 & $_{-0.11}^{+0.10}$ & 0.82 & 0.14 & $\gamma_{1}$ & 0.583 & $_{-0.050}^{+0.048}$ &  0.667 & $_{-0.053}^{+0.051}$ & 0.733 & 0.069 \\
$\gamma_{2}$\tablenotemark{\scriptsize c}  & $-0.68$ & $_{-0.47}^{+0.51}$ & $-0.17$ & $_{-0.54}^{+0.39}$ & $-0.45$ & $_{-0.66}^{+0.57}$ & $\gamma_{2}$ & $-0.22$ & $_{-0.42}^{+0.34}$ & 0.08 & $_{-0.27}^{+0.18}$ & $-0.41$ & $_{-0.49}^{+0.46}$ \\
\\
\multicolumn{2}{l}{\textit{Derived parameters}} & & & & & &  \multicolumn{2}{l}{\textit{Derived parameters}} & & & & &  \\
$D_{d}$ (ppm)\tablenotemark{\scriptsize d}  &  127 & $_{-32}^{+33}$ & 388 & $_{-65}^{+67}$ & 410 & $_{-120}^{+130}$ & $D_{d}$ (ppm) & 111 & $_{-36}^{+35}$ & 301 & $_{-42}^{+46}$ & 443 & $_{-85}^{+86}$  \\
$D_{n}$ (ppm)\tablenotemark{\scriptsize d}   &  14 & 32 & 39 & $_{-72}^{+70}$ & 70 & 130 & $D_{n}$ (ppm) & 18 & $_{-33}^{+36}$ & 0 & $_{-43}^{+48}$ & $-74$ & $_{-85}^{+90}$ \\
$i$ ($^{\circ}$)      &  84.6 & $_{-1.4}^{+2.0}$ & 85.8 & $_{-2.8}^{+2.7}$ & 84.8 & $_{-3.3}^{+3.0}$ & $i$ ($^{\circ}$)      & 86.21 & $_{-0.21}^{+0.23}$ & 88.5 & $_{-1.4}^{+1.1}$ & 83.7 & $_{-1.4}^{+1.7}$\\
$u_{1}$  & 0.18 & $_{-0.10}^{+0.11}$ & 0.26 & $_{-0.12}^{+0.09}$ & 0.23 & $_{-0.13}^{+0.12}$ & $u_{1}$  & 0.188 & $_{-0.095}^{+0.079}$ & 0.282 & $_{-0.060}^{+0.044}$ & 0.209 & $_{-0.092}^{+0.091}$ \\
$u_{2}$  & 0.43 & $_{-0.20}^{+0.19}$ & 0.22 & $_{-0.16}^{+0.21}$ & 0.34 & $_{-0.22}^{+0.28}$ & $u_{2}$  & 0.21 & $_{-0.13}^{+0.16}$ & 0.10 & $_{-0.07}^{+0.11}$ & 0.31 & $_{-0.18}^{+0.20}$ \\
$a$ (au) & 0.0365 & 0.0037 & 0.0244 & 0.0011 & 0.0203 & 0.0012 & $a$ (au) & 0.0549 & 0.0022 & 0.03585 & 0.00093 & 0.0224 & 0.0011 \\
$R_{p}$ ($R_{\mathrm{Jup}}$) & 1.38 & 0.13 & 1.081 & 0.028 & 1.454 & 0.068 & $R_{p}$ ($R_{\mathrm{Jup}}$) & 1.761 & 0.069 & 1.454 & 0.035 & 1.786 & 0.081\\
$M_{p}$ ($M_{\mathrm{Jup}}$)\tablenotemark{\scriptsize e} & \dots & \dots & 25.1 & $^{+3.7}_{-3.3}$ &\dots & \dots & $M_{p}$ ($M_{\mathrm{Jup}}$) & \dots & \dots & $3.9-11.2$ & \dots & 3.0 & 1.3 \\
\enddata
\textbf{Notes.}
\vspace{-0.25cm}\tablenotetext{\textrm{a}}{Mid-transit times derived from the MCMC and PB analyses.}
\vspace{-0.25cm}\tablenotetext{\textrm{b}}{Marginally detected phase-curve parameters are provided in parentheses. Square brackets denote applied Gaussian priors.}
\vspace{-0.25cm}\tablenotetext{\textrm{c}}{Modified limb-darkening parameters $\gamma_{1}\equiv 2u_{1}+u_{2}$ and $\gamma_{2}\equiv u_{1}-2u_{2}$.}
\vspace{-0.25cm}\tablenotetext{\textrm{d}}{$D_{d}$ and $D_{n}$ are the dayside and nightside fluxes, respectively. The dayside flux is equivalent to the secondary eclipse depth.}
\vspace{-0.25cm}\tablenotetext{\textrm{e}}{Companion masses derived from the measured ellipsoidal distortion, when applicable.}
\end{splitdeluxetable*}

\subsection{HAT-P-7}\label{hatp7}

The HAT-P-7 system consists of a highly irradiated 1.8~$M_{\mathrm{Jup}}$, 1.4~$R_{\mathrm{Jup}}$ gas giant that lies on a nearly pole-on 2.205~d orbit around an evolved F6 star with an effective temperature of 6350~K \citep{hatp7,winn2009,narita2009}. The brightness of the host star ($T=10.0$~mag, $V=10.5$~mag) has made HAT-P-7 an attractive candidate for both ground- and space-based atmospheric characterization. This system is also located within the Kepler field of view, and the full-orbit Kepler phase curve has been analyzed by several authors \citep{borucki2009,esteves2015,armstrong2016}. TESS observed HAT-P-7 in sectors 14 and 15.

From our phase-curve analysis, we obtained a strong detection of the secondary eclipse ($D_{d}=127^{+33}_{-32}$~ppm) and the atmospheric brightness modulation ($A_{\mathrm{atm}}=56^{+14}_{-13}$~ppm). The corresponding nightside flux is consistent with zero. No significant offset in the atmospheric phase-curve variation was measured ($\delta=5^{\circ}\pm 12^{\circ}$). This is consistent with the results from the Kepler phase-curve analysis, which found a small but statistically robust offset of $7\overset{\circ}{.}0\pm0\overset{\circ}{.}3$ \citep{esteves2015}. Meanwhile, phase curves of HAT-P-7b obtained in the Spitzer 3.6 and 4.5~$\mu$m bands show insignificant but formally consistent eastward shifts in the dayside hotspot of $7\overset{\circ}{.}0\pm7\overset{\circ}{.}5$ and $4\overset{\circ}{.}1\pm7\overset{\circ}{.}5$, respectively \citep{wong2016}. 

The ellipsoidal distortion and Doppler boosting signals were not detected in an unconstrained fit. For this and all other analogous cases, we plugged the measured values for $T_{\mathrm{eff}}$, $M_{p}$, $q$, $P$, $a/R_*$, and $i$ from the discovery papers into Equations~\eqref{ellip} and \eqref{dopp} to derive the predicted values $A_{\mathrm{ellip}}$ and $A_{\mathrm{Dopp}}$, which we used as priors in the final fit. For HAT-P-7, we obtained $A_{\mathrm{ellip}}=16\pm4$~ppm and $A_{\mathrm{Dopp}}=2.2\pm0.1$~ppm. The measured orbital ephemeris, transit-shape, and transit-depth parameter values are consistent with the measurements reported in the discovery paper \citep{hatp7} to within $1\sigma$. Our results also agree with the more precise planetary parameters from the Kepler phase-curve analysis in \citet{esteves2015}. The full set of results from our light-curve fit is provided in Table~\ref{tab:fit}. Marginal detections (i.e., those that yielded increases in the AIC upon inclusion in the model) are indicated with parentheses, while parameters that were constrained by Gaussian priors are shown with square brackets. We used the posteriors from our MCMC fit to compute various derived parameters: inclination $i$, standard quadratic limb-darkening coefficients $(u_1,u_2)$, orbital semimajor axis $a$, and planetary radius $R_{p}$.

\begin{figure}[t]
\includegraphics[width=\linewidth]{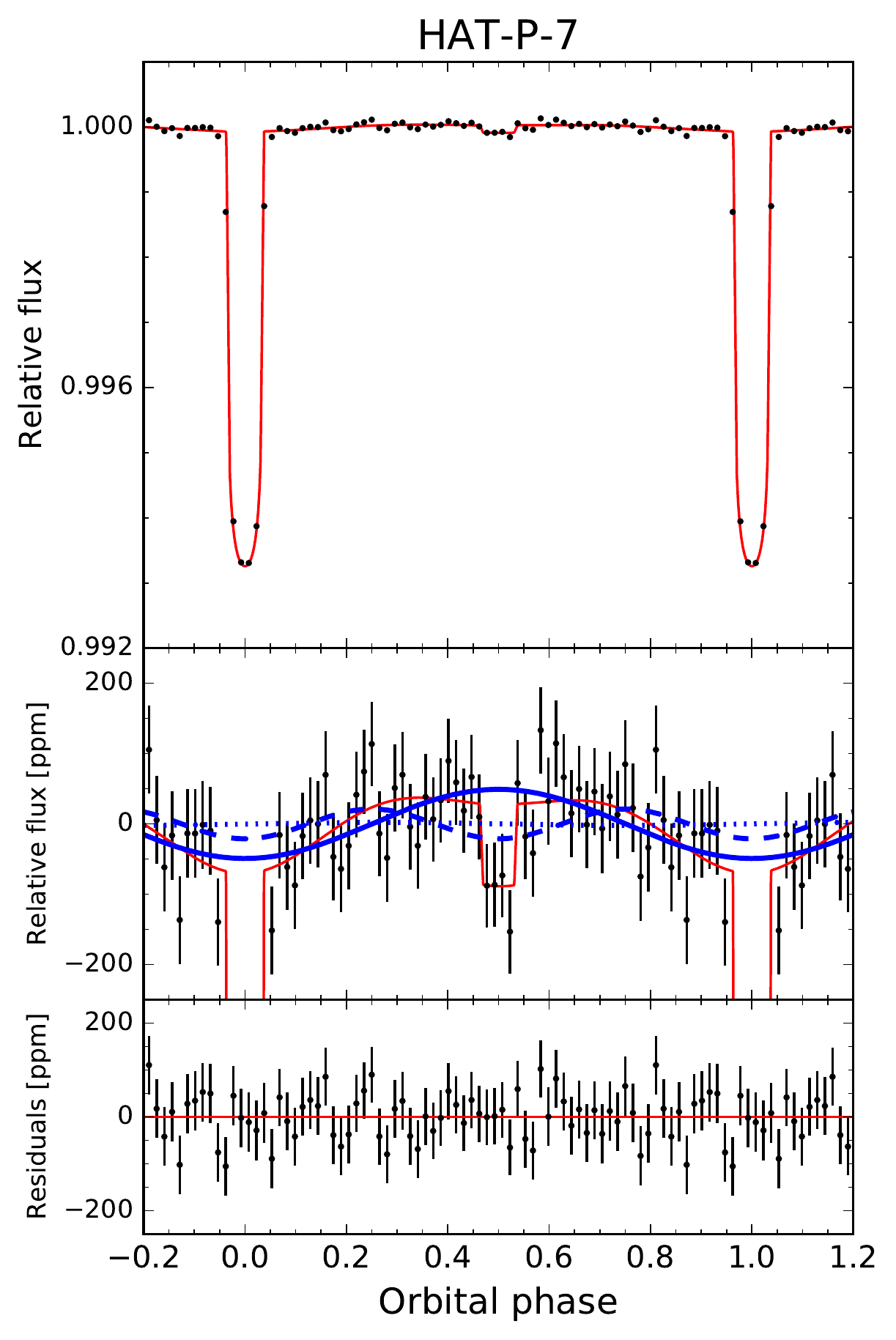}
\caption{Top: systematics-removed TESS light curve of HAT-P-7, phase-folded and binned in 40~minute intervals (black points). Here, and in all subsequent plots, orbital phase is given as a fraction of the orbital period. The best-fit full phase-curve model from the joint fit analysis is plotted in red. Middle: expanded view of the phase-curve variations, with the atmospheric modulation, ellipsoidal distortion, and Doppler boosting components overplotted in the solid, dashed, and dotted blue curves, respectively. Bottom: the corresponding residuals from the best-fit phase-curve model.}
\label{fig:first}
\end{figure}

Zoomed-in views of the systematics-corrected, phase-folded, and binned transit light curves for HAT-P-7 and all other systems are compiled in Figure~\ref{fig:transit}. Figure~\ref{fig:first} displays the full phase-folded TESS phase curve of HAT-P-7 and the corresponding residuals from the best-fit model; the middle panel shows the three components of the phase-curve model in blue. The binning interval is chosen so as to yield roughly 75 bins spanning the orbital period.

In their Kepler phase-curve analysis, \citet{esteves2015} found a significant phase-curve signal at the second harmonic of the orbital period (i.e., $\cos(3\phi)$ and $\sin(3\phi)$), which has a semiamplitude of around 2~ppm. This additional variability may be attributable to the spin-orbit misalignment, which causes the tidal bulge to traverse regions of the star's surface that have different surface gravities and temperatures. For the TESS light curve, the amplitude of this signal is dwarfed by the uncertainties on the phase-curve amplitudes.

Another notable result from the Kepler phase-curve study of HAT-P-7 was the detection of temporal variations in the offset between the location of peak brightness and the substellar point \citep[][but see also \citealt{lally2020}]{armstrong2016}. Specifically, it was reported that the direction of the phase-curve offset repeatedly shifts between westward and eastward on a timescale of tens to hundreds of days. The TESS observations of this system spanned two sectors ($\sim$55~d). While the photometric precision of the TESS light curves and the corresponding sensitivity of the phase-curve results are significantly lower than in the case of Kepler, we nevertheless carried out individual fits of each sector's light curve. We obtained mutually consistent amplitude and phase-shift values for sectors 14 and 15: $A_{\mathrm{atm},14}=44\pm20$~ppm, $A_{\mathrm{atm},15}=64\pm21$~ppm, $\delta_{14}=8^{\circ}\pm21^{\circ}$, and $\delta_{15}=2^{\circ}\pm16^{\circ}$. However, we note that even when jointly fitting $\sim$10 orbits of the system within each sector, the precision of our measured phase shifts still dwarfs the standard deviation of the individual Kepler-band offsets measured by \citet{armstrong2016}: $12^{\circ}$.

\subsection{KELT-1}\label{kelt1}

\begin{figure}[t]
\includegraphics[width=\linewidth]{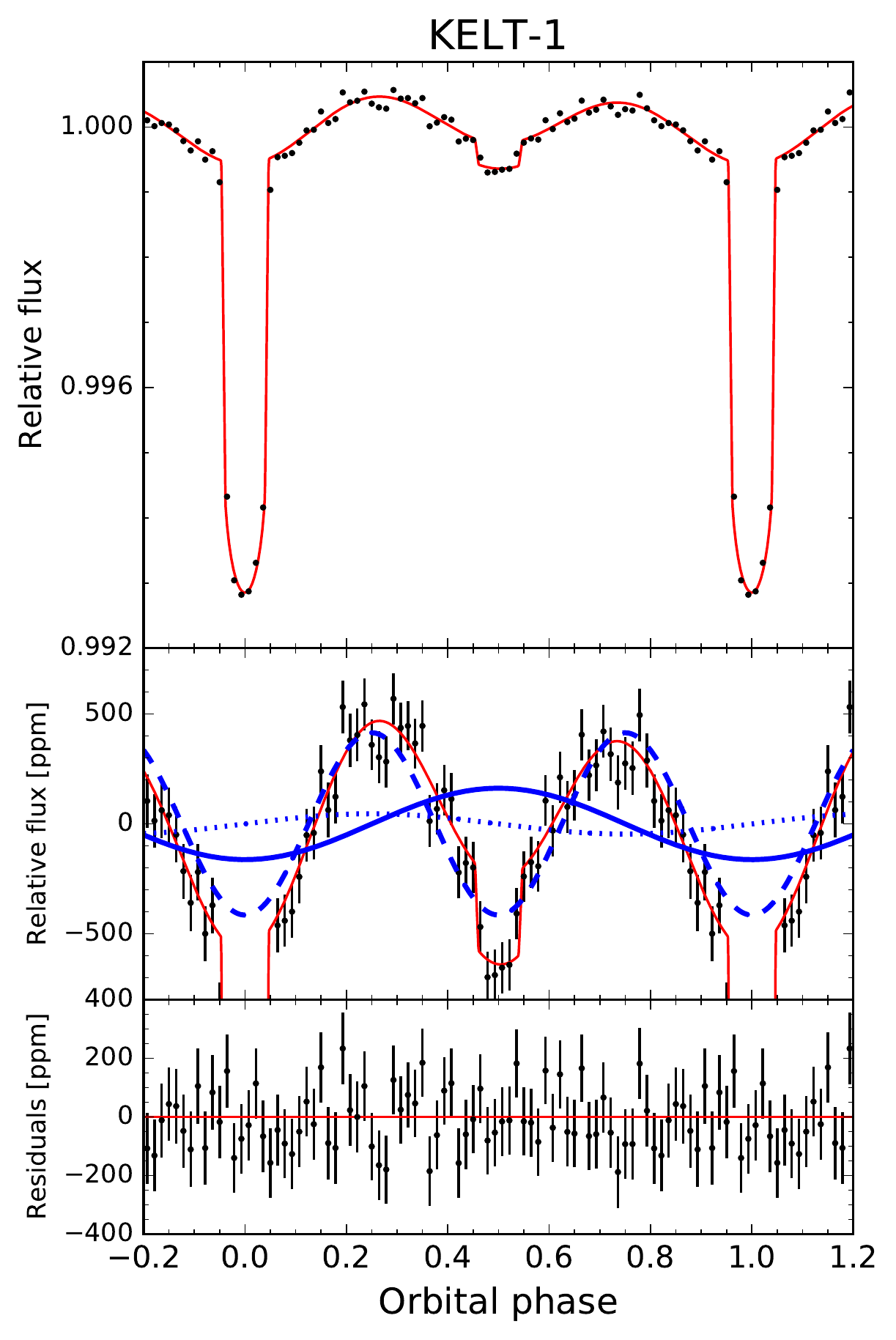}
\caption{Same as Figure~\ref{fig:first}, but for KELT-1. The phase-folded light curve is binned in 25~minute intervals.}
\label{fig:kelt1}
\end{figure}

KELT-1b is a 27~$M_{\mathrm{Jup}}$ brown dwarf on a 1.22~d low-obliquity orbit around an evolved mid-F star \citep{kelt1}. The system was observed by TESS in sector 17, and the full-orbit phase curve was previously studied in two independent analyses \citep{beatty2020,vonessen2020kelt1}. In this work, we utilized the updated photometry from the SPOC pipeline to reanalyze the light curve.

We measured a secondary eclipse depth of $388^{+67}_{-65}$~ppm, an atmospheric phase-curve modulation with a semiamplitude of $176^{+29}_{-30}$~ppm, and a nightside flux of $39^{+70}_{-72}$~ppm. Our eclipse depth is consistent with the ground-based $z'$-band ($\lambda_{\mathrm{eff}}=892$~nm) measurement of $490\pm230$ reported by \citet{beatty2014}. We did not find a significant phase-curve offset in the TESS phase curve. The strongest phase-curve component in the data is the ellipsoidal distortion signal, which has a semiamplitude of $416^{+25}_{-36}$~ppm. The Doppler boosting modulation was not detected in an unconstrained fit, and $A_{\mathrm{Dopp}}$ was constrained by a Gaussian prior in the final fit based on the predicted value: $43\pm2$~ppm. 

All of our phase-curve parameter values are consistent with those measured by the earlier analyses, demonstrating that the updated photometry and any differences in data-analysis methodology did not have any substantive effect on the conclusions of our phase-curve fit. In particular, the published secondary eclipse depths from \citet{beatty2020} and \citet{vonessen2020kelt1} --- $371^{+47}_{-49}$ and $320\pm69$~ppm, respectively --- agree with our value at much better than the $1\sigma$ level. Using Equation~\eqref{ellip} and the system parameters presented in \citet{kelt1}, we computed a predicted ellipsoidal distortion semiamplitude of $460\pm40$, which agrees with our measured value at the $0.9\sigma$ level. We utilized the same equation to arrive at an independent photometric constraint on the brown dwarf's mass based on the measured ellipsoidal distortion amplitude: $25.1^{+3.7}_{-3.3}$~$M_{\mathrm{Jup}}$.

The full results from our phase-curve analysis of KELT-1 are listed in Table~\ref{tab:fit}. The full-orbit phase-folded light curve is shown in Figure~\ref{fig:kelt1}.

Full-orbit Spitzer phase-curve observations of the KELT-1 system were presented in \citet{beatty2019}. Significant eastward phase offsets in the atmospheric brightness modulation signal were measured in both the 3.6 and 4.5~$\mu$m bandpasses --- $28\overset{\circ}{.}6\pm3\overset{\circ}{.}8$ and $18\overset{\circ}{.}5\pm5\overset{\circ}{.}1$, respectively. These values are larger than the marginal phase-curve offset that we measured from the TESS phase curve, suggesting that the infrared bandpasses are probing regions of the atmosphere with more efficient longitudinal heat transport and/or longer atmospheric radiative timescales than the optical wavelength observations \citep[e.g.,][]{showman2002,komacek2016}.

\subsection{KELT-16}\label{kelt16}

The ultra-short-period transiting system KELT-16, which contains a massive, inflated, 2.75~$M_{\mathrm{Jup}}$ hot Jupiter and an F7V star with $T_{\mathrm{eff}}=6236\pm54$~K \citep{kelt16}, was observed by the TESS spacecraft in sector 15. Similar to the case of HAT-P-7, we did not independently measure any phase-curve variability attributed to ellipsoidal distortion and Doppler boosting, and we applied priors to the corresponding amplitudes, which have predicted semiamplitudes of $72\pm10$ and $5.2\pm0.3$~ppm, respectively. We detected a secondary eclipse with a depth of $410^{+130}_{-120}$~ppm and an atmospheric brightness modulation signal with a semiamplitude of $175^{+64}_{-62}$~ppm and no offset. From the standpoint of statistical significance, this system displays the weakest signals among the 7 targets that had robust secondary eclipse and phase-curve detections.

\begin{figure}[t]
\includegraphics[width=\linewidth]{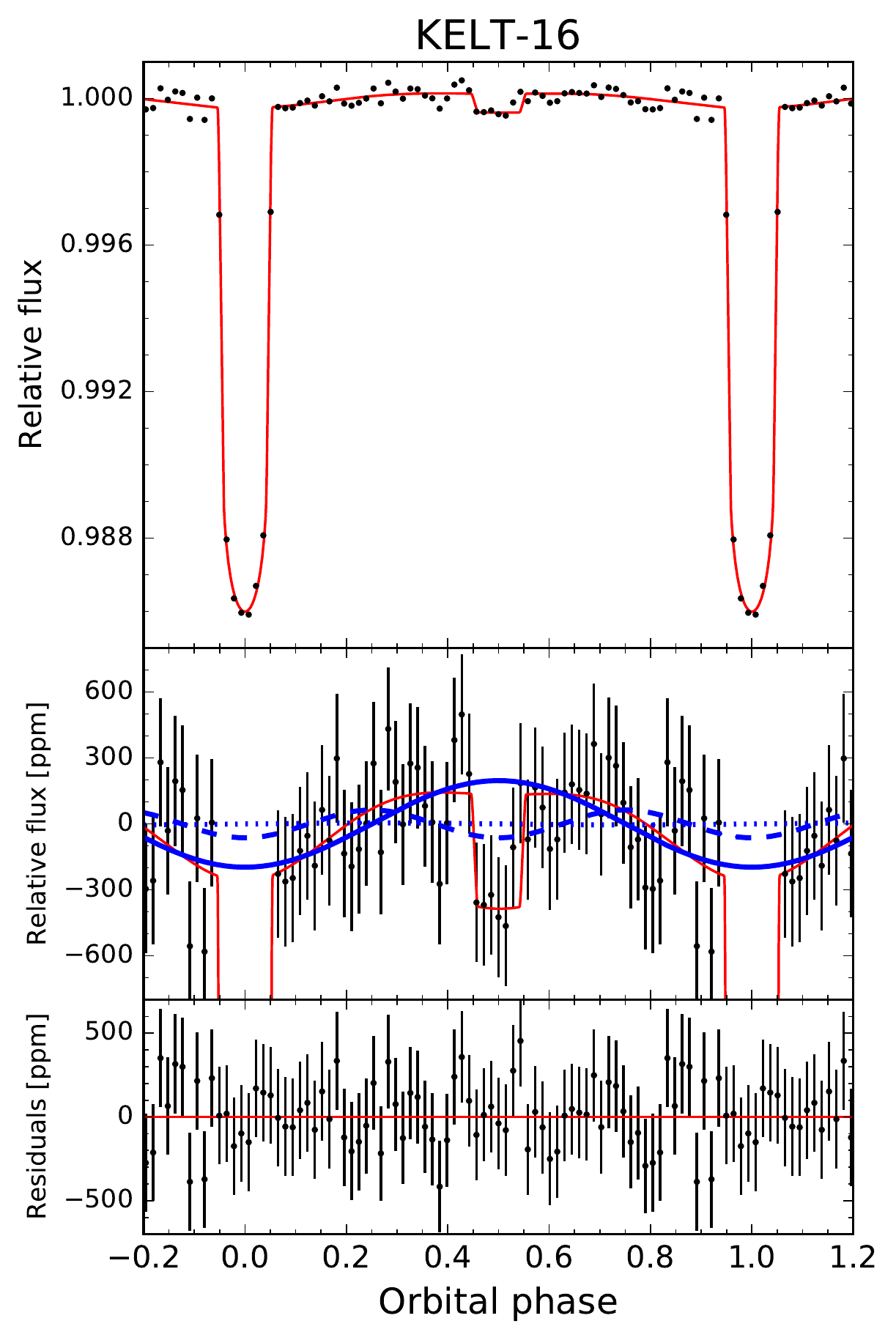}
\caption{Same as Figure~\ref{fig:first}, but for KELT-16. The phase-folded light curve is binned in 20~minute intervals.}
\label{fig:kelt16}
\end{figure}

Table~\ref{tab:fit} provides the full results from our light-curve analysis. The other system parameters are all consistent with the values from the discovery paper to well within $1\sigma$. Figure~\ref{fig:kelt16} shows the phase-folded light curve.

The Spitzer 4.5~$\mu$m full-orbit phase curve of KELT-16b was published by \citet{bell2020}. Dayside and nightside temperatures of $3030^{+150}_{-140}$ and $1520^{+410}_{-360}$~K were measured, while the atmospheric brightness modulation showed an unusual $30^{\circ}\pm13^{\circ}$ westward offset in the location of the dayside hotspot. The statistically insignificant phase-curve offset from our TESS light-curve analysis ($6^{+18}_{-19}$~deg) is formally consistent with the Spitzer value at the $1.6\sigma$ level.

\citet{mancini2021} published an independent analysis of the KELT-16 TESS phase curve. Using the PDC light curve, they modeled the brightness distribution across the planet's atmosphere as a dipole and derived a secondary eclipse depth of $434\pm42$~ppm, which is statistically identical to our value. From their analysis, they also obtained a marginal eastward offset in the dayside hotspot of $25^{\circ}\pm14^{\circ}$ --- consistent with our value at better than the $1\sigma$ level. We note that their analysis did not account for red noise or systematics modeling, which likely contributed to the significantly smaller uncertainty on the secondary eclipse depth.

\subsection{KELT-20}\label{kelt20}

\begin{figure}[t]
\includegraphics[width=\linewidth]{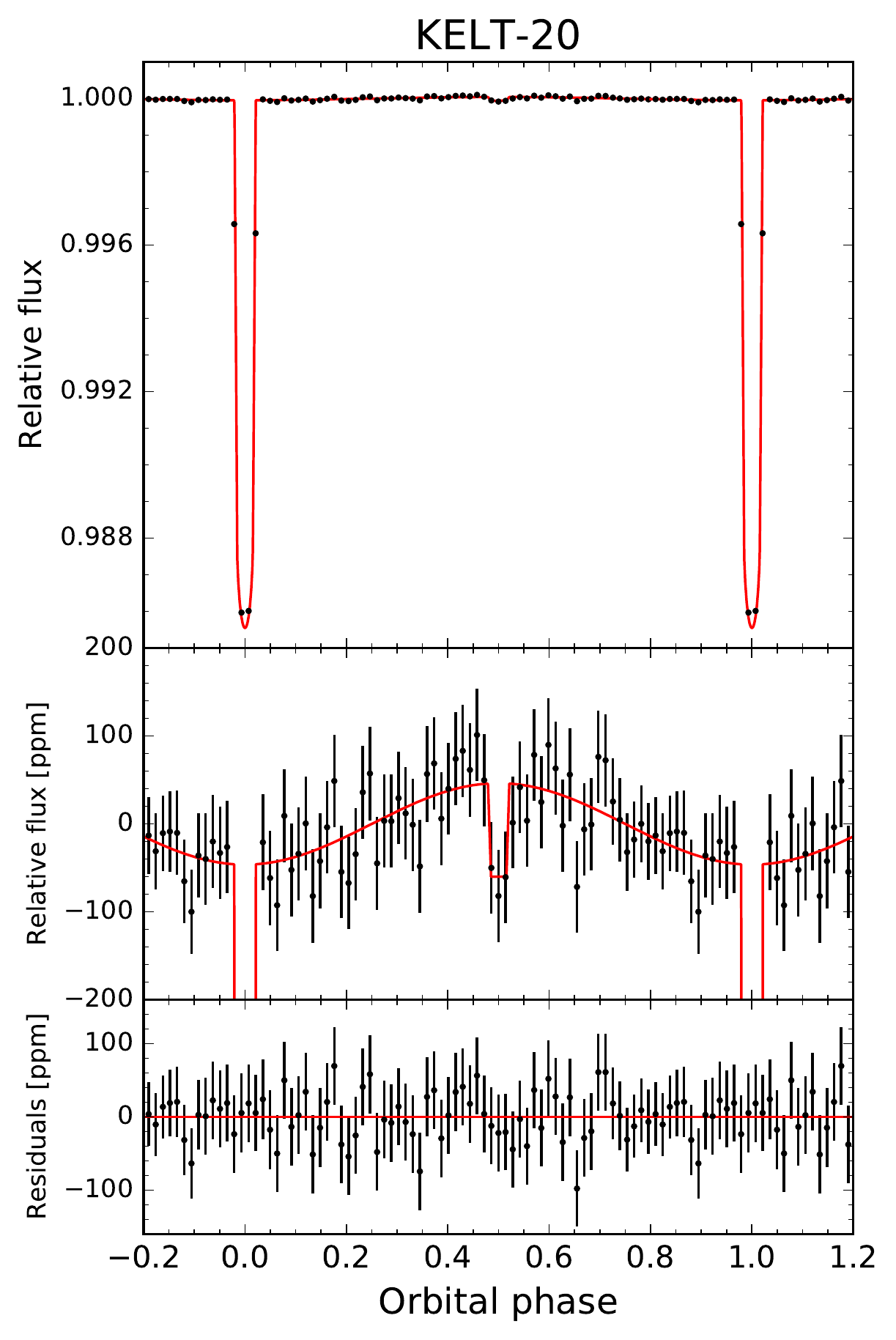}
\caption{Same as Figure~\ref{fig:first}, but for KELT-20. The phase-folded light curve is binned in 65~minute intervals. Due to the lack of rigid constraints on the planetary mass, no priors on the ellipsoidal distortion or Doppler boosting phase-curve amplitudes were applied.}
\label{fig:kelt20}
\end{figure}

This system was discovered independently by the Kilodegree Extremely Little Telescope (KELT) survey as KELT-20 \citep{kelt20} and by the Multi-site All-Sky CAmeRA (MASCARA) as MASCARA-2 \citep{mascara2}. The host star is an A star with an effective temperature of 8730~K and a mass of 1.8~$M_{\Sun}$. The highly irradiated transiting planet is a 1.7~$R_{\mathrm{Jup}}$ hot Jupiter on a 3.47~d orbit with a $3\sigma$ mass upper limit of about 3.5~$M_{\mathrm{Jup}}$. TESS observations of this bright $T=7.6$~mag target occurred during sector 14.

No discernible ellipsoidal distortion or Doppler boosting signal was found in the light curve, and given the poorly constrained mass from the discovery papers, we did not apply priors to the respective phase-curve amplitudes in the final fit. The measured secondary eclipse depth and atmospheric brightness modulation semiamplitude are $111^{+35}_{-36}$ and $43^{+13}_{-11}$~ppm, respectively. As with the previous targets, no statistically significant phase offset was detected.

The planet radius of $1.761\pm0.069$~$R_{\mathrm{Jup}}$ that we derived from the TESS light-curve fit is consistent with the values listed in both discovery papers. Our measurements of $i$ and $a/R_{*}$ are slightly more precise than the previously published values, while being consistent at better than $1\sigma$. The full results are given in Table~\ref{tab:fit}; the best-fit phase-curve model and phase-folded light curve are plotted in Figure~\ref{fig:kelt20}.

\subsection{Kepler-13A}\label{kepler13}

Kepler-13A was identified as a candidate planet host early on in the Kepler mission, with subsequent works confirming the existence of a highly-irradiated hot Jupiter on a 1.76~d orbit. The availability of long-baseline Kepler data also yielded a well-characterized phase-curve signal and an estimate of the mass ratio from the ellipsoidal distortion amplitude \citep[e.g.,][]{shporer2011,szabo2011,mazeh2012}. Follow-up imaging with adaptive optics uncovered a bound system of two A-type stars --- Kepler-13A and Kepler-13B --- with the brighter star Kepler-13A hosting the detected transiting planet and the fainter secondary orbited by a third late-type star Kepler-13BB \citep{santerne2012}. High-resolution spectra of the two binary components revealed that Kepler-13A has an effective temperature of $7650\pm250$~K, a roughly solar metallicity of $0.2\pm0.2$, and a mass of $1.72\pm0.10$~$M_{\Sun}$ \citep{shporer2014}.

An analysis of the full four-year Kepler phase curve was carried out by \citet{esteves2015}, who produced high signal-to-noise measurements of phase-curve amplitudes corresponding to all three processes --- atmospheric brightness modulation, ellipsoidal distortion, and Doppler boosting. In addition, \citet{shporer2014} presented secondary eclipse depths in the Spitzer 3.6 and 4.5~$\mu$m bands, as well as in the $K_{s}$ band (2.1~$\mu$m); they showed that the planet's dayside emission spectrum is consistent with a blackbody brightness temperature of $2750\pm160$~K and an elevated optical geometric albedo $A_{g}=0.33^{+0.04}_{-0.06}$.

TESS observed the Kepler-13A system in sectors 14, 15, and 26. The PDC photometry was corrected for the contamination from Kepler-13B by the SPOC pipeline. The transit light curves from the first two sectors were previously analyzed in \citet{szabo2020}, which presented a refined transit ephemeris. That work also confirmed earlier reports of a time-varying impact parameter --- a consequence of orbital precession excited by the oblate star and the significant spin-orbit misalignment of the system \citep{johnson2014,masuda2015}. In our TESS light-curve fit of all three sectors of data, we did not allow for a time-varying impact parameter, since the measured yearly drift of $\Delta b=-0.011$ is significantly smaller than the uncertainty on $b$ from the fit.

\begin{figure}[t]
\includegraphics[width=\linewidth]{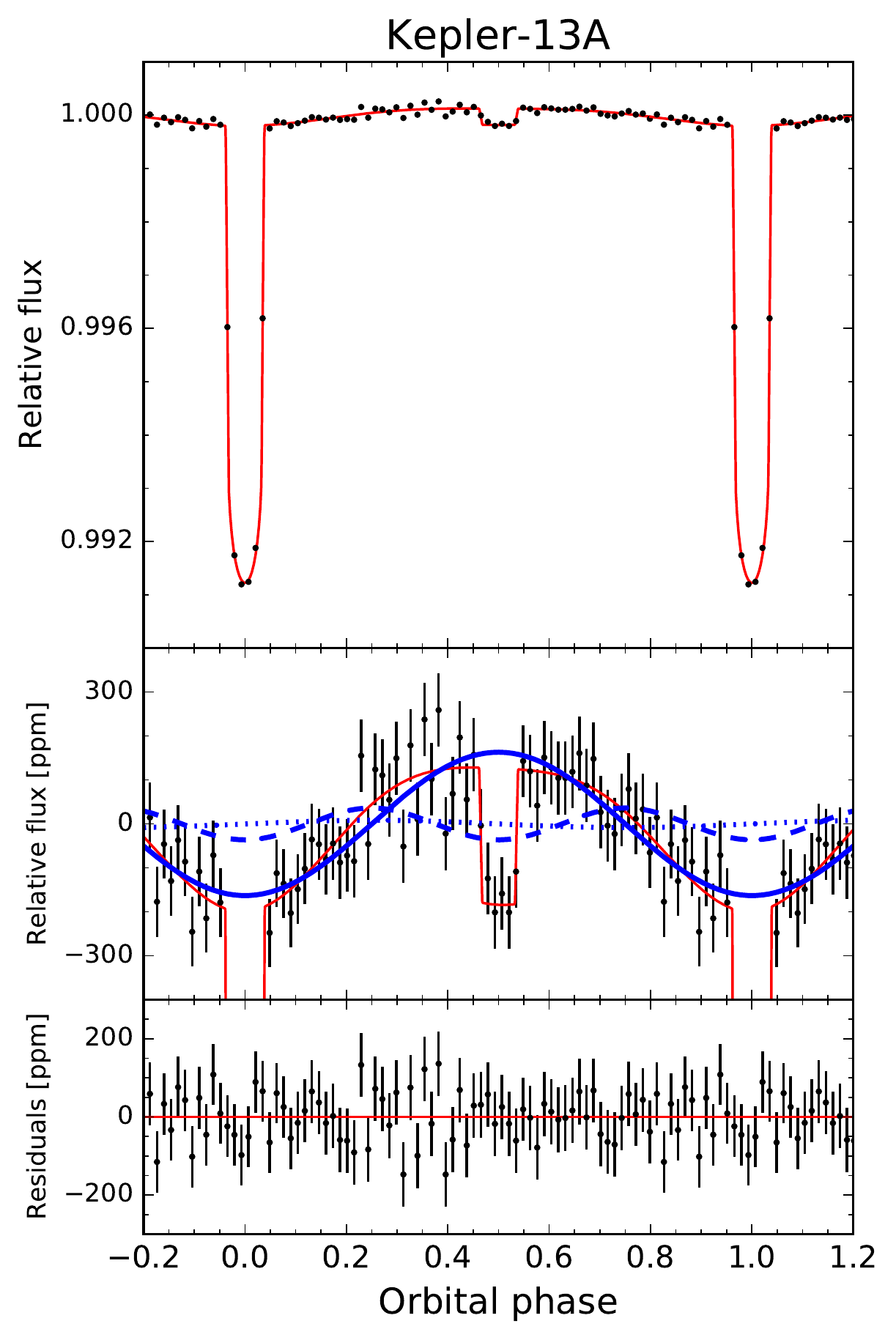}
\caption{Same as Figure~\ref{fig:first}, but for Kepler-13A. The phase-folded light curve is binned in 35~minute intervals.}
\label{fig:kepler13}
\end{figure}

We detected significant atmospheric brightness modulation and ellipsoidal distortion semiamplitudes of $151^{+15}_{-16}$ and $49^{+17}_{-16}$~ppm, respectively. Meanwhile, the Doppler boosting amplitude was constrained by a Gaussian prior based on the theoretical value: $6.8\pm1.7$~ppm. We measured a secondary eclipse depth of $301^{+46}_{-42}$~ppm and a nightside flux that is consistent with zero. Our phase-curve fit that included an offset in the atmospheric brightness modulation yielded a marginal eastward shift in the dayside hotspot of $8\overset{\circ}{.}9^{+5\overset{\circ}{.}0}_{-4\overset{\circ}{.}6}$. However, just as in the case of the Kepler phase curve, the model with zero offset is statistically favored. The values of $R_{p}/R_{*}$ and $a/R_{*}$ from our TESS light-curve fit (Table~\ref{tab:fit}) agree at much better than $1\sigma$ with the extremely precise values measured from the full Kepler light-curve fit in \citet{esteves2015}. The phase-folded TESS light curve is shown in Figure~\ref{fig:kepler13}.

Similar to the case of HAT-P-7, the Kepler phase curve of Kepler-13A showed an additional modulation of the host star's brightness at the second harmonic of the orbital period with a semiamplitude of around 7~ppm \citep{esteves2013,esteves2015,shporer2014}. This is well within the noise of the TESS photometry, and we did not retrieve any significant signal at this harmonic from an unconstrained fit.  

Across the temperature range spanned by the published stellar temperature and uncertainties, the gravity-darkening coefficient varies significantly \citep{claret2017}. As such, we were unable to derive a precise predicted ellipsoidal distortion amplitude. Likewise, the highly uncertain gravity-darkening profile of the star means that our measured ellipsoidal distortion amplitude is consistent with a wide range of planet masses: 4--11~$M_{\mathrm{Jup}}$.

\subsection{WASP-12}\label{wasp12}

TESS observations of WASP-12 took place during sector 20. This system consists of a 1.4~$M_{\mathrm{Jup}}$ hot Jupiter on a roughly one-day orbit around a late-F star with $T_{\mathrm{eff}}=6300$~K \citep{wasp12}. While WASP-12 is relatively faint ($T=11.1$~mag), the extreme dayside irradiation has made it an attractive target for atmospheric characterization, particularly in emission. Secondary eclipse observations have been carried out in all four channels of Spitzer/IRAC \citep{campo2011,madhusudhan2011,cowan2012,stevenson2014}, as well as with Hubble/WFC3 \citep{stevenson2014}. In addition, full-orbit Spitzer light-curve fits at 3.6 and 4.5~$\mu$m were published in \citet{cowan2012} and \citet{bell2019}. Long-term transit monitoring of WASP-12b has revealed significant orbital decay \citep[][see Section \ref{subsec:ephem}]{patra2017,yee2020}. In our TESS phase-curve analysis, we did not consider a time-varying ephemeris, given that the predicted period shortening across one 27-day sector of TESS observations is only $\sim$2.5~ms.

\begin{figure}[t]
\includegraphics[width=\linewidth]{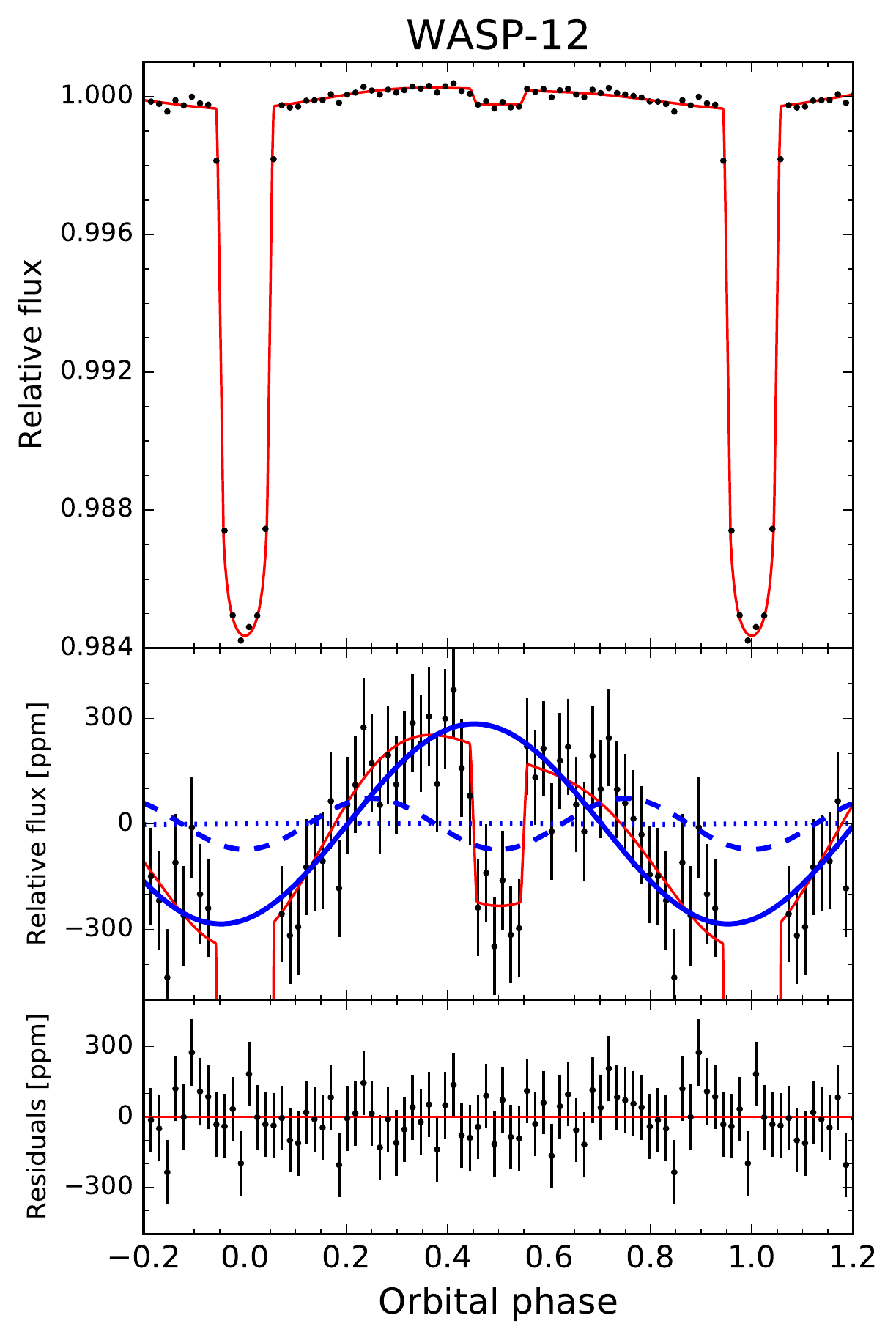}
\caption{Same as Figure~\ref{fig:first}, but for WASP-12. The phase-folded light curve is binned in 20~minute intervals.}
\label{fig:wasp12}
\end{figure}

We measured an atmospheric brightness modulation with a semiamplitude of $264^{+33}_{-30}$~ppm that is shifted eastward by $13\overset{\circ}{.}2\pm5\overset{\circ}{.}7$. The secondary eclipse depth is $443^{+86}_{-85}$~ppm, and the nightside flux is consistent with zero. We also detected a statistically significant ellipsoidal distortion component with $A_{\mathrm{ellip}}=80^{+33}_{-35}$~ppm. All of the transit-depth and orbital parameters ($R_{p}/R_{*}$, $a/R_{*}$, $b$; Table~\ref{tab:fit}) are consistent with the values in the discovery paper \citep{wasp12} at better than the $1\sigma$ level. Figure~\ref{fig:wasp12} shows the binned and phase-folded TESS light curve, from which the atmospheric brightness and ellipsoidal distortion phase-curve signals are clearly discernible.

In their analysis of the Spitzer phase curves, \citet{bell2019} measured an eastward offset in the dayside hotspot of $12\overset{\circ}{.}0\pm2\overset{\circ}{.}0$ at 4.5~$\mu$m (averaged between the 2010 and 2013 observations), which is consistent with the TESS-band value we derived from our analysis. Meanwhile, the individual phase-curve offset measurements at 3.6~$\mu$m differed significantly, with the 2010 epoch showing an eastward shift, while the 2013 epoch displayed a westward offset.

Comparing the predicted ellipsoidal distortion semiamplitude ($37\pm8$~ppm) with our measured value, we find a slight $1.2\sigma$ discrepancy. Similarly, we derived a planetary mass of $3.0\pm1.3$~$M_{\mathrm{Jup}}$ from the measured ellipsoidal distortion amplitude, which differs from the mass listed in the discovery paper ($1.41\pm0.10$~$M_{\mathrm{Jup}}$) at the same significance level. 

\citet{bell2019} reported a first-harmonic phase-curve modulation at 4.5~$\mu$m that is much larger than expected. Based on the lack of an analogous variation in the 3.6~$\mu$m phase curves, they concluded that the additional first-harmonic amplitude may be due to gas outflow from the atmosphere of WASP-12b into the host star. In this configuration, the stream of escaping gas is viewed edge-on during quadrature, with the thermal emission from heated CO gas contributing to the extra brightness modulation across the orbit primarily in the 4.5~$\mu$m bandpass. 

Given this mass-loss hypothesis, we propose that the somewhat larger-than-expected first-harmonic phase-curve signal we measured in the TESS light curve might be caused by scattered starlight off condensates and/or aerosols in the gas stream, or thermal emission from superheated gas accreting onto the host star. In order to effectively probe whether the deviation in the TESS-band first-harmonic phase-curve signal from the predicted amplitude is indeed significant, additional photometry from the TESS extended mission is needed.

In addition to the anomalous Spitzer 4.5~$\mu$m phase curve, several authors have reported possible time variability in the secondary eclipse depth. \citet{hooton2019} obtained two $i'$-band observations from two different ground-based telescopes and measured eclipse depths that differed from each other by more than $2\sigma$. Likewise, \citet{vonessen2019} obtained a pair of $V$-band secondary eclipse depths that are mutually discrepant at the $4.3\sigma$ level. While instrumental systematics and observing conditions may be the source of some or all of these eclipse-depth mismatches, some level of orbit-to-orbit variability could also be present, especially in the context of the aforementioned mass-loss hypothesis.

To explore the possibility of time-varying eclipse depths in the TESS light curve, we fit each secondary eclipse separately. First, we divided the best-fit systematics model from the light curve and removed the measured atmospheric brightness modulation and ellipsoidal distortion signals. Next, we constructed individual secondary eclipse light curves by selecting data points within 0.1 in orbital phase of each superior conjunction. There are 20 secondary eclipses that lie entirely within the time series and do not contain large gaps due to momentum dumps or trimmed flux ramps. We fit these light curves with the \texttt{batman} occultation model, while fixing all system parameters except the eclipse depth to the median values from the full MCMC light-curve fit.

\begin{figure}[t]
\includegraphics[width=\linewidth]{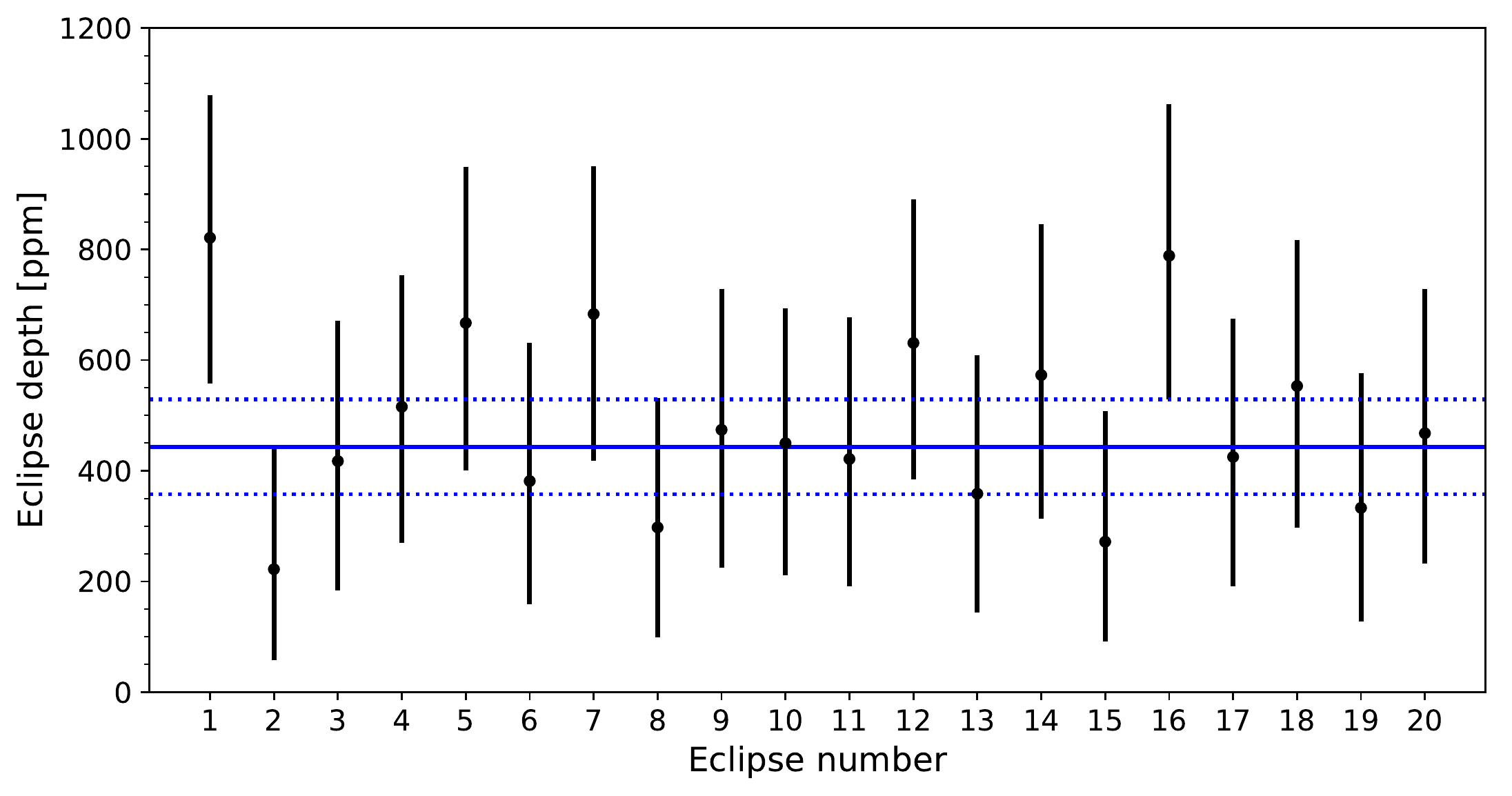}
\caption{Eclipse depth measurements for each of the 20 full secondary eclipses contained within the TESS light curve of WASP-12. The blue horizontal lines indicate the median global depth and $1\sigma$ bounds from the full light-curve fit (Table~\ref{tab:fit}). All individual depths are consistent with the global value to within $1.5\sigma$.}
\label{fig:wasp12ecl}
\end{figure}

Figure~\ref{fig:wasp12ecl} shows the measured eclipse depths for each epoch. All of the individual depths lie within $1.5\sigma$ of the global value from the full light-curve fit. The standard deviation of the individual eclipse depths is 160~ppm, while the mean uncertainty is 240~ppm. The highest and lowest eclipse depth measurements (from the first two epochs) are consistent at the $1.8\sigma$ level. Therefore, we do not find any evidence for significant orbit-to-orbit variability in the dayside brightness of WASP-12b to the level of precision in the TESS data. 

\citet{owens2021} carried out an independent analysis of the TESS phase curve of WASP-12, using both the SAP and the PDC light curves. While their methodology differed in several substantive ways, e.g., applying a Gaussian prior to the planet mass for the ellipsoidal distortion signal modeling and assuming a reflection-dominated atmospheric brightness modulation, their results are broadly consistent with our values. In the case of the PDC light-curve analysis, they obtained a secondary eclipse depth of $577^{+71}_{-72}$~ppm and a planetary phase curve semiamplitude of $265\pm30$~ppm, which agree with the corresponding values from our work ($443^{+86}_{-85}$ and $264^{+33}_{-30}$~ppm) at the $1.2\sigma$ and $<0.1\sigma$ levels, respectively. Meanwhile, the secondary eclipse depth they obtained from the SAP light-curve fit is $609^{+74}_{-73}$~ppm --- $1.5\sigma$ larger than our value. Notably, \citet{owens2021} also reported a marginal eastward phase shift in the dayside hotspot that agrees with our measurement at better than the $1\sigma$ level ($17\overset{\circ}{.}6\pm5\overset{\circ}{.}4$ vs. $13\overset{\circ}{.}2\pm5\overset{\circ}{.}7$).

\subsection{KELT-9}\label{kelt9}

The most extreme of the ultra-hot Jupiters discovered to date, KELT-9b is a massive $\sim$3~$M_{\mathrm{Jup}}$ gas giant on a near-polar 1.48~d orbit around a $\sim$10,000~K 2.3~$M_{\Sun}$ A0/B9 star \citep{kelt9}. An analysis of the Spitzer 4.5~$\mu$m full-orbit phase curve indicated a dayside brightness temperature of around 4500~K and a relatively low day--night temperature contrast attributed to the dissociation and recombination of H$_{2}$ \citep{mansfield2020}. Due to the rapid rotation of the host star and the large spin-axis misalignment, the transit light curves of this system show significant aberrations due to gravity darkening. A previous analysis by \citet{ahlers2020} was dedicated to the detailed modeling of the TESS transit light curves. 

\citet{wong2020kelt9} studied the initially released sector 14 and 15 TESS SPOC light curves of KELT-9 and uncovered several unusual features. First, they detected an unexpected signal at the first harmonic of the orbital period, with the overall modulation significantly offset from the expected phase alignment for ellipsoidal distortion. Using the gravity-darkening model derived by \citet{ahlers2020}, they proposed that this shift is caused by the rotational deformation of the host star and the near-polar orbit of KELT-9b, which results in time-varying insolation of the dayside hemisphere with two maxima and two minima per orbit. This interpretation was supported by numerical modeling of the time-varying stellar irradiation from the gravity-darkening analysis and an analogous signal measured from the planet's observed thermal phase curve \citep{mansfield2020}. The second peculiarity was a marginal detection of a phase-curve signal at the second harmonic of the orbital phase, similar to what was reported from the Kepler phase-curve analyses of HAT-P-7 and Kepler-13A. Lastly, they detected a significant sinusoidal stellar pulsation signal with a period of 7.59~hr.

As part of our systematic phase-curve analysis of northern targets, we revisited this system using the updated SPOC light curves. From a methodological standpoint, one difference between our current treatment and the analysis in \citet{wong2020kelt9} is the more conservative treatment of red noise in this work. Given the attested presence of an additional source of variability at the first harmonic of the orbital period, as well as the stellar pulsations, we modified the phase-curve modeling described in Section \ref{subsec:model} to fit the KELT-9 light curve. 

For the orbital photometric modulations, we applied two different models. In the first case (fit A), we followed the methods in \citet{wong2020kelt9} and assigned all variations at the first harmonic of the orbital phase to the host star:
\begin{align}
\label{fitaplanet}\psi_{p}^{A} &= \bar{f_{p}} - A_{\mathrm{atm}} \cos(\phi+\delta_{\mathrm{atm}}),\\
\psi_{*}^{A} &= 1+A_{\mathrm{Dopp}}\sin(\phi) \notag \\
\label{fitastar} & \qquad+A_{2}\sin(2\phi)+B_{2}\cos(2\phi).
\end{align}
Here, instead of a single cosine term for ellipsoidal distortion, we used a generic cosine--sine combination to capture the overall variability at the first harmonic; the planet's flux is modeled in the same way as before.

In the second case (fit B), we were motivated by the hypothesis described in \citet{wong2020kelt9}, wherein the additional signal in the first-harmonic modulation is due to temperature variations on the dayside hemisphere of KELT-9b stemming from time-variable stellar irradiation. As such, we added an additional irradiation term to the planet's flux, while keeping the star's flux model identical to the nominal case described in Equation~\eqref{star}:
\begin{align}
\psi_{p}^{B} &= \bar{f_{p}} - A_{\mathrm{atm}} \cos(\phi+\delta_{\mathrm{atm}}) \notag \\
\label{fitbplanet} & \qquad + A_{\mathrm{irrad}} \cos(2\left\lbrack\phi+\delta_{\mathrm{irrad}}\right\rbrack),\\
\label{fitbstar}\psi_{*}^{B} &= 1+A_{\mathrm{Dopp}}\sin(\phi)-A_{\mathrm{ellip}}\cos(2\phi).
\end{align}
The parameters $A_{\mathrm{irrad}}$ and $\delta_{\mathrm{irrad}}$ represent the semiamplitude and phase shift of the additional irradiation signal in the planet's flux. The shape of this model is almost identical to fit A, except for the mid-eclipse flux: in fit A, the shifted first harmonic signal on the star is still visible when the planet is occulted, while in fit B, the irradiation signal is not visible during mid-eclipse.

% Model fit parameters table
%-------------------------------------------------------------------------------
\begin{deluxetable}{lllll}
\tablewidth{0pc}
\tabletypesize{\scriptsize}
\tablecaption{
    Results from KELT-9 Phase-curve Fits
    \label{tab:kelt9}
}

\tablehead{\vspace{-0.2cm}\\ \vspace{-0.2cm}& \multicolumn{2}{c}{\underline{Fit A}\tablenotemark{\scriptsize a}} & \multicolumn{2}{c}{\underline{Fit B}\tablenotemark{\scriptsize a}} \\
    \colhead{Parameter} \vspace{-0.1cm}&
    \colhead{Value}                     &
    \colhead{Error}  &
    \colhead{Value}                     &
    \colhead{Error} 
}
\startdata
\multicolumn{4}{l}{\textit{Orbital and transit parameters}\tablenotemark{\scriptsize b}}   \\
$R_{p}/R_{*}$ & 0.0791 & $_{-0.0018}^{+0.0017}$ & 0.0790 & $_{-0.0019}^{+0.0018}$ \\
$T_{0}$ & 710.10518  & 0.00025 & 710.10522 & 0.00027 \\
$P$ (days) & 1.4811235 & 0.0000010 & 1.4811235 & 0.0000011 \\
$b$ & 0.14 & 0.04 & 0.14 & 0.04 \\
$a/R_{*}$ & 3.18 & 0.03 & 3.18 & 0.03\\
\\
\multicolumn{3}{l}{\textit{Phase-curve parameters}} & &  \\
$\bar{f_p}$ (ppm)    & 356 & $_{-18}^{+17}$ & 308 & 15 \\
$A_{\mathrm{atm}}$ (ppm)   &  271.6 & $^{+9.0}_{-9.2}$ & 271.9 & $^{+9.0}_{-8.9}$\\
$\delta_{\mathrm{atm}}$ ($^{\circ}$) & 2.6 & 1.4 & 2.6 & $_{-1.3}^{+1.4}$\\
$A_{\mathrm{ellip}}$ (ppm)  & \dots & \dots & $\left\lbrack44\right\rbrack$ & $\left\lbrack6\right\rbrack$ \\
$A_{\mathrm{Dopp}}$ (ppm)  & $\left\lbrack2.1\right\rbrack$\tablenotemark{\scriptsize c} & $\left\lbrack0.3\right\rbrack$\tablenotemark{\scriptsize c} & $\left\lbrack2.1\right\rbrack$ & $\left\lbrack0.3\right\rbrack$  \\
$A_{2}$ (ppm)  & $-30.3$ & $_{-6.1}^{+6.0}$ & \dots & \dots\\
$B_{2}$ (ppm)  & $7.7$ & $_{-8.4}^{+8.3}$ & \dots & \dots \\
$A_{\mathrm{irrad}}$ (ppm)  & \dots & \dots & 60.1 & $_{-9.1}^{+9.4}$\\
$\delta_{\mathrm{irrad}}$ ($^{\circ}$)  & \dots & \dots & 16.0 & $_{-3.2}^{+3.7}$\\
\\
\multicolumn{3}{l}{\textit{Stellar pulsation parameters}} & & \\
$\Pi$ (hr) & 7.5851 & 0.0012 & 7.5851 & 0.0011 \\
$\alpha$ (ppm) & $95.5$ & 5.8 & $96.6$ & $_{-6.2}^{+5.9}$\\
$\beta$ (ppm) & $87.3$ & $_{-5.9}^{+5.8}$ & $86.4$ & $_{-6.0}^{+6.2}$ \\
\\
\multicolumn{3}{l}{\textit{Derived parameters}} &  \\
$D_{d}$ (ppm)\tablenotemark{\scriptsize d}  &  627 & $_{-18}^{+17}$ & 630 & $_{-17}^{+18}$ \\
$D_{n}$ (ppm)\tablenotemark{\scriptsize d}   &  84 & $_{-23}^{+21}$ & 87 & 22 \\
\enddata
\textbf{Notes.}
\vspace{-0.25cm}\tablenotetext{\textrm{a}}{Fit A: the combined first harmonic photometric modulation, parameterized by $A_{2}$ and $B_{2}$, is attributed to the stellar flux. Fit B: the ellipsoidal distortion of the host star is assumed to occur according to predictions; the additional flux variation at the first harmonic is the planet's response to time-varying stellar irradiation.}
\vspace{-0.25cm}\tablenotetext{\textrm{b}}{In this fits, these parameters were constrained by Gaussian priors derived from \citet{ahlers2020}. $T_{0}$ is given in $\mathrm{BJD}_{\mathrm{TDB}}-2458000$.}
\vspace{-0.25cm}\tablenotetext{\textrm{c}}{Square brackets denote applied Gaussian priors.}
\vspace{-0.25cm}\tablenotetext{\textrm{d}}{$D_{d}$ and $D_{n}$ are the dayside and nightside fluxes, respectively. The dayside flux is equivalent to the secondary eclipse depth.}
\vspace{-0.7cm}
\end{deluxetable}

To account for the stellar pulsations, we multiplied the combined systematics and phase-curve model with the expression
\begin{equation}F_{\mathrm{puls}}(t) = 1+\alpha\sin(\xi)+\beta\cos(\xi),\end{equation}
where $\xi\equiv2\pi(t-T_{0})/\Pi$, $\Pi$ is the pulsation period, and $\alpha$ and $\beta$ are the coefficients of the sinusoidal pulsation.

Detailed modeling of the gravity-darkened transits is beyond the scope of this paper, and we trimmed the transits from the TESS light curve prior to fitting. We used the results from \citet{ahlers2020} as Gaussian priors to constrain the planet-to-star radius ratio, orbital parameters, and transit ephemeris. In order to measure the phase shift in the atmospheric brightness modulation at the fundamental of the orbital phase, we applied a prior to the Doppler boosting semiamplitude based on the predicted value: $2.1\pm0.3$~ppm. For fit B, we applied an additional prior to the host star's ellipsoidal distortion ($44\pm6$~ppm) to allow for the planet's irradiation signal at the same harmonic to be recovered.

\begin{figure}[t]
\includegraphics[width=\linewidth]{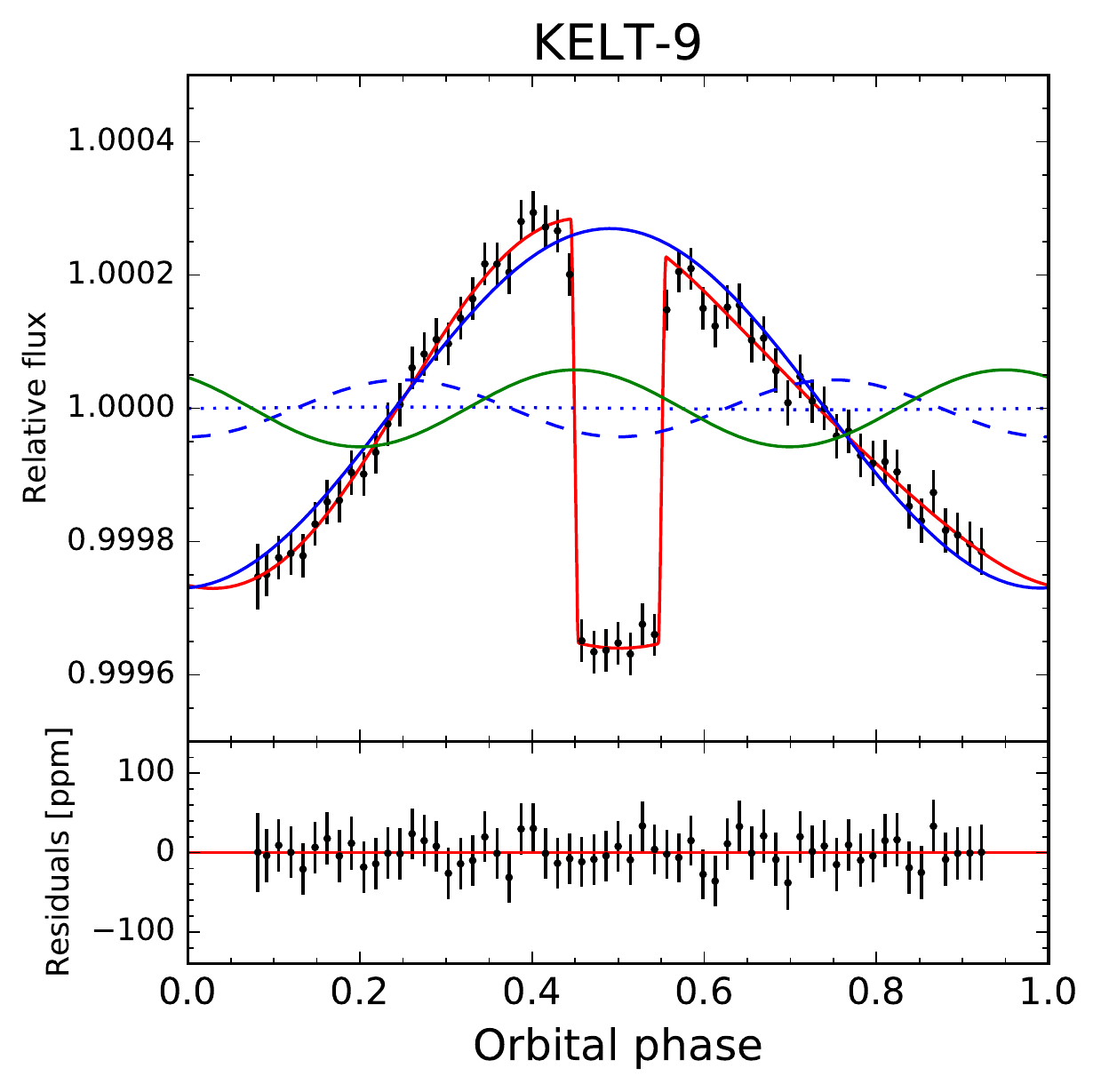}
\caption{Phase-folded and systematics-corrected TESS light curve of KELT-9, binned in 30 minute intervals. The transits have been removed. The solid, dashed, and dotted blue curves are the best-fit atmospheric brightness modulation, ellipsoidal distortion, and Doppler boosting signals. The solid green curve is the additional first-harmonic modulation in the planet's flux due to time-variable irradiation from the oblate host star over the course of its near-polar orbit.}
\label{fig:kelt9}
\end{figure}

The results of our two separate MCMC fits are shown in Table~\ref{tab:kelt9}. The semiamplitude and phase offset of the atmospheric brightness modulation, secondary eclipse depth, and nightside flux from fits A and B are statistically identical, as are the stellar pulsation parameter values. The log-probabilities of the two fits differ by less than 0.5. In the following, we designate fit B as the primary analysis, given the theoretically- and observationally-motivated explanation that the additional signal at the first harmonic stems from the time-varying irradiation of KELT-9b and should therefore be modeled separately from the stellar flux. We measured a secondary eclipse of $630^{+18}_{-17}$~ppm and a nightside flux of $87\pm22$~ppm. The atmospheric brightness modulation, with a semiamplitude of $271.9^{+9.0}_{-8.9}$~ppm, has a phase offset of $2\overset{\circ}{.}6^{+1\overset{\circ}{.}4}_{-1\overset{\circ}{.}3}$. Similar to the case of KELT-1, the phase shift in the TESS band is significantly smaller than the corresponding infrared measurement at 4.5~$\mu$m ($18\overset{\circ}{.}7^{+2\overset{\circ}{.}1}_{-2\overset{\circ}{.}3}$; \citealt{mansfield2020}). The additional irradiation signal in the planet's flux has a semiamplitude of $A_{\mathrm{irrad}}=60.1^{+9.4}_{-9.1}$~ppm and comes to maximum roughly 1.6~hr before mid-eclipse. The phase-folded TESS light curve and best-fit phase-curve model are plotted in Figure~\ref{fig:kelt9}.

Comparing the results of our analysis with those in \citet{wong2020kelt9}, we find that the secondary eclipse, nightside flux, and atmospheric brightness modulation semiamplitude values are mutually consistent at much better than the $1\sigma$ level. Meanwhile, the new phase shift $\delta_{\mathrm{atm}}$ presented here is $1.5\sigma$ smaller than the previous measurement. We obtained a stellar pulsation period of $7.5851\pm0.0011$~hr and a peak-to-peak pulsation amplitude of $260\pm12$~ppm, which differ from the corresponding values from \citet{wong2020kelt9} by $1.3\sigma$ and $2.1\sigma$, respectively. We note that TESS observations in sectors 14 and 15 were particularly affected by scattered light on the detectors, and the updated SPOC data contains significantly more flagged points than the initially released photometry (see, for example, the large gaps in the untrimmed SPOC data in Appendix~\ref{sec:rawplots}). The slight discrepancies between the current fit and the previous results in \citet{wong2020kelt9} may be indicative of some systematic biases in the earlier version of the photometry.

When fitting for an additional phase-curve signal at the second harmonic of the orbital phase, we obtained a semiamplitude of $15\pm7$~ppm, consistent with our previous measurement of $16\pm4$~ppm. However, given the enhanced per-point uncertainty needed to account for red noise at longer timescales, the significance of the present detection is much lower, and including the second harmonic term in the model led to large increases in the AIC and BIC. Therefore, we cannot claim a detection of photometric variability at this harmonic. Future light curves from the TESS extended mission will provide improved sensitivity to low-amplitude phase-curve signals and allow us to definitely determine whether a second-harmonic signal exists in the KELT-9 system.

\subsection{Marginal Detections and Non-detections}\label{subsec:marginal}

\begin{deluxetable}{cccccc}
\tablewidth{0pc}
\tabletypesize{\scriptsize}
\tablecaption{
    Marginal Detections and Non-detections
    \label{tab:bad}
}
\tablehead{
    \colhead{Target} &
    \colhead{Sector}                     &
    \colhead{\textit{T}\tablenotemark{\scriptsize$\mathrm{a}$}} &
    \colhead{$D_{d,\mathrm{pred}}$\tablenotemark{\scriptsize$\mathrm{b}$}}  &
    \colhead{$D_{d,\mathrm{meas}}$\tablenotemark{\scriptsize$\mathrm{b}$}} &
    \colhead{$A_{\mathrm{atm}}$\tablenotemark{\scriptsize$\mathrm{c}$}}}
\startdata
HAT-P-36 & 22 & 11.6 & 170 & $150_{-90}^{+100}$ & $<70$ \\
KELT-23A & 14--17,21,23 & 9.8 & 70 & $< 50$ & $<10$\\
Qatar-1 & 17,21,24,25 & 11.8 & 80 & $88_{-47}^{+66}$ & $23^{+25}_{-24}$\\
TrES-3 & 25,26 & 11.6 & 160 & $140_{-70}^{+100}$ & $51^{+30}_{-31}$\\
WASP-3 & 26 & 10.1 & 170 & $150\pm70$ & $38^{+26}_{-23}$\\
WASP-92 & 23--25 & 12.4 & 120 & $160_{-110}^{+120}$ & $<70$ \\
WASP-93 & 17 & 10.6 & 110 & $140_{-80}^{+130}$ & $30_{-25}^{+24}$ \\
WASP-135 & 26 & 12.3 & 150 & $120_{-150}^{+210}$ & $<160$\\
\enddata
\textbf{Notes.}
\vspace{-0.25cm}\tablenotetext{\textrm{a}}{Apparent magnitude in the TESS bandpass.}
\vspace{-0.25cm}\tablenotetext{\textrm{b}}{Predicted and measured secondary eclipse depths, in parts-per-million. Predictions assume $A_{g}=0.1$ and no day-night heat recirculation.}
\vspace{-0.25cm}\tablenotetext{\textrm{c}}{Measured semiamplitudes (or $2\sigma$ upper limits) of the atmospheric brightness modulation, in parts-per-million.}
\vspace{-0.4cm}
\end{deluxetable}

% Transit-only fit parameters table
%-------------------------------------------------------------------------------
\begin{splitdeluxetable*}{lllllllllBlllllllll}
\tablewidth{0pc}
\tabletypesize{\scriptsize}
\tablecaption{
    Results from Transit-only Light-curve Fits
    \label{tab:transitonly}
}
\vspace{-0.3cm}
\tablehead{& \multicolumn{2}{c}{\underline{HAT-P-36}} \vspace{-0.2cm}& \multicolumn{2}{c}{\underline{KELT-23A}} &
\multicolumn{2}{c}{\underline{Qatar-1}} & \multicolumn{2}{c}{\underline{TrES-3}} & & \multicolumn{2}{c}{\underline{WASP-3}} &  \multicolumn{2}{c}{\underline{WASP-92}} &
\multicolumn{2}{c}{\underline{WASP-93}} &
\multicolumn{2}{c}{\underline{WASP-135}} \\
    \colhead{Parameter} \vspace{-0.1cm}&
    \colhead{Value}                     &
    \colhead{Error}  &
    \colhead{Value}                     &
    \colhead{Error}  & 
    \colhead{Value}                     &
    \colhead{Error}  &  
    \colhead{Value}                     &
    \colhead{Error}  &  
    \colhead{Parameter} &
    \colhead{Value}                     &
    \colhead{Error}  &
    \colhead{Value}                     &
    \colhead{Error}  & 
    \colhead{Value}                     &
    \colhead{Error}  &  
    \colhead{Value}                     &
    \colhead{Error}  
}
\startdata
\multicolumn{2}{l}{\textit{Fitted parameters}} & & & & & & & & \multicolumn{2}{l}{\textit{Fitted parameters}} & & & & &\\
$R_p/R_*$     & 0.1226 & $_{-0.0012}^{+0.0011}$ & 0.13276 & $_{-0.00050}^{+0.00045}$ & 0.1458 & $_{-0.0019}^{+0.0018}$ & 0.1675 & $_{-0.0030}^{+0.0043}$ & $R_{p}/R_{*}$ & 0.1051 & $_{-0.0016}^{+0.0012}$ & 0.1052 & $_{-0.0034}^{+0.0019}$ & 0.1039 & $_{-0.0032}^{+0.0038}$ & 0.1402 & $_{-0.0033}^{+0.0018}$ \\
$T_{0,\mathrm{MCMC}}$ ($\mathrm{BJD}_{\mathrm{TDB}}-2458000$)\tablenotemark{\scriptsize a}  &  911.42294 & $_{-0.00013}^{+0.00014}$ & 769.612269 & $_{-0.000033}^{+0.000034}$ & 959.129856 & $_{-0.000079}^{+0.000080}$ & 1008.35093 & 0.00010 & $T_{0,\mathrm{MCMC}}$ ($\mathrm{BJD}_{\mathrm{TDB}}-2458000$) & 1023.18948 & 0.00015 & 971.31989 & $_{-0.00034}^{+0.00031}$ & 779.31211 & $_{-0.00045}^{+0.00046}$ & 1021.71852 & $_{-0.00031}^{+0.00028}$ \\
$T_{0,\mathrm{PB}}$ ($\mathrm{BJD}_{\mathrm{TDB}}-2458000$)\tablenotemark{\scriptsize a}  & 911.42299 & $^{+0.00030}_{-0.00031}$ & 769.612278 & $^{+0.000060}_{-0.000058}$ & 959.12981 & $^{+0.00014}_{-0.00013}$ & 1008.35101 & $^{+0.00022}_{-0.00020}$ & $T_{0,\mathrm{PB}}$ ($\mathrm{BJD}_{\mathrm{TDB}}-2458000$) & 1023.18947 & $^{+0.00029}_{-0.00025}$ & 971.32011 & $^{+0.00058}_{-0.00054}$ & 779.31199 & 0.00062 & 1021.71876 & $^{+0.00072}_{-0.00048}$ \\
$P$ (days)    & 1.327355 & $_{-0.000022}^{+0.000021}$ &  2.25528773 & $_{-0.00000073}^{+0.00000077}$ & 1.4200228 & $_{-0.0000013}^{+0.0000012}$ & 1.3061842 & $_{-0.0000085}^{+0.0000087}$ & $P$ (days) & 1.846866 & $_{-0.000037}^{+0.000036}$ & 2.174663 & $_{-0.000029}^{+0.000030}$ & 2.73253 & $_{-0.00015}^{+0.00014}$ & 1.401403 & $_{-0.000045}^{+0.000052}$ \\
$b$           & 0.08 & $_{-0.20}^{+0.17}$ & 0.523 & 0.011 & 0.603 & $_{-0.025}^{+0.023}$ & 0.836 & $_{-0.020}^{+0.026}$ & $b$ & 0.40 & $_{-0.12}^{+0.07}$ & 0.53 & $_{-0.18}^{+0.08}$ & 0.889 & $_{-0.029}^{+0.023}$ & 0.706 & $_{-0.061}^{+0.037}$ \\
$a/R_*$       & 4.762 & $_{-0.080}^{+0.063}$ & 7.614 & $_{-0.046}^{+0.043}$ & 6.41 & $_{-0.10}^{+0.11}$ &  5.83 & 0.10 & $a/R_*$ & 5.40 & $_{-0.18}^{+0.19}$ & 5.81 & $_{-0.34}^{+0.51}$ & 6.12 & $_{-0.27}^{+0.31}$ & 5.63 & $_{-0.23}^{+0.31}$ \\
$\gamma_{1}$\tablenotemark{\scriptsize b}  &  0.881 & $_{-0.085}^{+0.094}$ & 0.841 & 0.026 & 1.15 & $_{-0.11}^{+0.09}$ & 1.07 & $_{-0.35}^{+0.39}$ & $\gamma_{1}$ & 0.73 & $_{-0.09}^{+0.10}$ &  0.84 & 0.19 & 0.86 & $_{-0.38}^{+0.50}$ & 0.68 & $_{-0.29}^{+0.35}$ \\
$\gamma_{2}$\tablenotemark{\scriptsize b}  & $-0.61$ & 0.56 & $-0.01$ & $_{-0.25}^{+0.23}$ & $-0.45$ & $_{-0.75}^{+0.66}$ & $-0.31$ & $_{-0.75}^{+0.65}$ & $\gamma_{2}$ & $-0.38$ & $_{-0.62}^{+0.51}$ & $-0.53$ & $_{-0.68}^{+0.64}$ & $-0.37$ & $_{-0.66}^{+0.57}$ & $-0.28$ & $_{-0.66}^{+0.42}$ \\
\\
\multicolumn{2}{l}{\textit{Derived parameters}} & & & & & & & &  \multicolumn{2}{l}{\textit{Derived parameters}} & & & & &  \\
$i$ ($^{\circ}$)      &  89.1 & $_{-2.1}^{+2.3}$ & 86.06 & 0.10 & 84.60 & $_{-0.28}^{+0.30}$ & 81.77 & $_{-0.40}^{+0.30}$ & $i$ ($^{\circ}$)      & 85.8 & $_{-1.0}^{+1.4}$ & 84.7 & $_{-1.2}^{+2.1}$ & 81.65 & $_{-0.61}^{+0.66}$ & 82.81 & $_{-0.70}^{+0.94}$ \\
$u_{1}$  & 0.23 & 0.11 & 0.335 & $_{-0.056}^{+0.053}$ & 0.37 & $_{-0.17}^{+0.15}$ & 0.34 & $_{-0.22}^{+0.28}$ & $u_{1}$  & 0.21 & $_{-0.12}^{+0.10}$ & 0.22 & 0.14 & 0.23 & $_{-0.16}^{+0.29}$ & 0.18 & $_{-0.13}^{+0.18}$ \\
$u_{2}$  & 0.42 & 0.23 & 0.170 & $_{-0.091}^{+0.097}$ & 0.41 & $_{-0.25}^{+0.29}$ & 0.33 & $_{-0.23}^{+0.29}$ &  $u_{2}$  & 0.29 & $_{-0.20}^{+0.25}$ & 0.37 & $_{-0.25}^{+0.30}$ & 0.32 & $_{-0.23}^{+0.28}$ & 0.23 & $_{-0.16}^{+0.32}$ \\
$a$ (au) & 0.0243 & 0.0013 & 0.03527 & 0.00057 & 0.02453 & 0.00085 & 0.0217 & 0.0013 & $a$ (au) & 0.0329 & 0.0024 & 0.0362 & 0.0031 & 0.0434 & 0.0023 & 0.0251 & 0.0018 \\
$R_{p}$ ($R_{\mathrm{Jup}}$) & 1.308 & 0.068 & 1.287 & 0.020 & 1.168 & 0.038 & 1.307 & 0.080 & $R_{p}$ ($R_{\mathrm{Jup}}$) & 1.340 & 0.089 & 1.373 & 0.069 & 1.541 & 0.066 & 1.310 & 0.072\\
\enddata
\textbf{Notes.}
\vspace{-0.25cm}\tablenotetext{\textrm{a}}{Mid-transit times derived from the MCMC and PB analyses.}
\vspace{-0.25cm}\tablenotetext{\textrm{b}}{Modified limb-darkening parameters $\gamma_{1}\equiv 2u_{1}+u_{2}$ and $\gamma_{2}\equiv u_{1}-2u_{2}$.}
\end{splitdeluxetable*}

No statistically significant phase-curve amplitudes or secondary eclipse depths were measured in 8 of the 15 systems selected for detailed analysis. Table~\ref{tab:bad} lists these targets, along with the predicted secondary eclipse depths from Equation~\eqref{depth}. Also provided are the marginal detections or upper limit constraints on the secondary eclipse depths and atmospheric brightness modulation semiamplitudes that were measured from phase-curve fits that included only those components in the astrophysical model. In all cases, the eclipse-depth measurements are broadly consistent with the predicted values. 

Given the absence of any robust phase-curve signals in the TESS photometry, we carried out simplified fits with a flat out-of-transit model flux (i.e., transit-only fits). The results of these fits are presented in Table~\ref{tab:transitonly}. When comparing the parameter values with those from the respective discovery papers, we generally find good agreement and a moderate increase in the precision of the transit parameters ($R_{p}/R_{*}$, $b$, and $a/R_{*}$) in some cases, yielding improved constraints on the derived parameters, such as orbital semimajor axis $a$, inclination $i$, and planetary radius $R_{p}$. The systematics-corrected and phase-folded transit light curves are included in the compilation plot in Figure~\ref{fig:transit}.

The handful of instances where the new parameter values differ somewhat significantly ($>2\sigma$) from previous results are described below. KELT-23A was observed during 6 TESS sectors, and the high data volume produced the most precise transit shape and limb-darkening parameters of any target studied in this paper. We measured $b=0.523\pm0.011$, $a/R_{*}=7.614^{+0.043}_{-0.046}$, which are discrepant from the discovery paper values ($b=0.576^{+0.024}_{-0.027}$, $a/R_{*}=7.13^{+0.16}_{-0.15}$; \citealt{kelt23}) at the $1.8\sigma$ and $2.9\sigma$ levels, respectively. For Qatar-1, our impact parameter $b=0.603^{+0.023}_{-0.025}$ is $2.8\sigma$ larger than the value $b=0.696^{+0.021}_{-0.024}$ from \citet{qatar1}. Meanwhile, the more recent transit light-curve analysis in \citet{m15} presented a more consistent set of parameter values: $b=0.63\pm0.02$ and $a/R_{*}=6.319^{+0.070}_{-0.068}$. 

\begin{deluxetable*}{cccccccc}
\tablewidth{0pc}
\tabletypesize{\scriptsize}
\tablecaption{
    Dayside Blackbody Brightness Temperatures and Geometric Albedos
    \label{tab:temps}
}
\tablehead{
    \colhead{Planet} &
    \colhead{$D_{d,\mathrm{TESS}}$ (ppm)\tablenotemark{\scriptsize$\mathrm{a}$}}                     &
    \colhead{$D_{d,3.6}$ (ppm)\tablenotemark{\scriptsize$\mathrm{a}$}}                     &
    \colhead{$D_{d,4.5}$ (ppm)\tablenotemark{\scriptsize$\mathrm{a}$}} &
    \colhead{$T_{\mathrm{day}}$ (K)}  &
    \colhead{$A_{g}$\tablenotemark{\scriptsize$\mathrm{b}$}} &
    \colhead{$\chi^{2}_{r}$ \tablenotemark{\scriptsize$\mathrm{c}$}} & 
    \colhead{Reference\tablenotemark{\scriptsize$\mathrm{d}$}}
}
\startdata
\multicolumn{2}{l}{\textit{Year 1}} & & & & & & \\
WASP-4b & $120^{+80}_{-70}$ & $3190\pm310$ & $3430\pm270$ & $1954\pm67$ & $0.09\pm0.09$ & 2.39 & \citet{beerer2011}\\
WASP-5b  & $31_{-55}^{+73}$ & $1970\pm280$ & $2370\pm240$ & $2000\pm90$ & $<0.32$ & 0.17 & \citet{baskin2013}\\
WASP-18b & $339\pm21$  & $3040\pm190$ & $3790\pm150$ & $3046\pm66$ & $<0.03$ & 1.32 & \citet{maxted2013} \\
WASP-19b & $470^{+130}_{-110}$ & $4850\pm240$ & $5840\pm290$ & $2204\pm49$ & $0.17\pm0.07$ & 0.49 & \citet{wong2016}\\
WASP-36b & $90_{-70}^{+100}$  & $914\pm578$ & $1953\pm544$ & $1440\pm160$ & $0.16\pm0.15$ & 0.38 & \citet{garhart2020}\\
WASP-43b & $170\pm70$ & $3230\pm60$ & $3830\pm80$ & $1655\pm38$ & $0.13\pm0.06$ & 38.3 & \citet{stevenson2017}\\
WASP-46b & $230_{-110}^{+140}$ & $1360\pm701$ & $4446\pm589$ & $1880\pm120$ & $0.38\pm0.27$ & 5.87 & \citet{garhart2020}\\
WASP-64b & $230_{-110}^{+130}$ & $2859\pm270$ & $2071\pm471$ & $1989\pm86$ & $0.38\pm0.26$ & 8.01 & \citet{garhart2020}\\
WASP-77Ab & $53_{-22}^{+32}$ & $2016\pm94$ & $2487\pm127$ & $1840\pm33$ & $0.06\pm0.05$ & 1.49 & \citet{garhart2020}\\
WASP-78b & $210_{-90}^{+100}$ & $2001\pm218$ & $2013\pm351$ & $2550\pm130$ & $<0.56$ & 0.76 & \citet{garhart2020}\\
WASP-100b & $94\pm17$ & $1267\pm98$ & $1720\pm119$ & $2356\pm67$ & $0.22\pm0.08$ & 1.22 & \citet{garhart2020}\\
WASP-121b &  $486\pm59$ & $3685\pm114$ &  $4684\pm121$ & $2592\pm44$ & $0.26\pm0.06$ & 2.89 & \citet{garhart2020}\\
\hline
\multicolumn{2}{l}{\textit{Year 2}} & & & & & & \\
HAT-P-7b & $127^{+33}_{-32}$ & $1560\pm90$ & $1900\pm60$ & $2692\pm62$ & $<0.28$ & 0.24 & \citet{wong2016}\\
KELT-1b & $388^{+67}_{-65}$ & $1877\pm58$ & $2083\pm70$ & $2978\pm56$ & $0.45\pm0.16$ & 0.62 & \citet{beatty2019} \\
Kepler-13Ab & $301^{+46}_{-42}$ & $1560\pm310$ & $2220\pm230$ & $2786\pm160$ & $0.53\pm0.15$ & 0.84 & \citet{shporer2014} \\
Qatar-1b & $88^{+66}_{-47}$ & $2100\pm200$ & $3000\pm200$ & $1539\pm41$ & $0.14\pm0.11$ & 0.07 & \citet{keating2020} \\
TrES-3b & $140^{+130}_{-80}$ & $3450\pm350$ & $3470\pm540$ & $1737\pm70$ & $0.14\pm0.13$ & 1.84 & \citet{fressin2010} \\
WASP-3b & $150\pm70$ & $2090^{+400}_{-280}$ & $2820\pm120$ & $2372\pm66$ & $<0.55$ & 0.44 &  \citet{rostron2014} \\
%WASP-12 & $443^{+86}_{-85}$ & $4210\pm110$ & $4280\pm120$ & $2673\pm71$ & $0.13\pm0.06$ & 13.4 & \citet{stevenson2014wasp12} \\
WASP-12b & $443^{+86}_{-85}$ & $3854\pm88$\tablenotemark{\scriptsize$\mathrm{e}$} & $4160\pm100$\tablenotemark{\scriptsize$\mathrm{e}$} & $2710\pm55$ & $0.13\pm0.06$ & 7.11 & \citet{bell2019} \\
WASP-33b\tablenotemark{\scriptsize$\mathrm{f}$} & $320\pm37$ & $3506\pm173$ & $4250\pm160$ & $3145\pm65$ & $<0.08$ & 1.09 & \citet{zhang2018} \\
\enddata
\textbf{Notes.}
\vspace{-0.25cm}\tablenotetext{\textrm{a}}{Secondary eclipse depths measured in the TESS bandpass and the 3.6 and 4.5~$\mu$m Spitzer/IRAC bandpasses.}
\vspace{-0.25cm}\tablenotetext{\textrm{b}}{For marginal cases, $2\sigma$ upper limits are provided.}
\vspace{-0.25cm}\tablenotetext{\textrm{c}}{Reduced chi-squared value of the best-fit model to the three secondary eclipse measurements.}
\vspace{-0.25cm}\tablenotetext{\textrm{d}}{Literature references for the Spitzer secondary eclipse measurements.}
\vspace{-0.25cm}\tablenotetext{\textrm{e}}{Weighted averages of eclipse depth measurements from two epochs.}
\vspace{-0.25cm}\tablenotetext{\textrm{f}}{TESS secondary eclipse depth taken from \citet{vonessen2020wasp33}.}
\vspace{-0.7cm}
\end{deluxetable*}

\section{Discussion}\label{sec:dis}

Our systematic light-curve analysis of targets from the second year of the TESS mission yielded 7 systems with robust secondary eclipse and phase-curve signals. Some overarching observations include the following: (1) the only orbiting companion to show significant nightside flux in the TESS bandpass is KELT-9b, (2) KELT-9b and WASP-12b are the only systems in the list for which statistically significant phase shifts in the atmospheric brightness modulation were detected, and (3) the strengths of the ellipsoidal distortion modulations measured for KELT-1, Kepler-13A, and WASP-12 are broadly consistent with theoretical predictions. In this section, we use the results of our phase-curve fits to explore the atmospheric properties of these systems and derive updated transit ephemerides.

\subsection{Dayside Temperatures and Geometric Albedos}\label{subsec:temps}

The dayside temperature of an orbiting planet cannot be reliably determined from the system's brightness ratio in a single bandpass without making assumptions about the amount of reflected starlight and/or the level of heat redistribution across the atmosphere. In order to break this degeneracy, additional secondary eclipse measurements at other wavelengths are needed, preferably in the thermal infrared, where the emission from the companion's atmosphere dominates any reflected starlight. 

Spitzer secondary eclipse measurements have been published for several of the systems studied in our TESS phase-curve analysis. Just as in our Year 1 analysis, we only considered systems for which secondary eclipse depths were obtained in both the 3.6 and 4.5~$\mu$m bandpasses. Seven targets satisfy this criterion --- HAT-P-7, KELT-1, Kepler-13A, Qatar-1, TrES-3, WASP-3, and WASP-12. We list the eclipse depths and references in Table~\ref{tab:temps}. Whenever possible, we chose Spitzer eclipse depths that were obtained from combined analyses of the 3.6 and 4.5~$\mu$m full-orbit phase-curve fits, given the possibility of a significant bias in the measured values when not properly accounting for the variable out-of-eclipse flux \citep[e.g.,][]{bell2019}. We also included WASP-33, combining the TESS-band secondary eclipse depth measured by \citet{vonessen2020wasp33} with the Spitzer results from \citet{zhang2018}. 

For each system, we assumed that the thermal emission from the companion's dayside hemisphere in the three bandpasses is consistent with a single blackbody (i.e., the three wavelength ranges probe similar pressure levels within the atmosphere) and carried out a simple simultaneous fit of the three secondary eclipse depths to Equation~\eqref{depth}. Here, the measured planetary temperature $T_{p}$ is designated as the dayside blackbody brightness temperature $T_{\mathrm{day}.}$ To account for possible excess flux at short wavelengths due to reflected starlight off clouds and/or hazes, we allowed the geometric albedo $A_{g}$ to vary freely when modeling the TESS-band secondary eclipse depth; meanwhile, $A_g$ was fixed to zero at the Spitzer wavelengths. 

Following the methods in Paper 1, we used PHOENIX stellar models \citep{husser2013} and derived a best-fit interpolation polynomial as a function of $(T_{\mathrm{eff}},\log g,[\mathrm{Fe}/\mathrm{H}])$ for the band-integrated stellar flux in each of the three bandpasses. These polynomials were sampled in a Monte Carlo fashion in order to propagate the stellar parameter uncertainties to our temperature and albedo estimates. In the MCMC fitting procedure, we applied Gaussian priors to $R_p/R_*$, $a/R_*$, $T_{\mathrm{eff}}$, $\log g$, and [Fe/H] based on literature values and our light-curve fit results (Tables~\ref{tab:fit} and \ref{tab:transitonly}).

There was a minor issue in our implementation of the TESS transmission function in Paper 1 due to the erroneous inclusion of the photon-to-energy unit conversion factor $\lambda/hc$ in Equation~\eqref{depth}; this factor is needed for computing the CoRoT, Kepler, and Spitzer band-integrated fluxes, as the corresponding transmission functions are provided in photon units. In this paper, we have corrected the calculations and report updated values for all planets in our Year 1 and Year 2 TESS phase-curve sample. The alterations to the best-fit values are negligible (typically $<$0.1--0.2$\sigma$).

Table \ref{tab:temps} presents the dayside brightness temperature and TESS-band geometric albedo estimates; $2\sigma$ upper limits are provided in cases where the albedo value is consistent with zero to within $1\sigma$. We also list the reduced $\chi^{2}$ values for each fit. Most systems show $\chi_{r}^{2}\lesssim2$. Meanwhile, four planets --- WASP-12b, WASP-43b, WASP-46b, and WASP-64b --- have secondary eclipse depths that are not well-described by our simple blackbody+reflectivity model. These discrepancies may indicate that the three photometric bandpasses are probing pressure levels within the atmosphere that have significantly different temperatures. For all of these cases except WASP-46b, the brightness temperature derived from the Spitzer 4.5~$\mu$m eclipse depth alone is significantly lower than the 3.6~$\mu$m brightness temperature. 

One plausible explanation for this discrepancy is CO absorption. This scenario is particularly applicable to WASP-12b. Within the framework of the atmospheric mass-loss hypothesis (see Section \ref{wasp12} and \citealt{bell2019}), the column of escaped gas is oriented along the line of sight at superior conjunction. CO, which has strong absorption features within the 4.5~$\mu$m bandpass, may block a significant portion of the thermal emission from the planet, resulting in a lower apparent brightness temperature. Detailed modeling of the Hubble/WFC3 and Spitzer secondary eclipse depths of WASP-43b revealed strong absorption from CO within the Spitzer 4.5~$\mu$m bandpass, leading to a significant deviation from a blackbody emission spectrum \citep{stevenson2017}.

While most of the TESS-band geometric albedos we measured lie within $2\sigma$ of zero, KELT-1b and Kepler-13Ab show very high reflectivity. Previous atmospheric modeling of the dayside emission of KELT-1b, including the TESS-band secondary eclipse depth, corroborates our conclusion of a significantly nonzero albedo \citep{beatty2020}. For Kepler-13Ab, a joint fit of the Spitzer and Kepler secondary eclipse observations yields a Kepler-band geometric albedo of $0.35^{+0.04}_{-0.05}$ and a dayside brightness temperature of $2770\pm170$~K (Paper 1), which are statistically consistent with our values derived from the Spitzer+TESS fit. Through an analogous calculation for HAT-P-7b using the Kepler-band secondary eclipse depth measurement from \citet{esteves2015}, we obtained $T_{\mathrm{day}}=2666\pm47$~K and a Kepler-band geometric albedo of $A_{g} = 0.06\pm0.02$. 

\subsection{Revisiting the Albedo--Dayside Temperature Trend}\label{subsec:trends}

\begin{figure}[t]
\includegraphics[width=\linewidth]{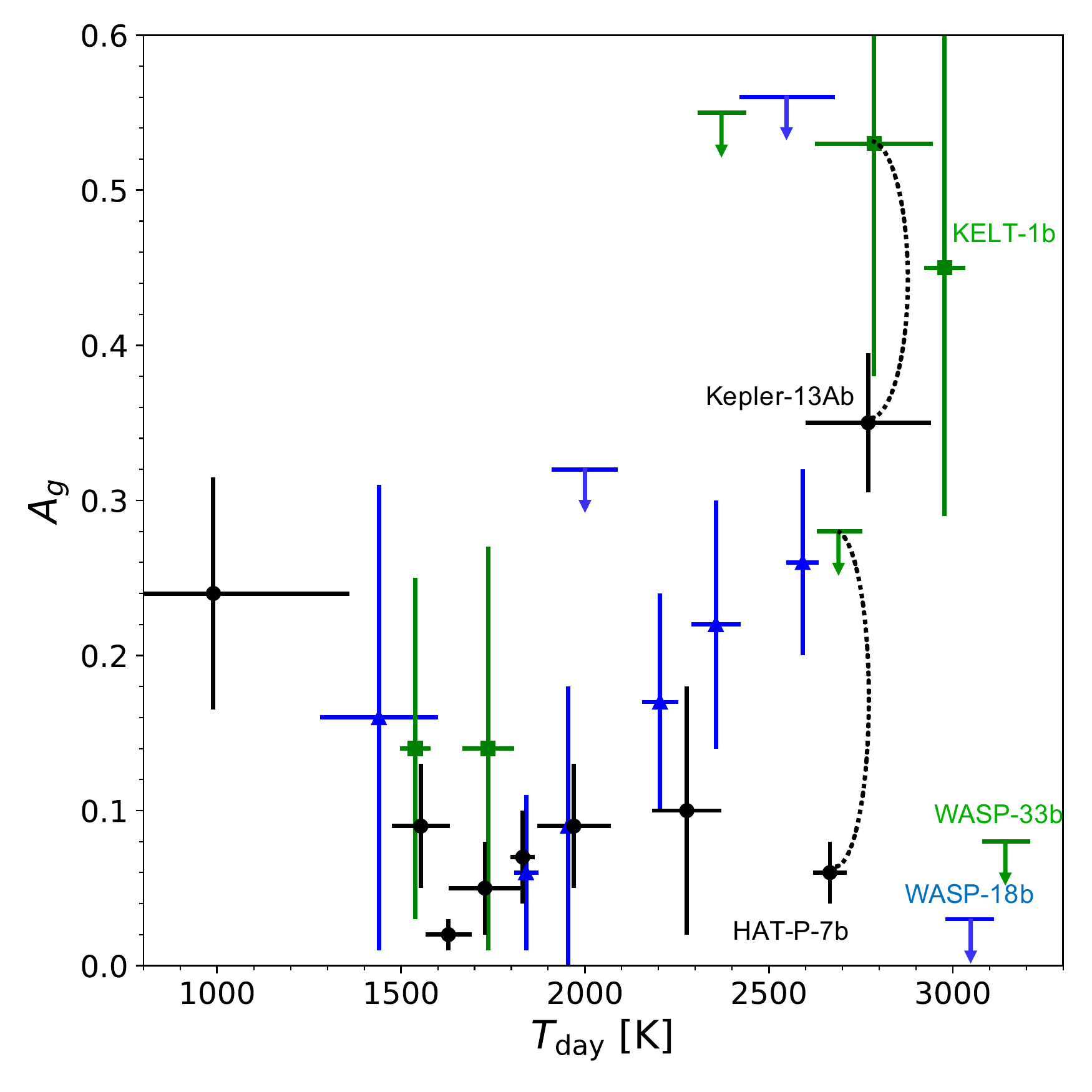}
\caption{Two-parameter plot showing the relationship between geometric albedo $A_{g}$ and dayside blackbody brightness temperature $T_{\mathrm{day}}$, as derived from our joint fits of visible wavelength secondary eclipse depths and Spitzer measurements. The blue triangles and green squares indicate the systems from the first and second year of the TESS primary mission, respectively; the black circles indicate the Kepler/CoRoT-band geometric albedos for the targets that were observed by those missions. For HAT-P-7b and Kepler-13Ab, which were observed by both Kepler and TESS, dashed curves connect the corresponding Kepler- and TESS-band geometric albedos and $T_{\mathrm{day}}$ measurements from the Spitzer+Kepler and Spitzer+TESS fits. For several objects, $2\sigma$ upper limits on the geometric albedo are shown by the arrows. Objects with poorly-fit emission spectra ($\chi^{2}_{r}>5$) are not shown. Some notable planets discussed in the text are labeled.}
\label{fig:trends}
\end{figure}

In Paper 1, we calculated dayside brightness temperatures and TESS-band geometric albedos following the methodology outlined in the previous subsection. We then searched for trends between albedo and various other system parameters, such as dayside temperature, stellar metallicity, and planetary surface gravity. A marginal $2.2\sigma$ positive correlation between the TESS-band geometric albedo and dayside temperature was detected among planets with $1500<T_{\mathrm{day}}<3000$~K, with the significance increasing to $5.5\sigma$ when including additional data points derived from Kepler and CoRoT secondary eclipses. Meanwhile, the very low albedo of WASP-18b ($T_{\mathrm{day}}=3046\pm66$~K) indicates a sharp break in the emergent trend for the most extremely-irradiated hot Jupiters.

Having expanded the body of self-consistent dayside temperatures and TESS-band geometric albedos with the results of our Year 2 analysis, we revisited the apparent trend between these two parameters. Figure~\ref{fig:trends} plots the geometric albedo as a function of dayside blackbody brightness temperature; we have omitted targets whose secondary eclipses are poorly fit by the blackbody+reflectivity model ($\chi^{2}_{r}>5$). The colored points indicate the measurements we obtained from TESS-band secondary eclipses, while the black points denote the analogously-derived values from Kepler or CoRoT eclipse depths (see Table~4 in Paper 1). For objects with both TESS and Kepler observations (HAT-P-7b and Kepler-13Ab), we plot both the TESS-band and Kepler-band geometric albedos derived from the Spitzer+TESS and Spitzer+Kepler fits, respectively.

The positive correlation between $A_{g}$ and $T_{\mathrm{day}}$ remains apparent in the larger dataset. The addition of the KELT-1b data point anchors the high-temperature end of the trend. At even higher temperatures, the low TESS-band geometric albedo of WASP-33b supports the previous suggestion that the dayside atmospheres of the hottest ultra-hot Jupiters strongly deviate from the aforementioned correlation and are consistent with zero reflectivity.

Just as in our Year 1 analysis, we carried out a MCMC linear fit to the expanded sample, focusing on systems in the range $1500<T_{\mathrm{day}}<3000$~K. While this temperature range was primarily selected to allow one-to-one comparison with the results in Paper 1, the choice was also motivated by both theoretical and empirical considerations. Previous cloud modeling in the context of optical atmospheric reflectivity has shown that while cooler planets can host a wide array of condensate species, resulting in complicated trends in predicted geometric albedo, silicate clouds are expected to largely disappear across the dayside hemisphere at temperatures above 1500--1700~K \citep[e.g.,][]{parmentier2016}. Visual inspection of the $A_g$ vs. $T_{\mathrm{day}}$ trend in Figure~\ref{fig:trends} clearly indicates a local minimum in geometric albedo near zero at around 1500--1600~K, with the small handful of cooler planets having somewhat higher, but poorly constrained albedo values, and a possible reversal in the trend seen among hotter targets. Meanwhile, the two hottest planets with $T_{\mathrm{day}}>3000$~K are clear outliers. We experimented with fitting higher-order polynomials to the albedos in both the constrained and full temperature ranges and found that the linear fit to objects in the range $1500<T_{\mathrm{day}}<3000$~K yielded the smallest reduced $\chi^2$ and the lowest BIC.

When using only the TESS-derived albedos, the addition of the Year 2 targets increases the significance of the positive correlation between TESS-band geometric albedo and dayside temperature from 2.2$\sigma$ to $3.1\sigma$. Here, the inclusion of KELT-1b as an additional high-albedo object is crucial in strengthening the detection of an overall trend. We also considered the Kepler/CoRoT-band albedos independently and obtained a similar trend between $A_g$ and $T_{\mathrm{day}}$ at $4.6\sigma$ significance. The TESS, Kepler, and CoRoT bandpasses overlap considerably, with the latter two having almost identical effective wavelengths and comparable bandwidths, and the TESS bandpass situated roughly $150$~nm redder on average. Following our previous Year 1 analysis, we carried out a combined MCMC fit to the TESS-, Kepler-, and CoRoT-derived albedos and retrieved a very robust $5.7\sigma$ trend. We note that geometric albedo varies with wavelength, and the Kepler and CoRoT bandpasses are more sensitive to bluer wavelengths, where the effect of Rayleigh scattering on the dayside reflectivity is more pronounced. Nevertheless, the presence of a significant trend between $A_g$ and $T_{\mathrm{day}}$ in both the TESS-derived and the Kepler/CoRoT-derived datasets suggests a broader correlation between atmospheric reflectivity and dayside temperature that holds across the visible wavelength range.

HAT-P-7b stands out as an outlier in the overall trend of $A_{g}$ vs. $T_{\mathrm{day}}$, particularly when considering the Kepler-derived geometric albedo value. A recent reanalysis of the Spitzer 4.5~$\mu$m phase curve of this planet has revealed some unusual properties \citep{bell2020}: (1) the day-night temperature contrast is very low ($\sim$400~K), implying much more efficient heat transport between the day- and nightside hemispheres than other hot Jupiters with similar levels of stellar irradiation, (2) the nightside brightness temperature of roughly 2500~K is significantly hotter than all other hot Jupiters with phase curve measurements, making HAT-P-7b comparable to the extremely irradiated KELT-9b, and (3) the corresponding inferred Bond albedo, derived from simple thermal balance considerations, is negative (see Table~\ref{tab:phaseintegral}), suggesting an additional source of heat across the planet's surface. It is evident that HAT-P-7b presents a challenging case for interpreting the global atmospheric thermal energy budget. We experimented with omitting this object from the TESS-band albedo--dayside temperature trend analysis and obtained a much stronger positive correlation at $3.7\sigma$ significance.

The addition of targets from the second year of the TESS mission primarily serves to solidify the detection of a positive correlation between $A_g$ and $T_{\mathrm{day}}$ reported in Paper 1. Here, we briefly reiterate some possible explanations for the trend that were postulated in that work. While most condensate species are expected to be in the vapor phase across the dayside hemisphere for temperatures above $\sim$2000~K, some highly refractory molecules such as TiO$_2$ and Al$_2$O$_3$ may survive near the western limb and poles at these higher temperatures \citep{powell2019}, before finally vaporizing completely for the most extreme cases: WASP-18b and WASP-33b. Such a scenario can be probed with higher-precision visible-light phase curves, which may reveal a westward shift in the location of maximum dayside brightness that is indicative of a cloudy western limb. Alternatively, additional sources of \textit{thermal} emission due to optical absorbers such as TiO/VO, atomic iron, and dissociated hydrogen may contribute to excess flux at short wavelengths \citep[e.g.,][]{arcangeli2018,lothringer2018}. High-resolution dayside emission is needed to address this possibility.

\subsection{Secondary Eclipse Spectrum Modeling of Kepler-13Ab}\label{subsec:modeling}
The albedo--temperature trend seen in Figure~\ref{fig:trends} is strongly driven by the two high-albedo objects with $2700<T_{\mathrm{day}}<3000$~K: KELT-1b and Kepler-13Ab. These orbiting companions are simultaneously among the most massive objects in the geometric albedo sample ($>$5~$M_{\mathrm{Jup}}$). When removing these objects from the sample, the significance of the albedo--temperature trend falls to $3.0\sigma$. Understanding these apparent high-reflectivity outliers is crucial for probing the underlying physical processes that may be responsible for the putative albedo--temperature trend. The simultaneous calculation of the dayside temperatures and geometric albedos in Section \ref{subsec:temps} assumes that the dayside emission spectra closely resemble blackbodies. Any significant wavelength-dependent deviation in the true emission spectra would induce systematic biases in the retrieved albedo values. 

\citet{beatty2017k13} obtained spectroscopic secondary eclipse observations of Kepler-13Ab using the Wide Field Camera 3 instrument on the Hubble Space Telescope (HST/WFC3) and found that the resultant emission spectrum indicates a non-inverted vertical temperature profile that decreases monotonically with increasing altitude. With this medium-resolution dataset in hand, we can carry out more sophisticated modeling of the secondary eclipse spectrum and probe whether or not the high geometric albedo inferred from our blackbody fits is an artifact of our simplified approach.

%From the shape of the emission spectrum in the near-infrared, they inferred that the temperature--pressure profile of the dayside atmosphere decreases to low pressures. The corollary of this observation is that the interior isotherm of Kepler-13Ab is higher than in the case of an isothermal blackbody, resulting in higher planetary thermal emission at optical wavelengths. It follows that when including the near-infrared emission spectrum, the calculated geometric albedo in the Kepler and TESS bandpasses (roughly 0.12--0.15) is significantly lower than the value derived from a blackbody fit of the Spitzer and Kepler eclipse depths only ($0.35^{+0.04}_{-0.05}$). This updated value places Kepler-13Ab well within the range spanned by the other cooler hot Jupiters.

To constrain the value of the geometric albedo and the atmospheric properties of Kepler-13Ab, we performed a suite of retrievals using the \textsc{Helios-r2} model \citep{kitzmann2020}. The model was updated to perform retrievals on emission spectra, including the option to use the corresponding filter response functions for photometric measurements and the additional contribution from reflected starlight. The compiled eclipse depths consist of the TESS and Kepler measurements from this work and \citet{shporer2014} --- $301^{+46}_{-42}$ and $173.7\pm1.8$~ppm, respectively --- the HST/WFC3 dataset \citep{beatty2017k13}, and the Spitzer 3.6 and 4.5~$\mu$m secondary eclipse depths \citep{shporer2014}. Due to its poor precision, we did not include the ground-based K-band secondary eclipse from \citet{shporer2014}.

\textsc{Helios-r2} calculates the eclipse depth at each wavelength according to Equation~\eqref{depth}. The thermal emission from the planet's dayside atmosphere is generated from a free temperature--pressure (TP) profile, with scattering neglected (see \citealt{kitzmann2020} for full details). The stellar spectrum of Kepler-13A is interpolated from the PHOENIX library of theoretical stellar spectra \citep{husser2013} and considered fixed in the retrievals.

\begin{deluxetable}{lll}[t]
	\tablecaption{Free Parameters and Prior Distributions Used for the Kepler-13Ab Secondary Eclipse Spectrum Retrievals \label{tab:retrieval_parameter}}
	\tabletypesize{\scriptsize}
	%\tablecolumns{4}
	%\tablewidth{0pt}
	\tablehead{
		\colhead{Parameter} &
		\colhead{Prior Type} &
		\colhead{Prior Values}
	}
	\startdata
	$\log g_p$   & Gaussian    & $3.92 \pm 0.03$\tablenotemark{\scriptsize$\mathrm{a}$} \\
	$R_p/R_*$  & Gaussian    & $0.087373 \pm 0.000024$\tablenotemark{\scriptsize$\mathrm{b}$} \\
	$a/R_*$    & Gaussian    & $4.5007 \pm 0.004$\tablenotemark{\scriptsize$\mathrm{b}$} \\
	$A_g$      & uniform     & 0--1  \\
	$T_1$      & uniform     & 1000--5000 \\
	$b_i$      & uniform     & 0.1--1.5 \\
	$x_i$      & log-uniform & $10^{-12}$--$10^{-2}$ \\
	\enddata
	\textbf{Notes.}
	\vspace{-0.15cm}\tablenotetext{\textrm{a}}{$g$ is given in cgs units. The value is based on the photometric mass estimate from \citet{shporer2014} derived from measured the ellipsoidal distortion amplitude.} 
	\vspace{-0.25cm}\tablenotetext{\textrm{b}}{From the Kepler light curve analysis in \citet{esteves2015}.}
	\vspace{-0.8cm}
\end{deluxetable}

The free parameters and their corresponding prior distributions are listed in Table~\ref{tab:retrieval_parameter}. The mixing ratios of the chemical species $x_i$ were assumed to be constant throughout the atmosphere. The geometric albedo contribution to the secondary eclipse depth was only considered for the TESS and Kepler photometric points. For the TP profile, the finite element approach in \textsc{Helios-r2} divides the atmosphere into discrete layers in log-pressure space, within which the temperature variation is modeled as a discretized polynomial of a predetermined order; continuity between adjacent layers is enforced. In our implementation, we used three second-order elements to model the planet's TP profile: the free parameters are the base temperature of the bottom layer $T_1$ and the relative scaling ratios $b_i$ that yield the temperatures of each successive grid point. The planet's atmospheric scale height was determined by the surface gravity $\log g_p$. That parameter, along with $R_{p}/R_{*}$ and $a/R_{*}$, was constrained by Gaussian priors based on literature measurements.

\begin{figure*}[t!]
\centering
\includegraphics[width=\linewidth]{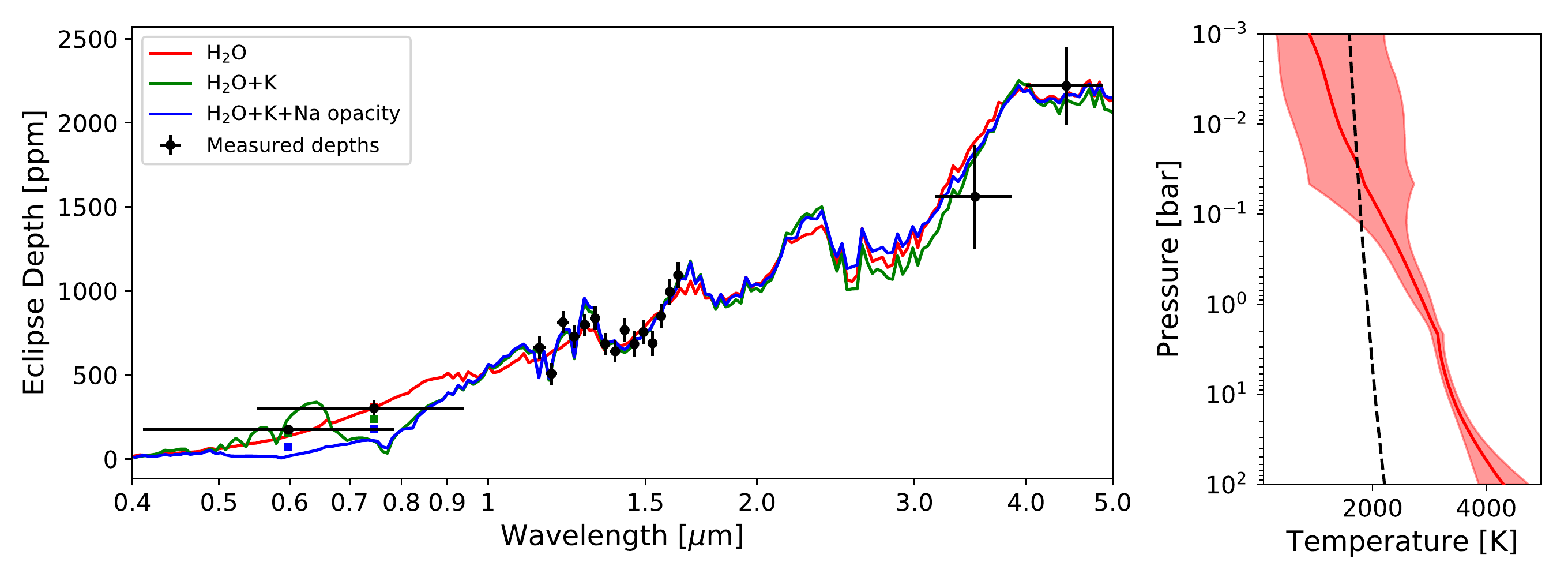}
\caption{Left panel: the median retrieved emission spectra of Kepler-13Ab from three atmospheric retrievals: (1) a model with only H$_2$O opacity, (2) a model containing H$_2$O and K, and (3) the best-performing model, which includes H$_2$O, K, and Na, with the abundance of Na scaled to the potassium abundance assuming a solar Na/K ratio. The retrieved reflected light component is not plotted. The colored squares at optical wavelengths indicate the band-integrated model fluxes for the Kepler and TESS bandpasses. The black data points are the measured eclipse depths. Right panel: the temperature--pressure profile from the H$_2$O+K+Na opacity retrieval. The shaded region indicates the 1$\sigma$ confidence interval. The black dashed line shows the condensation curve of TiO.}
\label{fig:models}
\end{figure*}

We first ran a free chemistry retrieval that fit for the abundances of a large number of atmospheric species, including H$_2$O, CO, Fe, FeH, Ti, TiO, VO, Na, K, Ca, and SH. In this and all subsequent retrievals, the remainder of the atmosphere was composed of H$_2$ and He, assuming the solar ratio for their relative elemental abundances and accounting for collision-induced absorption. From this superset, we identified those species that had significant detections by comparing the Bayesian evidence of the full retrieval and those with one or several species removed. We arrived at a small subset of species --- H$_2$O, K, Na, and TiO --- that we considered for subsequent retrievals, with no additional opacity sources.

\begin{deluxetable}{lllll}[t]
	\tablecaption{Comparison of Kepler-13Ab Secondary Eclipse Spectrum Retrievals \label{tab:comparison}}
	\tabletypesize{\scriptsize}
	%\tablecolumns{4}
	%\tablewidth{0pt}
	\tablehead{
		\multicolumn{1}{l}{Model} &
		\colhead{} &
		\multicolumn{1}{c}{$\ln Z$\tablenotemark{\scriptsize$\mathrm{a}$}} &
		\multicolumn{1}{c}{$B_1$\tablenotemark{\scriptsize$\mathrm{a}$}} &
		\multicolumn{1}{c}{$A_{g}$}
	}
	\startdata
	H$_2$O, K (+Na opacity) & & $-123.47\pm0.05$ & --- & $0.26^{+0.05}_{-0.06}$\\
	H$_2$O, K, TiO & & $-124.39\pm0.05$ & 2.51 & $0.34^{+0.06}_{-0.08}$\\
	H$_2$O, K, Na & & $-124.40\pm0.05$ & 2.53 & $0.20^{+0.07}_{-0.08}$ \\
	H$_2$O, K & & $-124.70\pm0.05$ & 3.42 & $0.05^{+0.04}_{-0.03}$\\
	H$_2$O & & $-132.56\pm0.06$ & 8870 & $0.01^{+0.02}_{-0.01}$\\
	\enddata
	\textbf{Note.}
	\vspace{-0.15cm}\tablenotetext{\textrm{a}}{$\ln Z$: logarithm of the Bayesian evidence; $B_1$: Bayes factor relative to the best-performing model.} 
	\vspace{-0.4cm}
\end{deluxetable}

In addition to retrievals where the abundances of the four aforementioned species were allowed to vary freely, we also ran models in which some species varied jointly with others, with the relative abundances fixed to the solar values (indicated in the table with the ``+opacity'' designation). This was done, for example, with Na, where the sodium abundance was scaled to the fitted potassium abundance in each iteration of the model. Such retrieval runs reduced the number of free parameters, while still accounting for the contributions of both species to the overall atmospheric opacity.

Table~\ref{tab:comparison} presents the Bayesian evidences and inferred geometric albedos for a range of retrievals. Figure~\ref{fig:models} shows the median emission spectrum models for three representative retrievals, alongside the measured TP profile from the best-performing retrieval, i.e., the run that includes H$_2$O, K, and opacity from Na assuming a solar Na/K abundance ratio. The two-dimensional posteriors and TP profiles are provided in Appendix~\ref{sec:kepler13_retrieval} for the same three retrievals. All of the models yield a non-inverted TP profile, which is necessary to match the shape of the H$_2$O absorption feature in the HST/WFC3 emission spectrum (see discussion in \citealt{beatty2017k13}). 

However, we find that the model with only H$_2$O opacity is significantly outperformed by models that include potassium, which has a broad absorption feature between 0.7 and 0.9~$\mu$m as well as a series of narrow absorptions between 1.1 and 1.3~$\mu$m. The latter features in particular provide a superior fit to the 4--5 data points at the short-wavelength end of the HST/WFC3 spectrum. Including additional species that absorb strongly in the optical (Na, TiO) led to further slight improvements to the Bayesian evidence. As illustrated in Figure~\ref{fig:models}, these species alter the detailed shape of the emission spectrum shortward of $\sim$1~$\mu$m, where only broadband photometric measurements are available; therefore, we are unable to discern any spectroscopic features, and the addition of Na or TiO primarily serves to adjust the planetary flux ratio between the Kepler and TESS bandpasses. Formally, the best-performing retrieval is the model that includes freely-varying H$_2$O and K abundance, with Na abundance scaled to the corresponding solar Na/K ratio.

Looking at the corresponding geometric albedo posteriors, we find a stark difference between models that include optical absorbers and those that do not. For the retrievals that include only H$_2$O and K, the modelled emission level at optical wavelengths is significantly higher than the corresponding best-fit blackbody ($T_{\mathrm{day}}=2786\pm160$~K; Table~\ref{tab:temps}), resulting in very low geometric albedos near zero. This is consistent with the findings in \citet{beatty2017k13}, where the forward models did not include any optical absorbers. Meanwhile, the addition of Na or TiO lowers the planetary emission in the TESS and Kepler bandpasses, thereby necessitating a non-negligible reflected light component in the overall dayside flux; the resulting best-fit albedo values range from 0.20 to 0.34. In short, the question of whether Kepler-13Ab has enhanced dayside reflectivity hinges upon the presence of optical absorbers.

The lack of a thermal inversion, combined with the relatively cool upper atmosphere, disfavors the presence of vapor-phase TiO on the dayside. In the right panel of Figure~\ref{fig:models}, we show the TiO condensation curve. At pressures lower than $\sim$10--100~mbar, the atmosphere may not be hot enough to support vapor-phase TiO. Furthermore, as discussed in depth by \citet{beatty2017k13}, the high surface gravity of Kepler-13Ab makes gravitational settling of condensed species very efficient, which may facilitate a cold trap process wherein TiO condenses out on the cooler nightside and becomes locked deep in the atmosphere \citep{parmentier2016}.

\begin{deluxetable*}{cccccccc}
\tablewidth{0pc}
\tabletypesize{\scriptsize}
\tablecaption{
    Thermal Energy Budget and Reflectivity Properties
    \label{tab:phaseintegral}
}
\tablehead{
    \colhead{Planet} &
    \colhead{$T_{\mathrm{irrad}}$ (K)}                   &
    \colhead{$T_{\mathrm{day}}$ (K)\tablenotemark{\scriptsize$\mathrm{a}$}}                     &
    \colhead{$T_{\mathrm{night}}$ (K)\tablenotemark{\scriptsize$\mathrm{a}$}}                     &
    \colhead{$\epsilon$}  &
    \colhead{$A_{B}$} &
    \colhead{$A_{g}$\tablenotemark{\scriptsize$\mathrm{b}$}} &
    \colhead{$q$} 
}
\startdata
CoRoT-2b & $2173\pm47$ & $1756^{+44}_{-43}$ & $873^{+51}_{-41}$ & $0.15\pm0.03$ & $0.29^{+0.08}_{-0.10}$ & $0.07\pm0.03$ & $4.1\pm2.2$\\
HAT-P-7b & $3058\pm76$ & $2930\pm100$ & $2520^{+240}_{-290}$ & $0.75^{+0.15}_{-0.19}$ & $-1.41^{+0.51}_{-0.57}$ &   $<0.28$ & --- \\
%KELT-1b & $3230^{+130}_{-120}$ & $1360^{+220}_{-250}$ & $0.11^{+0.08}_{-0.05}$ & $-0.26^{+0.21}_{-0.24}$ & $0.42\pm0.16$ & \\
KELT-1b & $3439\pm73$ & $3240\pm140$ & $1350^{+230}_{-260}$ & $0.11^{+0.08}_{-0.05}$ & $<$ 0.21 & $0.45\pm0.16$ & $<$ 0.5 \\
% KELT-9b & $4285^{56}_{-55}$ & $3182^{+98}_{-110}$ & $0.54\pm0.05$ & $0.25^{+0.13}_{-0.16}$ & & \\
%KELT-16b & $3030^{+150}_{-140}$ & $1520^{+410}_{-360}$ & $0.24^{+0.19}_{-0.13}$ & $-0.04^{+0.21}_{-0.27}$ & & \\
Qatar-1b & $1920\pm52$ & $1535\pm61$ & $900\pm180$ & $0.33^{+0.21}_{-0.17}$ & $0.20^{+0.15}_{-0.20}$ & $0.14\pm0.11$ & $1.4\pm1.7$\\
WASP-12b\tablenotemark{\scriptsize$\mathrm{c}$} & $3600\pm94$ &  $2935\pm85$ & $1330\pm180$ & $0.12^{+0.07}_{-0.05}$  & $0.26^{+0.11}_{-0.12}$ & $0.13\pm0.06$ & $2.0\pm1.3$\\
%WASP-18b & $3100^{+48}_{-49}$ & $1154^{+71}_{-82}$ & $0.05^{+0.02}_{-0.01}$ & $-0.05\pm0.08$ & $<$0.03 & \\
WASP-18b & $3423\pm30$ & $3151^{+59}_{-58}$ & $960^{+140}_{-170}$ & $0.03^{+0.02}_{-0.01}$ & $<$ 0.08 & $<$ 0.03 & ---\\
WASP-19b & $2942\pm48$ & $2291^{+67}_{-66}$ & $1380^{+120}_{-140}$ & $0.30^{+0.11}_{-0.09}$ & $0.31^{+0.09}_{-0.10}$ & $0.17\pm0.07$ & $1.8\pm0.9$\\
WASP-33b & $3916\pm53$ & $3232\pm49$ & $1559\pm39$ & $0.13\pm0.02$ & $0.24\pm0.06$ & $<$ 0.08 & $>$ 3.0 \\
WASP-43b & $2022\pm93$ & $1476^{+47}_{-46}$ & $640^{+100}_{-110}$ & $0.11^{+0.08}_{-0.05}$ & $0.52^{+0.10}_{-0.13}$ & $0.13\pm0.06$ & $4.0\pm2.0$\\
\enddata
\textbf{Notes.}
\vspace{-0.25cm}\tablenotetext{\textrm{a}}{Dayside and nightside brightness temperatures measured from analyses of Spitzer 4.5~$\mu$m phase curves \citep{bell2020}.}
\vspace{-0.25cm}\tablenotetext{\textrm{b}}{Geometric albedos derived from the TESS- or CoRoT-band secondary eclipse depths.}
\vspace{-0.25cm}\tablenotetext{\textrm{c}}{Weighted averages from the two full-orbit phase curve observations in 2010 and 2013.}
\vspace{-0.4cm}
\end{deluxetable*}

Meanwhile, there is no clear mechanism for cold-trapping sodium. The most prominent condensate species containing sodium is Na$_2$S, which condenses at 700--1200~K for pressures between 1~$\mu$bar and 100~bar \citep[e.g.,][]{visscher2006}. These temperatures are much lower than those found across both the dayside and nightside ($T_{\mathrm{night}}=2537\pm45$~K; \citealt{shporer2014}) of Kepler-13Ab. Therefore, vapor-phase sodium is expected to be present on the dayside of the planet and contribute significantly to the opacity at short wavelengths.

To summarize, we find that the measured secondary eclipse spectrum of Kepler-13Ab shows strong evidence for H$_2$O and K absorption, with Na opacity at optical wavelengths requiring significant reflected light across the dayside hemisphere. The retrieved geometric albedo from the H$_2$O+K+Na opacity model ($0.26^{+0.05}_{-0.06}$) is broadly consistent with the TESS-derived geometric albedo ($0.53\pm0.15$). Likewise, the modeled value is in good agreement with the Kepler-band geometric albedo we derived from the joint blackbody+reflectivity fit of the Spitzer 3.6 and 4.5~$\mu$m and Kepler-band secondary eclipse depths ($0.35^{+0.04}_{-0.05}$; Section \ref{subsec:temps}). These results show that the high dayside reflectivity derived from our previous simplistic approach is also inferred from more detailed atmospheric modeling of Kepler-13Ab, lending strong support to the temperature--albedo trend in Figure~\ref{fig:trends}. Future spectroscopic observations of KELT-1b, the other high-reflectivity object in our albedo sample, will allow for similar intensive atmospheric characterization and test the accuracy of its high inferred TESS-band geometric albedo ($0.45\pm0.16$).

\subsection{Interpretation of Geometric Albedos and Phase Integrals}\label{subsec:clouds}

Formally, the geometric albedo is evaluated at zero phase angle (superior conjunction), either at a specific wavelength or, in the case of broadband photometry, integrated over a finite range of wavelengths (e.g., the TESS bandpass).  The spherical albedo is the geometric albedo evaluated over all orbital phase angles \citep{russell1916,sobolev,seager}.  The Bond albedo $A_B$ \citep{bond} is the spherical albedo integrated over all wavelengths, weighted by the spectrum of the star (e.g., \citealt{marley99}).  The phase integral $q$ is formally defined as the ratio between the spherical and geometric albedos \citep{russell1916,sobolev,seager}, but in practice, it is often defined as (e.g., \citealt{pearl90,pearl91})
\begin{equation}
q = \frac{A_B}{A_g}.
\end{equation}
There is a rich history of measuring the geometric albedo, Bond albedo, and phase integral for the planets and moons of our Solar System \citep{bond,russell1916,horak50,hapke63,vau64,hanel81,hanel83,pearl90,pearl91}.  It is worth noting that $A_g$, $A_B$, and $q$ of Jupiter were substantially revised between the Voyager \citep{hanel81} and Cassini \citep{li18} datasets.  Meanwhile, Cassini data of Saturn, Neptune, and Uranus have not been analyzed to produce estimates for $A_g$, $A_B$ and $q$ (L. Li, private communication). 

The Bond albedo of an exoplanet can be straightforwardly estimated from its thermal phase curve. The dayside and nightside brightness temperatures depend on both the Bond albedo $A_B$ and the efficiency of heat transport from the dayside to the nightside $\epsilon$ \citep[e.g.,][]{cowanagol}:
\begin{gather}
    T_{B,p,\mathrm{day}}=T_{*}\sqrt{\frac{R_{*}}{a}}(1-A_{B})^{1/4}\left(\frac{2}{3}-\frac{5}{12}\epsilon\right)^{1/4},\label{tday}\\
    T_{B,p,\mathrm{night}}=T_{*}\sqrt{\frac{R_{*}}{a}}(1-A_{B})^{1/4}\left(\frac{\epsilon}{4}\right)^{1/4}.\label{tnight}
\end{gather}
Here, $\epsilon$ ranges from 0 to 1, corresponding to the extremes of no day--night heat recirculation and full recirculation (i.e., uniform temperature across the planet), respectively. 

To calculate the Bond albedos, we utilized dayside and nightside blackbody brightness temperatures derived from a recent uniform reanalysis of Spitzer 4.5~$\mu$m full-orbit phase curves published in \citet{bell2020}. For targets with both TESS and Spitzer secondary eclipse results, we carried out joint MCMC fits of the measured dayside and nightside temperatures using Equations~\eqref{tday} and \eqref{tnight}, while applying Gaussian priors to $T_{*}$ and $a/R_{*}$ based on values from the corresponding discovery papers and Table~\ref{tab:fit}. The resulting $A_B$ and $\epsilon$ values are listed in Table~\ref{tab:phaseintegral}, along with the published dayside and nightside brightness temperatures from the Spitzer 4.5~$\mu$m phase curve analysis and the irradiation temperatures, which is defined as $T_{\mathrm{irrad}}\equiv T_{*}\sqrt{R_{*}/a}$. In the case of KELT-1b and WASP-18b, 2$\sigma$ upper limits on $A_B$ are provided. The inferred negative Bond albedo of HAT-P-7b is a notable outlier, along with its very high nightside brightness temperature (see also discussion in \citealt{wong2016}). We note that the simple thermal balance arguments underpinning Equations~\eqref{tday} and \eqref{tnight} break down and can lead to biases in the inferred quantities when there are strong discrepancies in temperature--pressure profiles and/or compositional gradients between the dayside and nightside hemispheres. Future spectroscopically-resolved full-orbit thermal phase curves will help disentangle the various physical and chemical processes that affect the thermal energy budget on exoplanets.

\begin{figure}%[b]
\includegraphics[width=\linewidth]{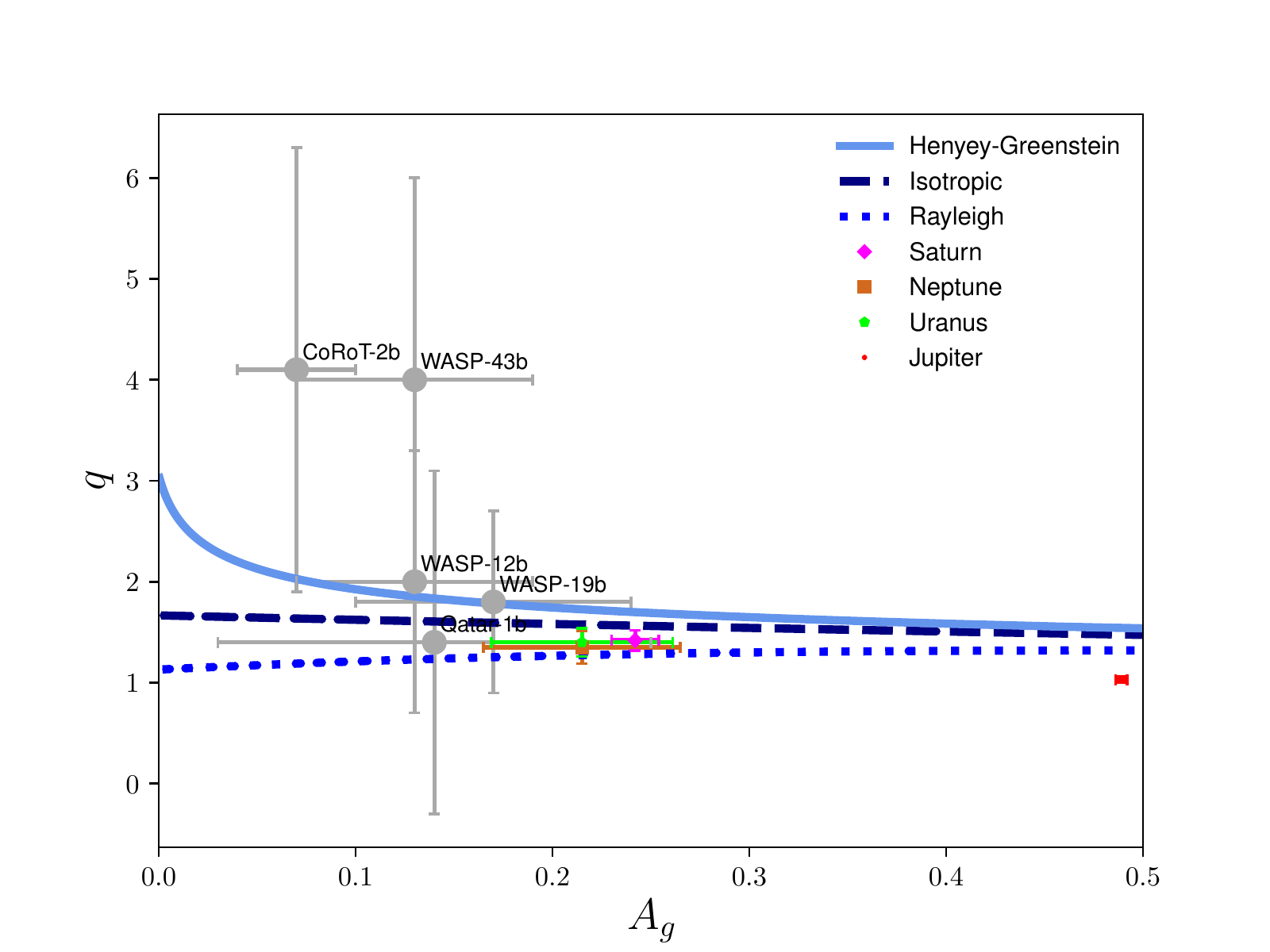}
\caption{Plot of phase integral versus geometric albedo for 5 objects in our sample. Also shown are entries for the gas and ice giants of the Solar System. The model curves corresponding to the Rayleigh, isotropic, and Henyey--Greenstein reflection laws were calculated using the theory of \cite{heng21} for homogeneous atmospheres, which relates $A_g$ and $q$ for any reflection law. For comparison with Jupiter, we show the curve corresponding to the Henyey--Greenstein reflection law with a scattering asymmetry factor of $g_0=0.4$. The uncertainties on $q$ for the planets in our TESS sample are too large to robustly exclude any reflection law.}
\label{fig:q}
\end{figure}

Combining the geometric albedos measured from TESS data in the current study and the derived Bond albedos, we estimated the values of the phase integral (Table~\ref{tab:phaseintegral}).  Figure \ref{fig:q} plots $q$ versus $A_g$ for the 5 objects in our sample with measured values of both quantities.  \cite{heng21} described an \textit{ab initio} theory for single and multiple scattering of radiation in a semi-infinite, homogeneous atmosphere that relates $A_g$, $A_B$, and $q$ in terms of fundamental physical parameters: the single-scattering albedo $\omega_0$ and the scattering asymmetry factor $g_0$.  The theory was developed for any law of reflection that depends only on the scattering angle.  For Rayleigh and isotropic scattering, $\omega_0$ is the only free parameter needed to calculate $A_g$, $A_B$, and $q$.  For the commonly used Henyey--Greenstein reflection law \citep{hg}, which describes anisotropic scattering, $-1 \le g_0 \le 1$ quantifies the degree of asymmetry; $g_0=0$ corresponds to the limit of isotropic scattering.  Isotropic and Rayleigh scattering correspond to the regime of small-particle scattering, where ``small" has a well-defined meaning: $2\pi r/\lambda \ll 1$, where $r$ is the radius of the (spherical) particle and $\lambda$ is the wavelength \citep{mie,pierrehumbert,kh18}.

\begin{figure}%[b]
\includegraphics[width=\linewidth]{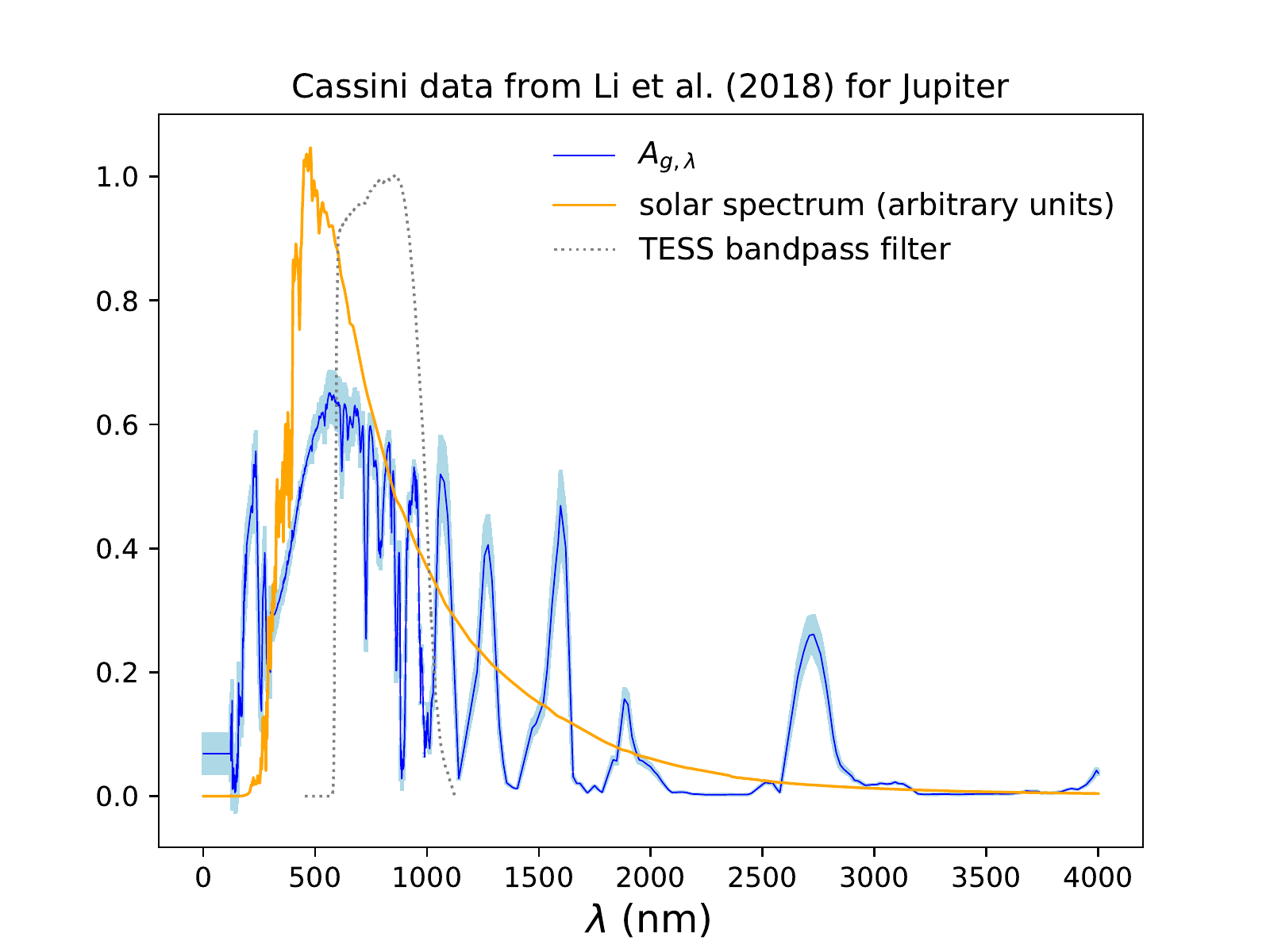}
\caption{The blue curve and shaded region denote the wavelength-dependent geometric albedo $A_{g,\lambda}$ and corresponding uncertainties derived from the Cassini data analysis of Jupiter, previously published as Figure 3 in \cite{li18}. The orange curve is the solar spectral irradiance taken from Supplementary Figure 1 of \cite{li18}. Overlaid in gray is the TESS bandpass filter. These three curves were used to estimate the geometric albedo of Jupiter integrated over the TESS bandpass: $A_g = 0.489 \pm 0.003$.}
\label{fig:li2018}
\end{figure}

\begin{figure*}
\includegraphics[width=0.5\linewidth]{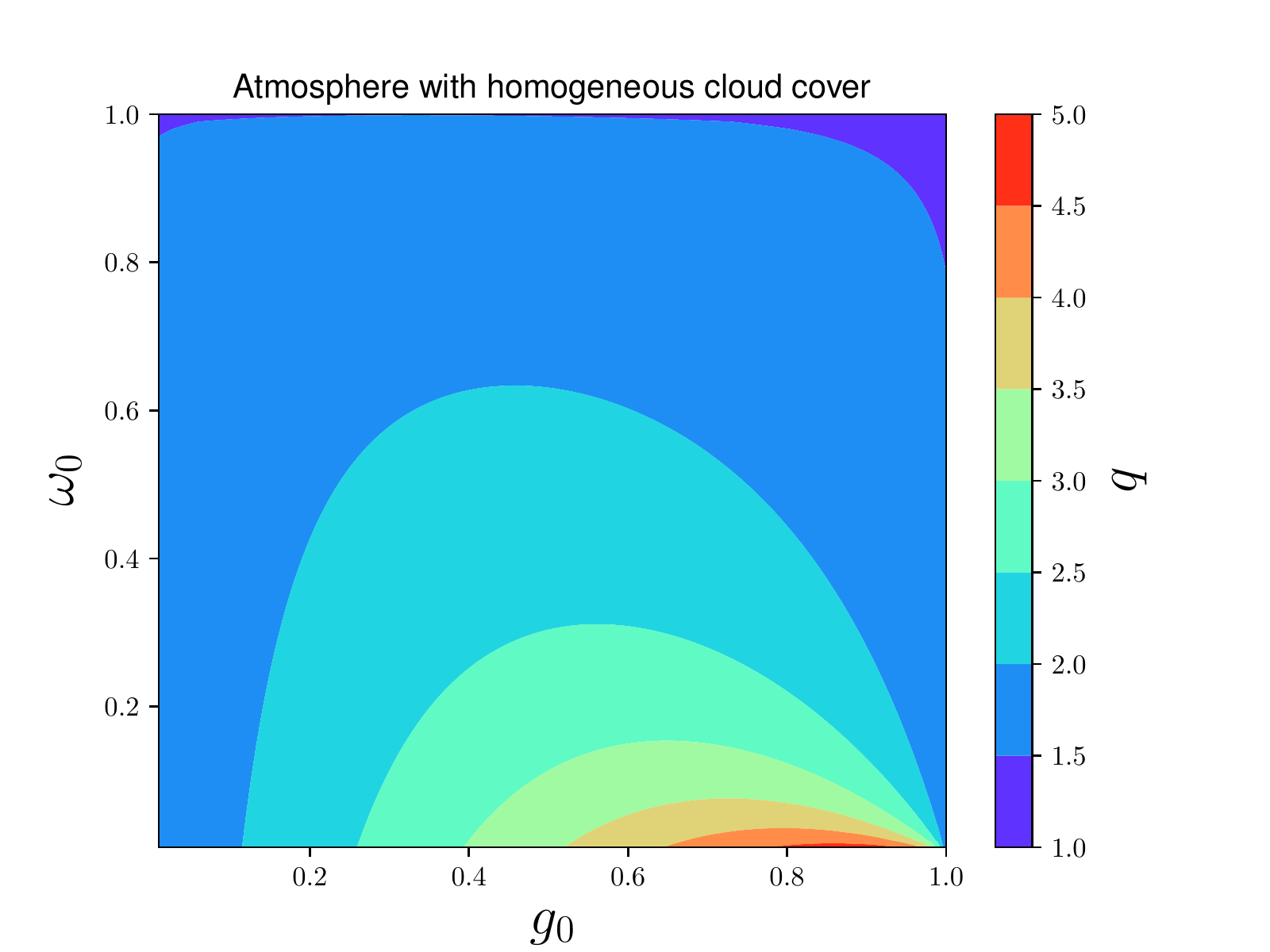}
\includegraphics[width=0.5\linewidth]{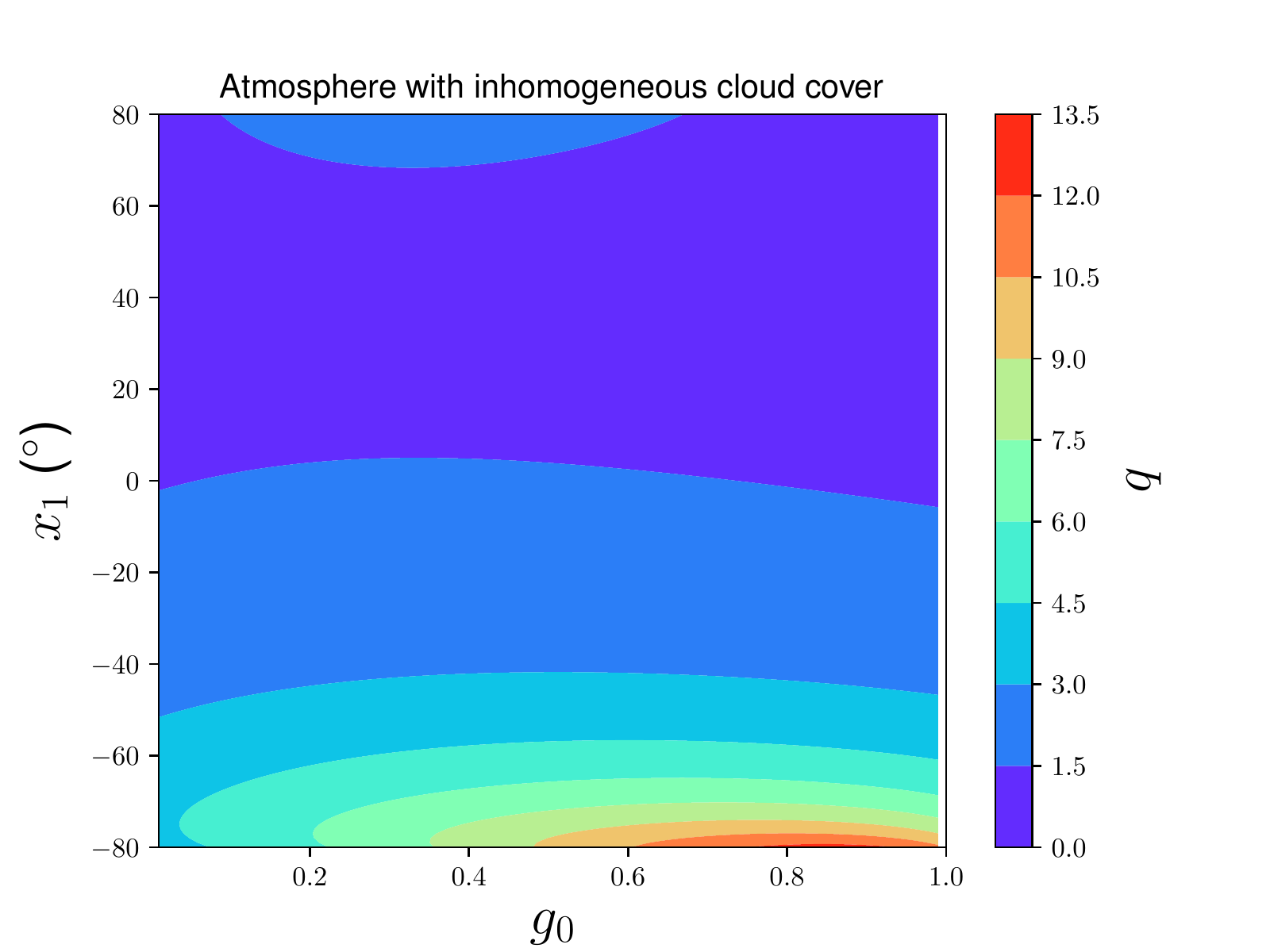}
\caption{Plots of phase-integral values $q$ computed for purely reflective, semi-infinite atmospheres with homogeneous (left panel) and inhomogeneous (right panel) cloud cover, assuming the Henyey--Greenstein reflection law \citep{heng21}. For inhomogeneous cloud cover, the atmosphere is assumed to be dark (with single-scattering albedo $\omega_0=0.01$) between the local latitudes of $x_1$ and $x_2=90^\circ$; outside of these local latitudes, the single-scattering albedo is unity. Atmospheres with high phase-integral values ($q>5$) require inhomogeneous cloud cover concentrated near the limb of the dayside hemisphere.}
\label{fig:q2}
\end{figure*}

The values for Saturn \citep{hanel83}, Neptune \citep{pearl91} and Uranus \citep{pearl90} in Figure~\ref{fig:q} were taken from Table 7 of \cite{pearl91}, which also lists Jupiter as having $A_g=0.274 \pm 0.013$, $A_B=0.343 \pm 0.032$, and $q=1.25 \pm 0.1$ \citep{hanel81}.  However, the more recent study of \cite{li18} used Cassini data to calculate $A_B=0.503 \pm 0.012$. Figure~\ref{fig:li2018} shows the geometric albedo $A_{g,\lambda}$ as a function of wavelength (taken from Figure 3 of \citealt{li18}).  To estimate $A_g$ in the TESS bandpass, we weighted $A_{g,\lambda}$ with the TESS transmission function and the solar spectral irradiance (i.e., the solar spectrum) from Supplementary Figure 1 of \cite{li18} and integrated over the TESS bandpass. This yielded $A_g = 0.489 \pm 0.003$.  The uncertainty on $A_g$ was estimated by randomly sampling the Gaussian-distributed $A_{g,\lambda}$, where the uncertainties are provided by Figure 3 of \cite{li18}. Using $A_B=0.503 \pm 0.012$, the phase integral is $q=1.03 \pm 0.03$, where we have assumed that the uncertainties on $A_g$ and $A_B$ are uncorrelated.  We checked that weighting $A_{g,\lambda}$ by a 5780~K blackbody spectrum instead of the solar spectral irradiance from \cite{li18} produces almost identical values for $A_g$ and $q$.

Figure~\ref{fig:q} illustrates how the current uncertainties on $A_g$ and $q$ are large enough that the atmospheres of Qatar-1b, WASP-12b, and WASP-19b are consistent with the Rayleigh, isotropic, and Henyey--Greenstein ($g_0=0.4$) reflection laws.  For CoRoT-2b and WASP-43b, the isotropic and Rayleigh reflection laws are weakly disfavored, but nonetheless, the measured phase-integral values lie within $1.5\sigma$ of all three theoretical curves.  In other words, the current constraints on $q$ for hot Jupiter atmospheres are uninformative.  The phase integral of Jupiter is inconsistent with Rayleigh scattering in the TESS bandpass, but robust conclusions cannot be drawn about the size of the scattering particles from analyzing $q$ and $A_g$ alone.  Detailed analysis of the Cassini Jupiter phase curves using a double Henyey--Greenstein reflection law indicates the presence of large ($g_0 \approx 0.4$ for 0.4--1~$\mu$m), possibly irregular, polydisperse aerosols in the Jovian atmosphere \citep{hl21}.  The chemistry of the clouds and hazes in the atmosphere of Jupiter remains an active area of debate and investigation (e.g., \citealt{sf02,baines19,braude20,dahl21}).

%Despite the $q$-value of Jupiter being consistent with isotropic scattering, Figure \ref{fig:q2} shows that both the $g_0=0$ and $g_0=0.4$ curves are consistent with the $q$-value of Jupiter, implying that 

The medians of the phase-integral posteriors for CoRoT-2b and WASP-43b lie around $q=4$. For homogeneous cloud cover, it is difficult to produce $q\sim 4$--5, unless $\omega_0 \sim 0$ and $g_0 \sim 1$ (see Figure \ref{fig:q2}). Using the theory developed by \cite{heng21}, we experimented with inhomogeneous cloud cover for reflective, semi-infinite atmospheres obeying the Henyey--Greenstein reflection law. This consideration was motivated by the unusual westward phase offset in the Kepler-7b phase curve \citep{demory2013,hu15,shporer2015}, as well as the general circulation models of \cite{oreshenko16} and \cite{rr17}.  We assumed that the atmosphere is dark (with a single-scattering albedo of $\omega_0=0.01$) between the local latitudes of $x_1$ and $x_2=90^\circ$, where $x_1$ was allowed to vary between $-80^\circ$ and $80^\circ$.  Outside of the region bounded by $x_1$ and $x_2$, the atmosphere is bright and perfectly reflective, with a single-scattering albedo of unity.  Figure~\ref{fig:q2} shows that such inhomogeneous atmospheres readily produce $q \sim 4$--5 and even $q \sim 10$.  The higher values of $q$ are due to a phase shift in the reflected light phase curve of a planet with inhomogeneous cloud cover, which results in a diminished value of $A_g$ while maintaining a high $A_B$ value.  The main prediction from these calculations is that high-$q$ dayside atmospheres are likely to have patchy, inhomogeneous clouds concentrated near the limb, which should produce a significant shift in the reflected light component of the visible-wavelength phase curve, as in the case of Kepler-7b.

%The high values of geometric albedo ($A_g \sim 0.4$--0.5) for KELT-1b and Kepler-13Ab remain an unsolved mystery, as such values of $A_g$ imply single-scattering albedos approaching unity if the scattering of radiation is isotropic and mediated by small particles \citep{heng21}.  Further progress on the nature of the scattering particles in KELT-1b and Kepler-13Ab will come from a detailed analysis of multi-wavelength phase curves.

\subsection{Updated Transit Ephemerides}\label{subsec:ephem}

\begin{deluxetable*}{cccccc}
\tablewidth{0pc}
\tabletypesize{\scriptsize}
\tablecaption{
    Updated Transit Ephemerides
    \label{tab:ephem}
}
\tablehead{
    \colhead{System} \vspace{-0.3cm}&
    \colhead{$N$\tablenotemark{\scriptsize$\mathrm{a}$}}                    &
    \colhead{$\Delta t$\tablenotemark{\scriptsize$\mathrm{a}$}} &
    \colhead{$T_{0}$}  &
    \colhead{$P$} &
    \colhead{$dP/dt$\tablenotemark{\scriptsize$\mathrm{b}$}} \\
    \colhead{} &
    \colhead{}                    &
    \colhead{(days)} &
    \colhead{(BJD$_{\mathrm{TDB}}-2450000$)}  &
    \colhead{(days)} &
    \colhead{(ms yr$^{-1}$)} }
\startdata
HAT-P-7 & 12 & 4021 & $5430.58175\pm0.00038$ & $2.20473639\pm0.00000047$ & $<110$\\
HAT-P-36 & 24 & 3356 & $6766.43049\pm0.00013$ & $1.32734686\pm0.00000024$ & $<20$\\
KELT-1 & 15 & 2879 & $7306.97607\pm0.00018$ & $1.21749412\pm0.00000023$ & $<36$ \\
KELT-16 & 52 & 1885 & $8056.357231\pm0.000078$ & $0.96899322\pm0.00000021$ & $<40$ \\
KELT-20 & 11 & 1397 & $8312.58576\pm0.00019$ & $3.4740985\pm0.0000016$ & $<600$ \\
KELT-23A\tablenotemark{\scriptsize$\mathrm{c}$} & \dots & \dots & $8769.612278\pm0.000060$ & $2.25528773\pm0.00000077$ & \dots \\
Qatar-1 & 81 & 3319 & $6458.466991\pm0.000042$ & $1.420024305\pm0.000000076$ & $<8.6$ \\
TrES-3  & 99 & 4822 & $5591.367536\pm0.000057$ & $1.306186358\pm0.000000060$ & $<5.9$ \\
WASP-3 & 66 & 4879 & $5362.762292\pm0.000093$ & $1.84683510\pm0.00000020$ & $<14$ \\
WASP-12 & 132 & 4338 & $6722.378206\pm0.000038$ & $1.091419740\pm0.000000029$ & $-29.1\pm2.0$\\
WASP-92 & 2 & 2590 & $6381.28419\pm0.00028$ & $2.17467334\pm0.00000054$ & \dots \\
WASP-93 & 2 & 2700 & $6079.56495\pm0.00046$ & $2.73253748\pm0.00000078$ & \dots \\
WASP-135 & 9 & 3791 & $8249.55918\pm0.00050$ & $1.4013776\pm0.0000016$ & $<150$ \\
\enddata
\textbf{Notes.}
\vspace{-0.25cm}\tablenotetext{\textrm{a}}{$N$: number of published transit-timing measurements included in fit; $\Delta t$: time baseline spanned by the timing measurements.}
\vspace{-0.25cm}\tablenotetext{\textrm{b}}{Period derivative. In the case of non-detections, the $2\sigma$ upper limit on the absolute value is given.}
\vspace{-0.25cm}\tablenotetext{\textrm{c}}{For KELT-23A, we list the updated ephemeris derived from our analysis of the long-baseline multisector full-orbit TESS light curve (Table~\ref{tab:transitonly}), which is significantly more precise than the ephemeris derived from an analogous global timing analysis using the measured mid-transit time and previous literature values.}
\vspace{-0.6cm}
\end{deluxetable*}

The new transit-timing measurements we obtained from the TESS light-curve fits provide additional time baseline to the body of published timings. To fit for updated transit ephemerides and probe for orbital period drift, we gathered timing measurements available in the peer-reviewed literature, following selection criteria analogous to those outlined in Paper 1. We included all timings that (i) have a well-specified time standard, from which we converted to $\mathrm{BJD}_{\mathrm{TDB}}$ when necessary, (ii) were derived from transit light curves that contain at least half of both ingress and egress, (iii) were fit without any applied priors on the mid-transit time, and (iv) were not affected by starspots, significant time-correlated noise, or other clearly discernible systematic features in the residuals. Many of our targets have benefited from extensive follow-up transit monitoring, and for the sake of maximizing uniformity, we relied on published global reanalyses of previous transit light curves whenever possible. The detailed light-curve vetting carried out in many of these more recent works was also used in our timing measurement selection process. Appendix~\ref{sec:timinglist} provides an exhaustive list of all transit timings included in our analysis.

For each system, we fit the transit timings to both a linear and a quadratic transit ephemeris model:
\begin{align}
T_{\mathrm{lin}}(E) &= T_{0}+PE, \label{linear}\\
T_{\mathrm{quad}}(E) &= T_{0}+PE+\frac{P}{2}\frac{dP}{dt}E^2, \label{quad}
\end{align}
where $E$, $T_{0}$, $P$, and $dP/dt$ are the transit epoch, zeroth epoch mid-transit time, orbital period, and period derivative, respectively. The zeroth epoch was set to the published transit epoch closest to the weighted average of all available transit timings.

The literature transit timings were combined with the measurements we obtained from our TESS light-curve fits using the residual permutation analysis, $T_{0,\mathrm{PB}}$. As with our light-curve fitting, we included a uniform per-point scatter scaling to ensure $\chi_r^2=1$ for cases in which the non-inflated uncertainties produced $\chi_r^2>1$. The full list of updated transit ephemerides is given in Table~\ref{tab:ephem}. For the systems where no significant period variation was detected, we provide $2\sigma$ upper limits on $|dP/dT|$. The observed minus calculated ($O-C$) timing residual plots for all systems with more than two published epochs are shown in Appendix~\ref{sec:ephemplots}. 

TESS observed KELT-23A for six sectors, and the transit ephemeris we derived from our light-curve fit significantly supersedes the result from the corresponding multiepoch transit-timing fit. As such, we simply list the TESS ephemeris for that system in the table. We also excluded Kepler-13A, which was observed throughout the second year of the TESS mission during sectors 14, 15, and 26 (Section \ref{kepler13}). The only previous source of transit timings for this system is the four-year Kepler light curve. A dedicated analysis of all individual Kepler and TESS transits (up through sector 15) was carried out in \citet{szabo2020}, resulting in an exquisite updated transit ephemeris: $T_0=2455101.708005\pm0.000013$~$\mathrm{BJD}_{\mathrm{TDB}}$, $P=1.76358762\pm0.00000003$~d. The best-fit transit ephemeris from our full TESS light-curve analysis (Table~\ref{tab:fit}) agrees with those values.

\citet{mancini2021} presented an independent transit-ephemeris analysis of the KELT-16 system, including the TESS transits, and obtained $P=0.96899340\pm0.00000018$~d, which agrees with our value at the $0.7\sigma$ level. Our set of fitted transit timings is a subset of the timings utilized in their work, due to our more stringent selection criteria.

The only system that displays significant period variation is WASP-12. This period drift has been confirmed by several previous transit-timing analyses \citep{patra2017,yee2020}, with the latter work strongly supporting the scenario of orbital decay over apsidal precession. Recently, \citet{turner2020} presented a dedicated analysis of transit and secondary eclipse light curves from TESS, reinforcing the conclusion that the orbit of WASP-12b is decaying and not precessing. The decay rate we calculated in this paper --- $dP/dt=-29.1\pm2.0$~ms~yr$^{-1}$ --- is statistically identical to the value published in \citet{yee2020}: $-29.0\pm2.0$~ms~yr$^{-1}$. Meanwhile, \citet{turner2020} obtained a slightly higher decay rate of $-32.5\pm1.6$~ms~yr$^{-1}$, which is still consistent with our value at the $1.3\sigma$ level.

Among the other targets, Qatar-1 and TrES-3 have the most precise updated ephemerides, with $2\sigma$ upper limits on period variations of less than 10~ms~yr$^{-1}$. The added time baseline that TESS will provide when it reobserves all of these systems in the extended mission will drastically tighten the current upper limits on orbital decay. Future transit-timing measurements of WASP-12 will yield even more exquisite precision on the decay rate and produce unprecedented constraints on the host star's tidal quality factor, with important implications for the study of star--planet interactions in this unique system.

\section{Conclusions}\label{sec:con}

In this paper, we presented the results from our systematic phase-curve study of previously-discovered transiting systems observed during the second year of the TESS primary mission, consisting of TESS sectors 14--26 from 2019 July 18 to 2020 July 4. We carried out a uniform data processing and light-curve fitting analysis on 15 systems that satisfied our target-selection criteria. The primary findings of our study are summarized below:
\begin{itemize}
\item Seven systems show statistically significant secondary eclipses: HAT-P-7b ($127^{+33}_{-32}$ ppm), KELT-1b ($388^{+67}_{-65}$ ppm), KELT-9b ($630^{+18}_{-17}$ ppm), KELT-16b ($410^{+130}_{-120}$ ppm), KELT-20b ($111^{+35}_{-36}$ ppm), Kepler-13Ab ($301^{+46}_{-42}$ ppm), and WASP-12b ($443^{+86}_{-85}$ ppm). The full results from our light-curve fits are provided in Table~\ref{tab:fit}.
\item All seven systems also display atmospheric brightness modulation, with measured semiamplitudes of $56^{+14}_{-13}$, $176^{+26}_{-30}$, $271.9^{+9.0}_{-8.9}$, $175^{+64}_{-62}$, $43^{+13}_{-11}$, $151^{+15}_{-16}$, and $264^{+33}_{-30}$ ppm, respectively. For two systems --- KELT-9b and WASP-12b --- we detected significant eastward offsets in the location of the dayside brightness maximum, with magnitudes of $2\overset{\circ}{.}6^{+1\overset{\circ}{.}4}_{-1\overset{\circ}{.}3}$ and $13\overset{\circ}{.}2\pm5\overset{\circ}{.}7$, respectively.
\item We measured significant ellipsoidal distortion modulation on KELT-1, Kepler-13A, and WASP-12. The amplitudes of these signals are in good agreement with the predicted values from theoretical models. For KELT-9, we repeated the analysis from \citet{wong2020kelt9} with the updated photometry from the TESS SPOC pipeline and recovered the additional time-varying irradiation signal caused by the planet's near-polar orbit, with has a semiamplitude of $60.1^{+9.4}_{-9.1}$~ppm (Table~\ref{tab:kelt9}).
\item For the remaining eight systems --- HAT-P-36, KELT-23A, Qatar-1, TrES-3, WASP-3, WASP-92, WASP-93, and WASP-135 --- no significant phase-curve signals were detected. The results of our transit light-curve fits of these systems are provided in Table \ref{tab:transitonly}, with upper limits and marginal secondary eclipse depths listed in Table~\ref{tab:bad}.
\item We self-consistently computed dayside blackbody brightness temperatures and TESS-band geometric albedos for objects in our sample with published Spitzer 3.6 and 4.5 $\mu$m secondary eclipse depths (Table \ref{tab:temps}). KELT-1b and Kepler-13Ab show enhanced albedos ($0.45\pm0.16$ and $0.53\pm0.15$). These high-reflectivity endmembers strengthen the statistical significance of the previously-reported trend between increasing dayside temperature and increasing TESS-band geometric albedo for objects with $1500<T_{\mathrm{day}}<3000$~K. 
\item For Kepler-13Ab, we sought to confirm the high inferred albedo by leveraging the HST/WFC3 spectrophotometric dataset obtained by \citet{beatty2017k13} and carrying out a detailed atmospheric retrieval analysis of the secondary eclipse spectrum. We found a decreasing temperature--pressure profile and strong detections of H$_2$O and K absorption, as well as likely Na opacity at optical wavelengths. The presence of Na requires additional reflected light on the dayside hemisphere to match the measured TESS and Kepler secondary eclipse depths, yielding a high geometric albedo consistent with the value we obtained from the simple blackbody model.
\item Using Spitzer-derived Bond albedos and TESS/CoRoT-derived geometric albedos, we estimated the phase integral for five objects in our combined primary mission target list and compared them to \textit{ab initio} calculations of reflective atmospheres with homogeneous cloud cover. The large uncertainties on the phase integrals of CoRoT-2b, Qatar-1b, WASP-12b, WASP-19b, and WASP-43b do not allow us to rule out any reflection law. Phase integrals with values of roughly 4--5 are indicative of atmospheres with inhomogeneous cloud cover --- a hypothesis that may be tested with future high-precision visible-wavelength phase curves.
\item Combining transit timings from our TESS light-curve fits and literature values, we calculated updated transit ephemerides. We obtained an orbital period decay rate of $29.1\pm2.0$~ms~yr$^{-1}$ for WASP-12b, consistent with previous measurements.
\end{itemize}

% concluding future-looking paragraphs
Over the course of our two-year TESS phase-curve study, we have measured statistically significant secondary eclipses and/or phase-curve signals for 17 known transiting systems, with a consistent analysis framework that ensures a uniform set of results. This number is comparable to the size of the Kepler phase-curve target list \citep{esteves2015} and firmly establishes the contribution of TESS in the realm of time-domain exoplanet science. As TESS continues its full-sky survey throughout the current extended mission, and possible additional extended missions, several avenues for further study promise to expand the utility of TESS phase curves and provide additional insight into ensemble-level trends.

Improving the signal-to-noise of the phase curve measurements will be a priority for follow-up study. With new sectors of data from the current extended mission, as well as possible additional photometry from further extensions of the TESS mission through the middle of the decade, the uncertainties on the secondary eclipse depths and atmospheric brightness modulation amplitudes will decrease significantly. For the brightest targets, the precision may rival that obtained for the Kepler phase curves, which benefited from four years of near-continuous observation. These high-precision phase-folded light curves will enable exquisite resolution of the longitudinal brightness distribution and, when combined with analogous high-quality phase-curve observations at infrared wavelengths, produce detailed cloud and temperature maps. 

Increased precision on the secondary eclipse depths will be especially consequential for our understanding of the trend between dayside temperature and geometric albedo. With the longer baseline from the extended mission, additional targets will populate the sample as currently marginal signals become robust detections and new near-ecliptic systems get observed in future sectors. Of particular interest is extending the TESS-band geometric albedo sample to cooler temperatures ($T_{\mathrm{day}}<1500$~K). Likewise, reducing the uncertainties on the dayside and nightside brightness temperatures will yield improved Bond albedo measurements and begin to unlock the explanatory potential of exoplanet phase-integral measurements for constraining atmospheric scattering laws.

The addition of more TESS data from extended missions will also enable a detailed study of variability in atmospheric properties. Such variability has been claimed by a few authors \citep[e.g.,][but see also \citealt{lally2020}]{jackson2019,armstrong2016}, and a low level of variability is expected from recent theoretical modeling \citep{komacek2020}.

Another fruitful avenue for further study is expanding the wavelength coverage of secondary eclipse spectra using spectroscopic measurements with current and near-future facilities, including HST and the James Webb Space Telescope (JWST). Our retrieval analysis of Kepler-13Ab (Section \ref{subsec:modeling}) offers a glimpse into the type of intensive atmospheric characterization that can be done when combining higher-resolution near-infrared emission spectra with optical and thermal infrared measurements. The exquisite capabilities of JWST in particular will produce detailed temperature--pressure profiles and chemical abundance constraints for a broad range of exoplanet atmospheres, allowing us to explore the interplay between atmospheric composition, heat transport, and cloud cover. By constructing a more complete picture of the dayside atmosphere, we can solidify the inferred optical geometric albedos and definitively assess the albedo--temperature trend that we have uncovered in our multi-year TESS phase-curve study.

%===============================================================================
\acknowledgments

Funding for the TESS mission is provided by NASA’s Science Mission directorate. This paper includes data collected by the TESS mission, which are publicly available from the Mikulski Archive for Space Telescopes (MAST). Resources supporting this work were provided by the NASA High-End Computing (HEC) Program through the NASA Advanced Supercomputing (NAS) Division at Ames Research Center for the production of the SPOC data products. I.W. is supported by a Heising-Simons \textit{51 Pegasi b} postdoctoral fellowship. K.H. acknowledges a honorary professorship from the University of Warwick. T.D. acknowledges support from MIT’s Kavli Institute as a Kavli postdoctoral fellow. We thank Liming Li for useful correspondence and providing data from \cite{li18} in electronic form. We also thank an anonymous referee for helpful comments that improved the manuscript.

%===============================================================================

\appendix 
\section{List of Light-curve Segments}\label{sec:segmentlist}
\restartappendixnumbering

Table~\ref{tab:segments} lists the light-curve segments for the 15 systems analyzed in this paper. The three-number sequence assigned to each segment in the second column denotes the TESS sector, spacecraft orbit (two per sector), and sequential data-segment number. Data segments lasting less than 1~day were excluded from our analysis and are not listed here. The third and fourth columns show the number of data points before and after removing flagged points, flux ramps, and outliers; the first and last time stamps of each data segment are also tabulated. The seventh column lists the order of the polynomial used in the detrending function (see Section \ref{subsec:model}). The final column describes any removed flux ramps, as well as segments that were excluded from the analysis due to severe systematics.

\startlongtable
\begin{deluxetable}{cccccccc}
\tablewidth{0pc}
\tabletypesize{\scriptsize}
\tablecaption{
    Summary of Light-Curve Segments
    \label{tab:segments}
}
\tablehead{
    \colhead{Target} &
    \colhead{Segment\tablenotemark{\scriptsize$\mathrm{a}$}}                     &
    \colhead{$n_{\mathrm{raw}}$\tablenotemark{\scriptsize$\mathrm{b}$}}                     &
    \colhead{$n_{\mathrm{trimmed}}$\tablenotemark{\scriptsize$\mathrm{b}$}} &
    \colhead{$T_{\mathrm{start}}$\tablenotemark{\scriptsize$\mathrm{c}$}}  &
    \colhead{$T_{\mathrm{end}}$\tablenotemark{\scriptsize$\mathrm{c}$}} & 
    \colhead{Order\tablenotemark{\scriptsize$\mathrm{d}$}} & 
    \colhead{Comments}
}
\startdata
HAT-P-7 & 14-1-1 & 3283 & 3188 &  683.356 &  687.899 & 2 & \\
        & 14-1-2 & 3240 & 3123 &  687.910 &  692.399 & 1 & \\
        & 14-1-3 & 2867 & 2805 &  692.410 &  696.391 & 1 & \\
        & 14-2-1 & 3095 & 2493 &  698.098 &  701.628 & 0 & removed 0.75~d from start\\
        & 14-2-2 & 3060 & 2992 &  701.639 &  705.878 & 0 & \\
        & 14-2-3 & 3109 & 3044 &  705.889 &  710.206 & 3 & \\
        & 15-1-1 & 3095 & 3004 &  711.367 &  715.649 & 2 & \\
        & 15-1-2 & 3060 & 3001 &  715.660 &  719.899 & 0 & \\
        & 15-1-3 & 1351 & 1155 &  719.910 &  721.585 & 1 & \\
        & 15-2-1 & 3100 & 2978 &  724.943 &  729.232 & 2 & \\
        & 15-2-2 & 3060 & 3004 &  729.243 &  733.482 & 3 & \\
        & 15-2-3 & 1602 & 1428 &  733.493 &  735.517 & 1 & \\
\hline
HAT-P-36 & 22-1-1 & 4187 & 3921 &  900.358 &  905.964 & 6 & \\
        & 22-1-2 & 4700 & 4598 &  905.975 &  912.507 & 5 & \\
        & 22-2-1 & 3758 & 2809 &  916.352 &  920.359 & 5 & removed 1.00~d from start\\
        & 22-2-2 & 4410 & 4354 &  920.370 &  926.497 & 7 & \\
\hline
KELT-1 & 17-1-1 & 2820 & 2700 &  764.689 &  768.589 & 6 & \\
        & 17-1-2 & 2790 & 2707 &  768.600 &  772.464 & 5 & \\
        & 17-2-1 & 2923 & 2672 &  777.986 &  781.776 & 7 & removed 0.25~d from start\\
        & 17-2-2 & 2880 & 2302 &  782.538 &  785.776 & 6 & removed 0.75~d from start\\
        & 17-2-3 & 1548 & 1022 &  785.788 &  787.235 & 0 & removed 0.50~d from end\\
\hline
KELT-9 & 14-1-1 & 3283 & 2636 &  683.562 &  687.782 & 1 & \\
 (no transits) & 14-1-2 & 3240 & 1387 &  688.005 &  690.243 & 3 & removed 0.50~d from end\\
        & 14-1-3 & 2867 & 945 &  694.576 &  696.142 & 2 &  removed 0.50~d from start\\
        & & & & & & & and 0.25~d from end\\
        & 14-2-1 & 3095 & 2357 &  697.597 &  701.629 & 3 & removed 0.25~d from start\\
        & 14-2-2 & 3060 & 2141 &  701.640 &  705.138 & 6 & \\
        & 14-2-3 & 3109 & 1125 &  708.186 &  709.999 & 3 & \\
        & 15-1-1 & 3095 & 2456 &  711.703 &  715.650 & 4 & removed 0.25~d from start\\
        & 15-1-2 & 3060 & 2509 &  715.661 &  719.900 & 2 & \\
        & 15-2-1 & 3100 & 2412 &  725.285 &  729.233 & 2 & removed 0.25~d from start\\
        & 15-2-2 & 3060 & 2508 &  729.246 &  733.483 & 4 & \\
        & 15-2-3 & 2822 & \dots &  733.494 &  734.408 & \dots & large gap\\
\hline
KELT-16 & 15-1-1 & 3095 & 2837 &  711.369 &  715.651 & 1 & removed 0.25~d from start\\
        & 15-1-2 & 3060 & 3006 &  715.662 &  719.901 & 3 & \\
        & 15-2-1 & 3100 & 2141 &  726.197 &  729.234 & 1 & removed 1.25~d from start\\
        & 15-2-2 & 3060 & 2996 &  729.245 &  733.484 & 2 & \\
\hline
KELT-20 & 14-1-1 & 3283 & 3209 &  683.357 &  687.901 & 1 & \\
        & 14-1-2 & 3240 & 1960 &  687.912 &  690.746 & 2 & \\
        & 14-2-1 & 3095 & 3019 &  697.348 &  701.630 & 1 & \\
        & 14-2-2 & 3060 & 2260 &  701.641 &  704.857 & 3 & \\
\hline
KELT-23A & 14-1-1 & 3283 & 3196 &  683.353 &  687.896 & 1 & \\
        & 14-1-2 & 3240 & 3121 &  687.907 &  692.396 & 2 & \\
        & 14-1-3 & 2867 & \dots &  692.407 &  696.387 & \dots & severe systematics\\
        & 14-2-1 & 3095 & 3024 &  697.343 &  701.625 & 1 & \\
        & 14-2-2 & 3060 & 2649 &  702.136 &  705.875 & 2 & removed 0.50~d from start\\
        & 14-2-3 & 1494 & 1141 &  705.886 &  707.508 & 1 & removed 0.25~d from end\\
        & 15-1-1 & 3095 & 3020 &  711.364 &  715.646 & 2 & \\
        & 15-1-2 & 3077 & 2164 &  715.657 &  718.727 & 1 & removed 1.00~d from end\\
        & 15-2-1 & 3100 & 2978 &  724.940 &  729.229 & 1 & \\
        & 15-2-2 & 3060 & 2287 &  729.240 &  732.480 & 1 & removed 1.00~d from end\\
        & 16-1-1 & 4238 & 3446 &  739.651 &  744.520 & 2 & removed 1.00~d from start\\
        & 16-1-2 & 2288 & \dots &  744.532 &  747.509 & \dots & severe systematics\\
        & 16-2-1 & 4235 & 4144 &  751.655 &  757.521 & 4 & \\
        & 16-2-2 & 2468 & \dots &  757.532 &  760.760 & \dots & severe systematics\\
        & 17-1-1 & 2820 & 2704 &  764.683 &  768.583 & 2 & \\
        & 17-1-2 & 2790 & 2713 &  768.594 &  772.458 & 3 & \\
        & 17-2-1 & 2923 & 2509 &  778.229 &  781.771 & 2 & removed 0.50~d from start\\
        & 17-2-2 & 2880 & 2826 &  781.782 &  785.771 & 5 & \\
        & 21-1-1 & 4530 & 4027 &  870.937 &  876.711 & 3 & removed 0.50~d from start\\
        & 21-1-2 & 4500 & 4403 &  876.722 &  882.961 & 1 & \\
        & 21-2-1 & 4717 & 4161 &  885.511 &  891.461 & 3 & removed 0.50~d from start\\
        & 21-2-2 & 4543 & 4078 &  891.472 &  897.282 & 1 & removed 0.50~d from end\\
        & 23-1-1 & 4342 & 2401 &  931.830 &  936.148 & 4 & removed 1.50~d from start\\
        & 23-1-2 & 3394 & 3310 &  936.159 &  940.872 & 1 & \\
        & 23-2-1 & 3584 & 2611 &  946.112 &  949.877 & 4 & removed 1.00~d from start\\
        & 23-2-2 & 3592 & 3535 &  949.888 &  954.876 & 2 & \\
\hline
Kepler-13A & 14-1-1 & 3283 & 3023 &  683.609 &  687.899 & 3 & removed 0.25~d from start\\
        & 14-1-2 & 3240 & 3129 &  687.910 &  692.399 & 0 & \\
        & 14-1-3 & 2867 & 2801 &  692.410 &  696.391 & 3 & \\
        & 14-2-1 & 3095 & 3003 &  697.346 &  701.628 & 2 & \\
        & 14-2-2 & 3060 & 2983 &  701.639 &  705.878 & 1 & \\
        & 14-2-3 & 3109 & 3045 &  705.889 &  710.206 & 2 & \\
        & 15-1-1 & 3095 & 3006 &  711.367 &  715.649 & 0 & \\
        & 15-1-2 & 3060 & 2987 &  715.660 &  719.899 & 1 & \\
        & 15-1-3 & 1345 & 1139 &  719.910 &  721.577 & 1 & \\
        & 15-2-1 & 3100 & 2966 &  724.943 &  729.232 & 0 & \\
        & 15-2-2 & 3060 & 3004 &  729.243 &  733.482 & 0 & \\
        & 15-2-3 & 1587 & 1414 &  733.493 &  735.496 & 0 & \\
        & 26-1-1 & 4350 & 3162 & 1010.769 & 1015.296 & 0 & removed 0.50~d from start\\
        & & & & & & & and 1.00~d from end \\
        & 26-1-2 & 4239 & 3095 & 1016.305 & 1020.692 & 1 & removed 1.50~d from end\\
        & 26-2-1 & 4355 & 3856 & 1023.617 & 1029.149 & 1 & removed 0.50~d from start\\
        & 26-2-2 & 4304 & 3838 & 1029.160 & 1034.635 & 0 & removed 0.50~d from end\\
%\hline
%MASCARA-5 & 15-1-1 & 3095 & 3016 &  711.367 &  715.649 & 4 & \\
%        & 15-1-2 & 3060 & 2975 &  715.660 &  719.899 & 2 & \\
%        & 15-1-3 & 2844 & 2763 &  719.910 &  723.859 & 4 & \\
%        & 15-2-1 & 3100 & 2308 &  725.945 &  729.233 & 2 & removed 1.00~d from start\\
%        & 15-2-2 & 3060 & 2990 &  729.244 &  733.483 & 0 & \\
%        & 15-2-3 & 2822 & 2421 &  733.494 &  736.912 & 3 & removed 0.50~d from end\\
%        & 16-1-1 & 4238 & 3433 &  739.405 &  744.274 & 0 & removed 0.75~d from start\\
%        & & & & & & & and 0.25~d from end \\
%        & 16-1-2 & 3284 & 2897 &  744.535 &  748.644 & 2 & removed 0.25~d from end\\
%        & 16-2-1 & 4235 & 3601 &  752.410 &  757.524 & 2 & removed 0.75~d from start\\
%        & 16-2-2 & 3691 & 2788 &  758.537 &  762.460 & 1 & removed 1.00~d from start\\
\hline
Qatar-1 & 17-1-1 & 2820 & 2352 &  765.189 &  768.586 & 2 & removed 0.50~d from start\\
        & 17-1-2 & 2790 & 2708 &  768.597 &  772.461 & 1 & \\
        & 17-1-3 & 1730 & 1196 &  772.472 &  774.172 & 1 & removed 0.50~d from end\\
        & 17-2-1 & 2923 & 2338 &  778.483 &  781.773 & 1 & removed 0.75~d from start\\
        & 17-2-2 & 2880 & 2820 &  781.784 &  785.773 & 2 & \\
        & 17-2-3 & 1636 & 1475 &  785.784 &  787.855 & 1 & \\
        & 21-1-1 & 4438 & 3477 &  871.757 &  876.709 & 0 & removed 1.00~d from start\\
        & 21-1-2 & 4500 & 4413 &  876.720 &  882.959 & 2 & \\
        & 21-2-1 & 4430 & 4161 &  885.517 &  891.458 & 2 & removed 0.25~d from start\\
        & 21-2-2 & 4543 & 4461 &  891.470 &  897.779 & 4 & \\
        & 24-1-1 & 9042 & 7753 &  957.294 &  968.346 & 0 & removed 1.50~d from start\\
        & 24-2-1 & 5622 & 4773 &  970.271 &  977.062 & 5 & removed 1.00~d from start\\
        & 24-2-2 & 3749 & 3675 &  977.074 &  982.279 & 0 & \\
        & 25-1-1 & 3997 & 3871 &  983.632 &  989.167 & 3 & \\
        & 25-1-2 & 4649 & 4472 &  989.178 &  995.634 & 2 & \\
        & 25-2-1 & 4542 & 4200 &  997.167 & 1003.209 & 2 & removed 0.25~d from start\\
        & 25-2-2 & 4382 & 4278 & 1003.220 & 1009.305 & 0 & \\
\hline
TrES-3 & 25-1-1 & 3997 & 3880 &  983.635 &  989.170 & 5 & \\
(SAP)   & 25-1-2 & 4649 & 4483 &  989.181 &  995.637 & 4 & \\
        & 25-2-1 & 4542 & 4379 &  996.920 & 1003.212 & 4 & \\
        & 25-2-2 & 4382 & 4275 & 1003.223 & 1009.308 & 6 & \\
        & 26-1-1 & 4350 & 4226 & 1010.270 & 1016.295 & 4 & \\
        & 26-1-2 & 4239 & 4134 & 1016.307 & 1022.193 & 5 & \\
        & 26-2-1 & 4355 & 3160 & 1024.619 & 1029.150 & 2 & removed 1.50~d from start\\
        & 26-2-2 & 4304 & 4176 & 1029.161 & 1035.137 & 2 & \\
\hline
WASP-3 & 26-1-1 & 4350 & 4046 & 1010.520 & 1016.295 & 4 & removed 0.25~d from start\\
        & 26-1-2 & 4239 & 3057 & 1017.805 & 1022.193 & 4 & removed 1.50~d from start\\
        & 26-2-1 & 4355 & 4208 & 1023.118 & 1029.150 & 4 & \\
        & 26-2-2 & 4304 & 4179 & 1029.161 & 1035.137 & 4 & \\
\hline
WASP-12 & 20-1-1 & 3910 & 3768 &  842.510 &  847.922 & 3 & \\
        & 20-1-2 & 3870 & 3788 &  847.935 &  853.298 & 0 & \\
        & 20-1-3 & 1119 & 1094 &  853.310 &  854.862 & 1 & \\
        & 20-2-1 & 3029 & 2641 &  858.200 &  861.944 & 1 & removed 0.25~d from start\\
        & 20-2-2 & 3959 & 3874 &  861.955 &  867.444 & 1 & \\
        & 20-2-3 & 989 & 952 &  867.455 &  868.827 & 1 & \\
\hline
WASP-92 & 23-1-1 & 4959 & 3709 &  929.971 &  936.148 & 0 & removed 0.50~d from start\\
        & 23-1-2 & 3394 & 2981 &  936.160 &  940.372 & 1 & removed 0.50~d from end\\
        & 23-2-1 & 4214 & 3955 &  944.235 &  949.878 & 3 & \\
        & 23-2-2 & 3592 & 3535 &  949.889 &  954.876 & 1 & \\
        & 24-1-1 & 9042 & 8803 &  955.797 &  968.349 & 0 & \\
        & 24-2-1 & 5622 & 5496 &  969.274 &  977.065 & 0 & \\
        & 24-2-2 & 3749 & 3655 &  977.077 &  982.282 & 0 & \\
        & 25-1-1 & 3997 & 3698 &  983.886 &  989.170 & 0 & removed 0.25~d from start\\
        & 25-1-2 & 4649 & 4476 &  989.181 &  995.636 & 0 & \\
        & 25-2-1 & 4542 & 3999 &  997.421 & 1003.211 & 0 & removed 0.50~d from start\\
        & 25-2-2 & 4382 & 4276 & 1003.222 & 1009.307 & 4 & \\
\hline
WASP-93 & 17-1-1 & 2820 & 2518 &  764.938 &  768.589 & 2 & removed 0.25~d from start\\
        & 17-1-2 & 2790 & 2704 &  768.600 &  772.464 & 0 & \\
        & 17-2-1 & 2923 & 2832 &  777.734 &  781.776 & 1 & \\
        & 17-2-2 & 2880 & 2823 &  781.787 &  785.776 & 2 & \\
        & 17-2-3 & 1226 & \dots &  785.787 &  787.289 & \dots & severe systematics\\
\hline
WASP-135 & 26-1-1 & 4350 & 4229 & 1010.271 & 1016.296 & 2 & \\
        & 26-1-2 & 4239 & 3582 & 1017.056 & 1022.193 & 3 & removed 0.75~d from start\\
        & 26-2-1 & 4355 & 4227 & 1023.118 & 1029.150 & 2 & \\
        & 26-2-2 & 4304 & 3308 & 1029.911 & 1034.636 & 1 & removed 0.75~d from start\\
        & & & & & & & and 0.50~d from end \\
\enddata
\textbf{\\}
\textbf{Notes.}
\vspace{-0.25cm}\tablenotetext{\textrm{a}}{The numbers indicate the TESS sector, spacecraft orbit (two per sector), and segment number, respectively.}
\vspace{-0.25cm}\tablenotetext{\textrm{b}}{Number of data points in each data segment before and after removing flagged points, outliers, and flux ramps.}
\vspace{-0.25cm}\tablenotetext{\textrm{c}}{Start and end times of each data segment ($\mathrm{BJD}_{\mathrm{TDB}}-2458000$).}
\vspace{-0.25cm}\tablenotetext{\textrm{d}}{Order of the polynomial systematics detrending model used in the final joint fits.}
\end{deluxetable}

\section{Raw and Corrected Light Curves}\label{sec:rawplots}
\restartappendixnumbering

Figures~\ref{fig:raw1} and \ref{fig:raw2} present the light-curve plots for the 15 systems studied in this paper. The raw photometry is shown in the top panels, with the spacecraft momentum dumps denoted by the vertical blue lines. The bottom panels display the photometry after outlier removal, ramp trimming, and systematics detrending.

\begin{figure*}[t!]
    \centering
    \includegraphics[width=0.9\linewidth]{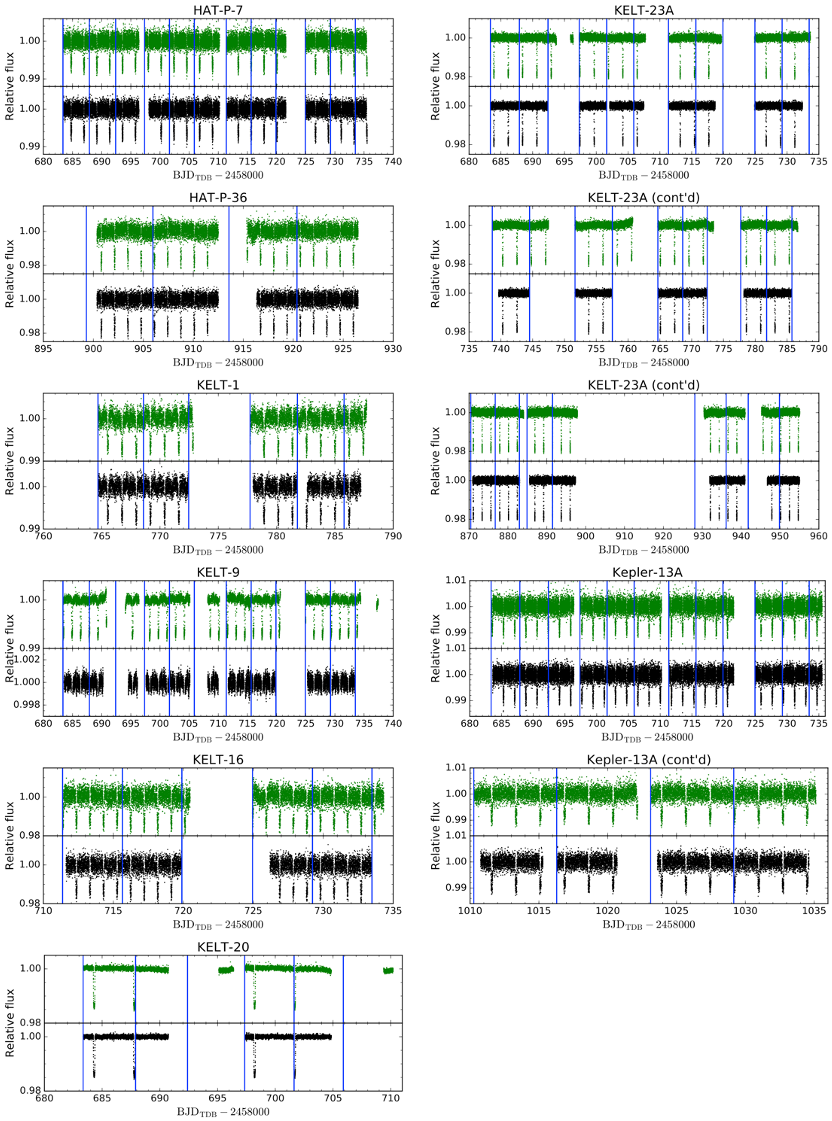}
    \caption{Light curves for 8 of the 15 targets analyzed in this paper. The top and bottom panels show the photometry before and after trimming flux ramps and correcting for systematics trends. The vertical blue lines indicate the momentum dumps. For several systems, the full light curves are split across multiple plots for clarity.}
    \label{fig:raw1}
\end{figure*}

\begin{figure*}[t!]
    \centering
    \includegraphics[width=0.9\linewidth]{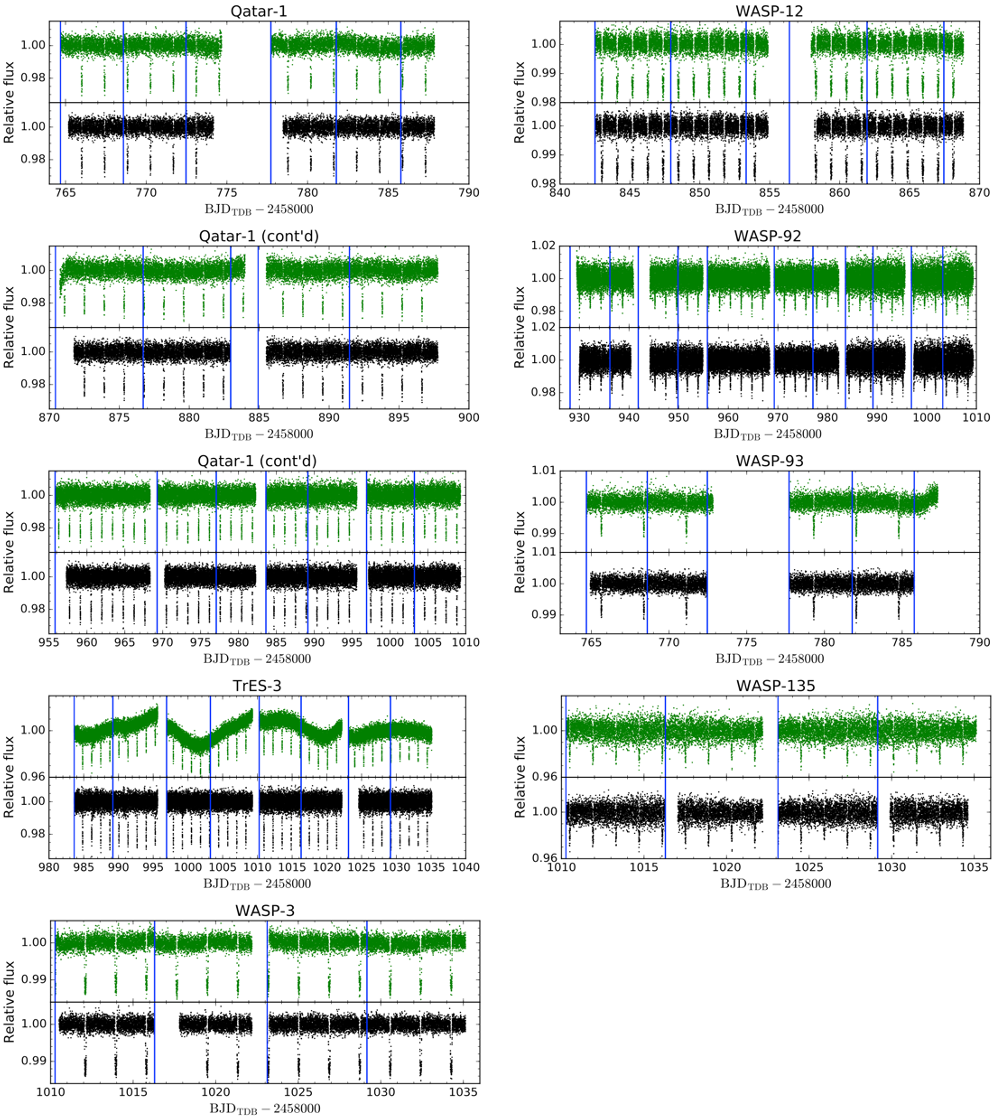}
    \caption{Continuation of Figure \ref{fig:raw1}. Note the stellar variability in the TrES-3 light curve.}
    \label{fig:raw2}
\end{figure*}

\pagebreak

\section{Retrieval Results for Kepler-13A\texorpdfstring{\MakeLowercase{b}}{b}}\label{sec:kepler13_retrieval}
\restartappendixnumbering

\begin{figure*}[t!]
  \centering
  \includegraphics[width=0.48\linewidth]{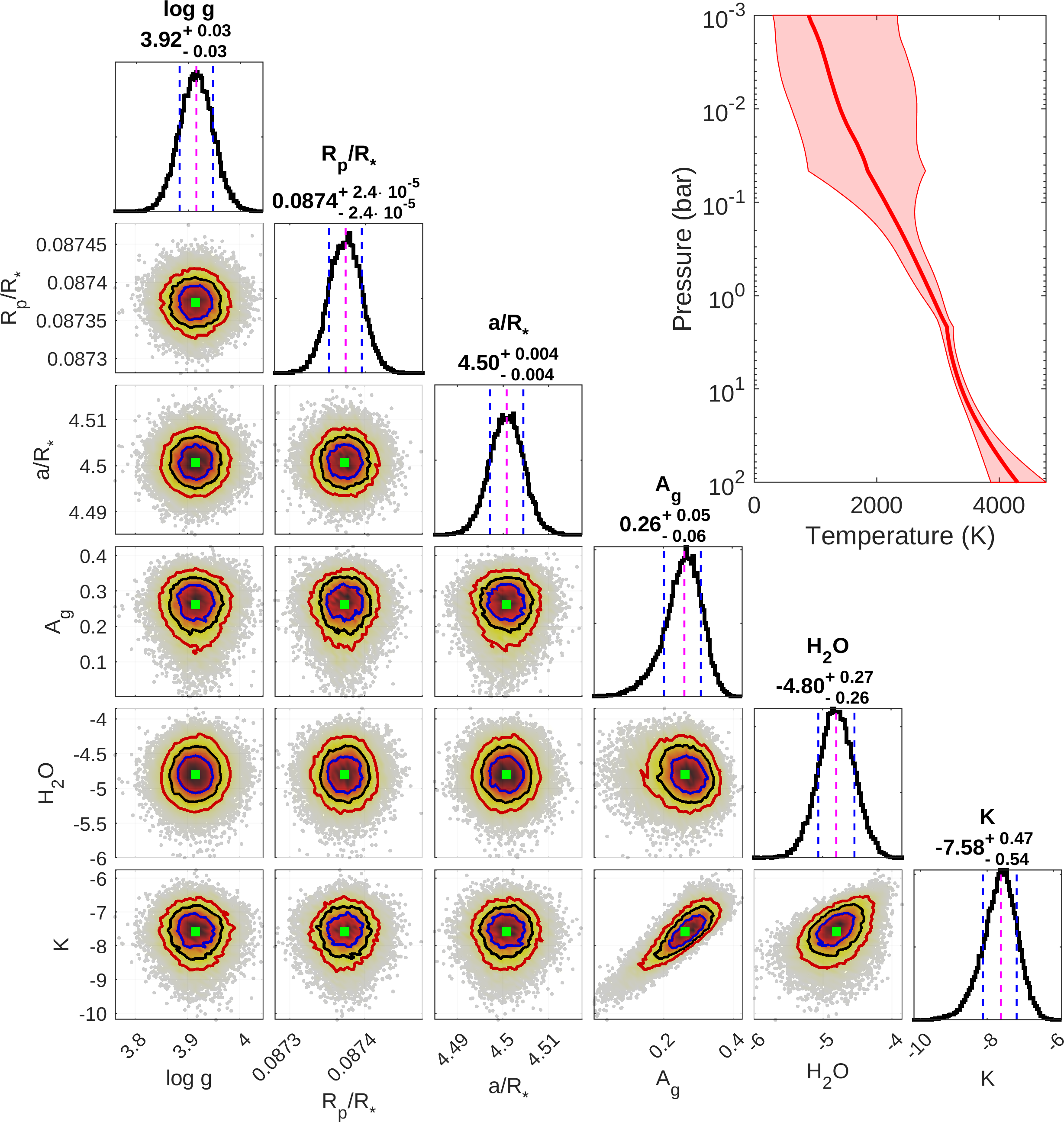}
  \includegraphics[width=0.48\linewidth]{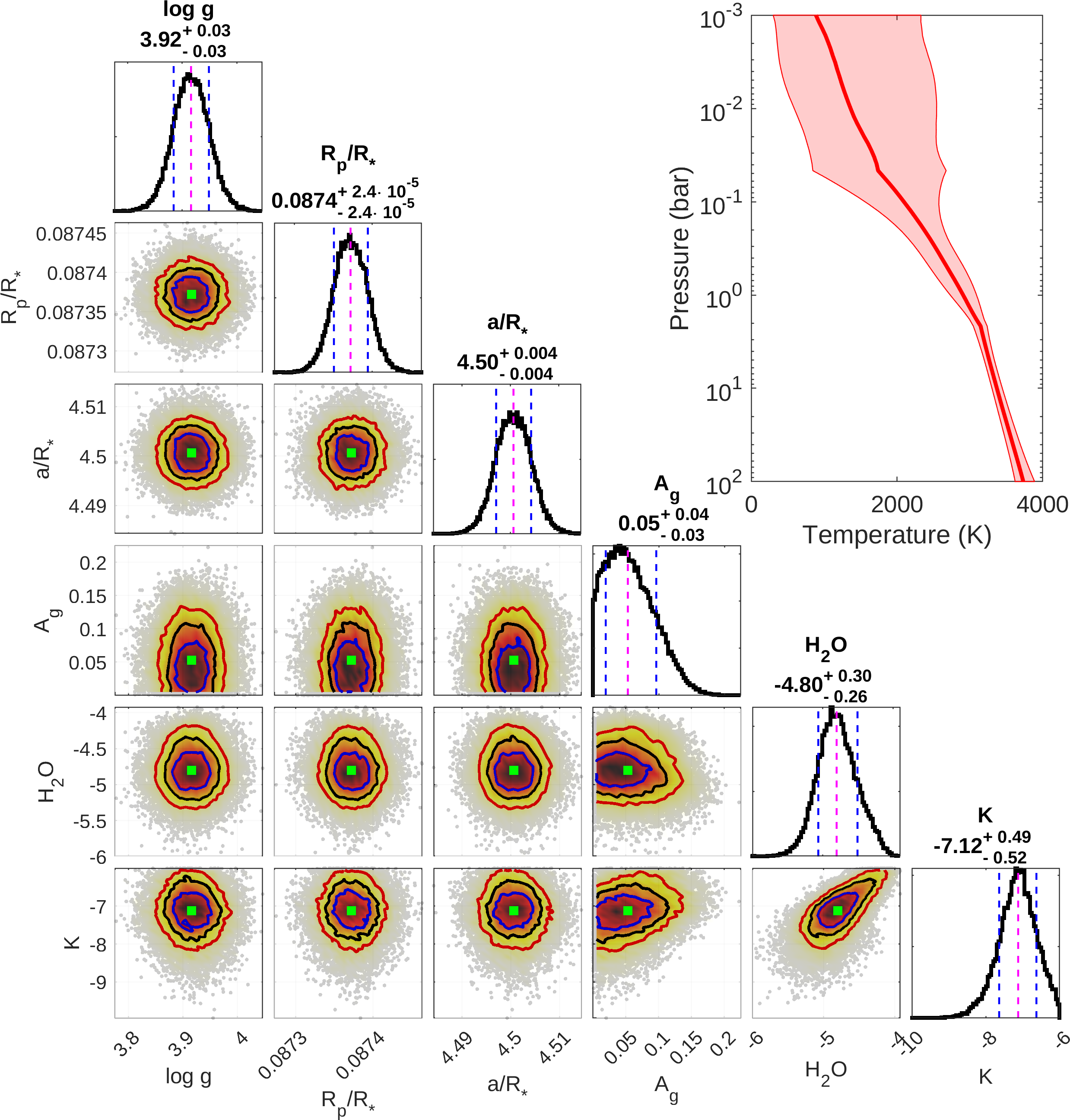}\\
  \includegraphics[width=0.42\linewidth]{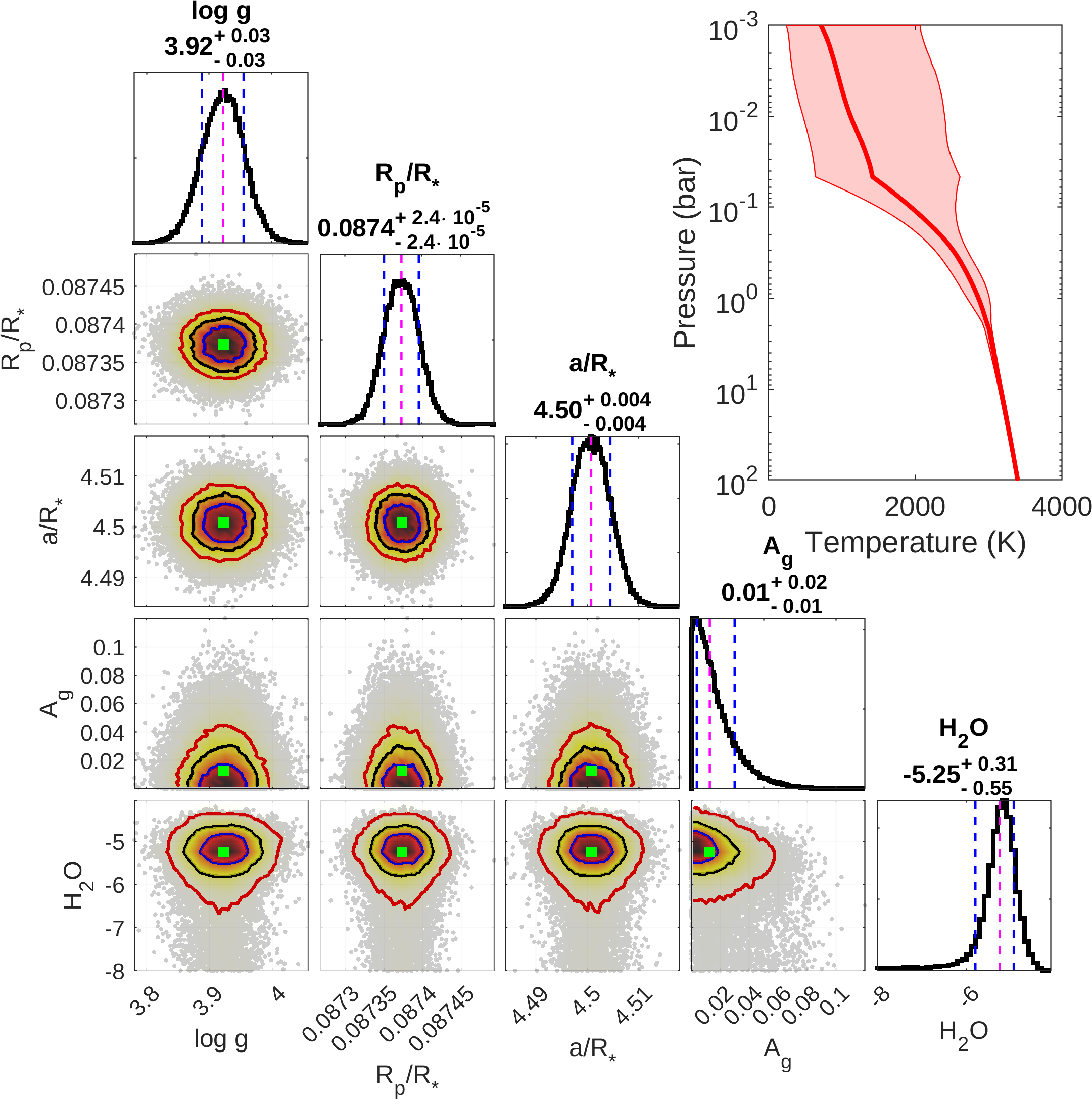}
  \caption{Posterior distributions and temperature--pressure profiles from three retrieval runs on the Kepler-13Ab secondary eclipse depths using different models. Upper left panel: the best-performing model, including free abundances of H$_2$O and K, with the Na abundance derived from the retrieved K abundance by assuming solar element-abundance ratios. Upper right panel: model with free abundances of H$_2$O and K. Lower panel: model with free abundance of H$_2$O. The solid blue, red, and yellow lines in the two-parameter correlation plots mark the 1$\sigma$, 2$\sigma$, and 3$\sigma$ bounds, respectively. The location of the median model is marked by the green squares. In the temperature--pressure profiles, the solid red line corresponds to the median profile, while the shaded region represents to the 1$\sigma$ confidence interval.}
  \label{fig:kepler13_posterior}
\end{figure*}

Figure~\ref{fig:kepler13_posterior} shows corner plots of the two-dimensional posteriors from three atmospheric retrievals of the Kepler-13Ab secondary eclipse spectrum: (1) the best-performing model, which includes H$_2$O, K, and Na, with the abundance of Na scaled to the potassium abundance assuming a solar Na/K ratio, (2) a model containing H$_2$O and K, and (3) a model with only H$_2$O opacity. The retrieved temperature--pressure profiles are also provided.

\pagebreak
\,
\pagebreak

\section{Compilation of Transit-Timing Measurements}\label{sec:timinglist}
\restartappendixnumbering

The full list of published transit timings used in our updated orbital ephemeris fits is provided in Table~\ref{tab:timings}. These entries were vetted, following the criteria outlined in Section \ref{subsec:ephem}. For each system, the zeroth epoch was assigned to the transit closest to the weighted average of all timing measurements. We separately list the sources of the original transit light curves and the calculated mid-transit times utilized in our ephemeris fits. The references in these two columns differ whenever earlier transit light curves were systematically reanalyzed in subsequent works.

\startlongtable
\begin{deluxetable*}{cccccc}
\tablewidth{0pc}
\tabletypesize{\scriptsize}
\tablecaption{
    List of Transit Timings
    \label{tab:timings}
}
\tablehead{
    \colhead{Target} &
    \colhead{$T_{0}$ ($\mathrm{BJD}_{\mathrm{TDB}}$)}                     &
    \colhead{$\sigma$ (d)}                     &
    \colhead{Epoch} & 
    \colhead{Light curve source} & 
    \colhead{Timing source}
}
\startdata
HAT-P-7 & 2454687.58230 & 0.00120 & $-$337 & \citet{christiansen2010} & \citet{christiansen2010} \\
 & 2454696.40488 & 0.00075 & $-$333 & \citet{christiansen2010} & \citet{christiansen2010} \\
 & 2454698.60720 & 0.00120 & $-$332 & \citet{christiansen2010} & \citet{christiansen2010} \\
 & 2454700.81470 & 0.00110 & $-$331 & \citet{christiansen2010} & \citet{christiansen2010} \\
 & 2454703.01588 & 0.00099 & $-$330 & \citet{christiansen2010} & \citet{christiansen2010} \\
 & 2454705.22294 & 0.00086 & $-$329 & \citet{christiansen2010} & \citet{christiansen2010} \\
 & 2454707.43152 & 0.00074 & $-$328 & \citet{christiansen2010} & \citet{christiansen2010} \\
 & 2454709.63137 & 0.00092 & $-$327 & \citet{christiansen2010} & \citet{christiansen2010} \\
 & 2454731.67993 & 0.00041 & $-$317 & \citet{winn2009} & \citet{southworth2011} \\
 & 2455419.55820 & 0.00070 & $-$5 & \citet{wong2016} & \citet{wong2016} \\
 & 2455430.58280 & 0.00047 & 0 & \citet{wong2016} & \citet{wong2016} \\
 & 2458709.02466 & 0.00034 & 1487 & this work & this work \\
 \hline
HAT-P-36 & 2455555.89060 & 0.00043 & $-$912 & \citet{hatp36} & \citet{mancini2015} \\
 & 2455608.98390 & 0.00030 & $-$872 & \citet{hatp36} & \citet{mancini2015} \\
 & 2456007.18909 & 0.00071 & $-$572 & \citet{wang2019} & \citet{wang2019} \\
 & 2456015.15110 & 0.00034 & $-$566 & \citet{wang2019} & \citet{wang2019} \\
 & 2456356.28178 & 0.00104 & $-$309 & \citet{wang2019} & \citet{wang2019} \\
 & 2456372.20849 & 0.00078 & $-$297 & \citet{wang2019} & \citet{wang2019} \\
 & 2456397.42894 & 0.00031 & $-$278 & \citet{mancini2015} & \citet{mancini2015} \\
 & 2456721.30085 & 0.00084 & $-$34 & \citet{wang2019} & \citet{wang2019} \\
 & 2456725.28100 & 0.00068 & $-$31 & \citet{wang2019} & \citet{wang2019} \\
 & 2456729.26439 & 0.00095 & $-$28 & \citet{wang2019} & \citet{wang2019} \\
 & 2456749.17580 & 0.00108 & $-$13 & \citet{wang2019} & \citet{wang2019} \\
 & 2456753.15819 & 0.00085 & $-$10 & \citet{wang2019} & \citet{wang2019} \\
 & 2456762.44834 & 0.00018 & $-$3 & \citet{mancini2015} & \citet{mancini2015} \\
 & 2456766.43055 & 0.00028 & 0 & \citet{mancini2015} & \citet{mancini2015} \\
 & 2457070.39280 & 0.00086 & 229 & \citet{wang2019} & \citet{wang2019} \\
 & 2457138.08645 & 0.00073 & 280 & \citet{wang2019} & \citet{wang2019} \\
 & 2457398.24830 & 0.00043 & 476 & \citet{wang2019} & \citet{wang2019} \\
 & 2457402.22948 & 0.00041 & 479 & \citet{wang2019} & \citet{wang2019} \\
 & 2457406.21179 & 0.00092 & 482 & \citet{wang2019} & \citet{wang2019} \\
 & 2457447.35849 & 0.00053 & 513 & \citet{wang2019} & \citet{wang2019} \\
 & 2457459.30494 & 0.00098 & 522 & \citet{wang2019} & \citet{wang2019} \\
 & 2457491.16331 & 0.00072 & 546 & \citet{wang2019} & \citet{wang2019} \\
 & 2457864.14544 & 0.00163 & 827 & \citet{wang2019} & \citet{wang2019} \\
 & 2458911.42299 & 0.00031 & 1616 & this work & this work \\
 \hline
KELT-1 & 2455899.55385 & 0.00071 & $-$1156 & \citet{kelt1} & \citet{m18a} \\
 & 2455905.63900 & 0.00160 & $-$1151 & \citet{kelt1} & \citet{m18a} \\
 & 2455911.72593 & 0.00075 & $-$1146 & \citet{kelt1} & \citet{m18a} \\
 & 2455927.55589 & 0.00057 & $-$1133 & \citet{kelt1} & \citet{m18a} \\
 & 2455933.64303 & 0.00064 & $-$1128 & \citet{kelt1} & \citet{m18a} \\
 & 2457306.97602 & 0.00030 & 0 & \citet{beatty2019} & \citet{beatty2019} \\
 & 2457959.55263 & 0.00054 & 536 & \citet{m18a} & \citet{m18a} \\
 & 2457981.46777 & 0.00078 & 554 & \citet{m18a} & \citet{m18a} \\
 & 2458015.55731 & 0.00046 & 582 & \citet{m18a} & \citet{m18a} \\
 & 2458020.42711 & 0.00055 & 586 & \citet{m18a} & \citet{m18a} \\
 & 2458026.51676 & 0.00074 & 591 & \citet{m18a} & \citet{m18a} \\
 & 2458081.30196 & 0.00085 & 636 & \citet{m18a} & \citet{m18a} \\
 & 2458126.34908 & 0.00085 & 673 & \citet{m18a} & \citet{m18a} \\
 & 2458367.41290 & 0.00140 & 871 & \citet{m18a} & \citet{m18a} \\
 & 2458778.92707 & 0.00050 & 1209 & this work & this work \\
 \hline
KELT-16 & 2457165.85142 & 0.00101 & $-$919 & \citet{kelt16} & \citet{m18a} \\
 & 2457166.82179 & 0.00089 & $-$918 & \citet{kelt16} & \citet{m18a} \\
 & 2457168.75660 & 0.00200 & $-$916 & \citet{kelt16} & \citet{m18a} \\
 & 2457198.79802 & 0.00074 & $-$885 & \citet{kelt16} & \citet{m18a} \\
 & 2457228.83690 & 0.00100 & $-$854 & \citet{kelt16} & \citet{m18a} \\
 & 2457238.52790 & 0.00180 & $-$844 & \citet{kelt16} & \citet{m18a} \\
 & 2457328.64440 & 0.00140 & $-$751 & \citet{kelt16} & \citet{m18a} \\
 & 2457329.61146 & 0.00094 & $-$750 & \citet{kelt16} & \citet{m18a} \\
 & 2457330.58151 & 0.00046 & $-$749 & \citet{kelt16} & \citet{m18a} \\
 & 2457363.52676 & 0.00101 & $-$715 & \citet{kelt16} & \citet{m18a} \\
 & 2457714.30206 & 0.00071 & $-$353 & \citet{m18a} & \citet{m18a} \\
 & 2457914.88456 & 0.00051 & $-$146 & \citet{patra2020} & \citet{patra2020} \\
 & 2457915.85370 & 0.00062 & $-$145 & \citet{patra2020} & \citet{patra2020} \\
 & 2457924.57245 & 0.00043 & $-$136 & \citet{mancini2021} & \citet{mancini2021} \\
 & 2457925.54315 & 0.00034 & $-$135 & \citet{mancini2021} & \citet{mancini2021} \\
 & 2457926.51262 & 0.00024 & $-$134 & \citet{mancini2021} & \citet{mancini2021} \\
 & 2457926.51262 & 0.00024 & $-$134 & \citet{mancini2021} & \citet{mancini2021} \\
 & 2457927.48073 & 0.00023 & $-$133 & \citet{mancini2021} & \citet{mancini2021} \\
 & 2457927.48156 & 0.00048 & $-$133 & \citet{m18a} & \citet{m18a} \\
 & 2457930.38847 & 0.00049 & $-$130 & \citet{mancini2021} & \citet{mancini2021} \\
 & 2457945.89131 & 0.00106 & $-$114 & \citet{mancini2021} & \citet{mancini2021} \\
 & 2457946.86111 & 0.00069 & $-$113 & \citet{mancini2021} & \citet{mancini2021} \\
 & 2457957.51989 & 0.00030 & $-$102 & \citet{mancini2021} & \citet{mancini2021} \\
 & 2457958.48844 & 0.00026 & $-$101 & \citet{m18a} & \citet{m18a} \\
 & 2457959.45852 & 0.00051 & $-$100 & \citet{mancini2021} & \citet{mancini2021} \\
 & 2457960.42762 & 0.00036 & $-$99 & \citet{mancini2021} & \citet{mancini2021} \\
 & 2457962.36544 & 0.00105 & $-$97 & \citet{mancini2021} & \citet{mancini2021} \\
 & 2457986.58974 & 0.00034 & $-$72 & \citet{mancini2021} & \citet{mancini2021} \\
 & 2457988.52797 & 0.00039 & $-$70 & \citet{m18a} & \citet{m18a} \\
 & 2457988.52836 & 0.00045 & $-$70 & \citet{mancini2021} & \citet{mancini2021} \\
 & 2457989.49700 & 0.00053 & $-$69 & \citet{mancini2021} & \citet{mancini2021} \\
 & 2458021.47285 & 0.00059 & $-$36 & \citet{mancini2021} & \citet{mancini2021} \\
 & 2458021.47346 & 0.00039 & $-$36 & \citet{m18a} & \citet{m18a} \\
 & 2458022.44176 & 0.00063 & $-$35 & \citet{mancini2021} & \citet{mancini2021} \\
 & 2458022.44219 & 0.00047 & $-$35 & \citet{m18a} & \citet{m18a} \\
 & 2458023.41090 & 0.00069 & $-$34 & \citet{mancini2021} & \citet{mancini2021} \\
 & 2458026.31752 & 0.00074 & $-$31 & \citet{m18a} & \citet{m18a} \\
 & 2458056.35704 & 0.00044 & 0 & \citet{mancini2021} & \citet{mancini2021} \\
 & 2458301.51280 & 0.00030 & 253 & \citet{mancini2021} & \citet{mancini2021} \\
 & 2458302.48200 & 0.00026 & 254 & \citet{mancini2021} & \citet{mancini2021} \\
 & 2458303.44940 & 0.00054 & 255 & \citet{mancini2021} & \citet{mancini2021} \\
 & 2458334.45858 & 0.00063 & 287 & \citet{m18a} & \citet{m18a} \\
 & 2458365.46578 & 0.00064 & 319 & \citet{m18a} & \citet{m18a} \\
 & 2458368.37232 & 0.00048 & 322 & \citet{m18a} & \citet{m18a} \\
 & 2458401.31876 & 0.00028 & 356 & \citet{m18a} & \citet{m18a} \\
 & 2458677.48078 & 0.00071 & 641 & \citet{mancini2021} & \citet{mancini2021} \\
 & 2458710.42937 & 0.00123 & 675 & \citet{mancini2021} & \citet{mancini2021} \\
 & 2458719.14831 & 0.00045 & 684 & this work & this work \\
 & 2458743.37244 & 0.00090 & 709 & \citet{mancini2021} & \citet{mancini2021} \\
 & 2458744.34073 & 0.00056 & 710 & \citet{mancini2021} & \citet{mancini2021} \\
 & 2458990.46559 & 0.00084 & 964 & \citet{mancini2021} & \citet{mancini2021} \\
 & 2459050.54472 & 0.00026 & 1026 & \citet{mancini2021} & \citet{mancini2021} \\
 \hline
KELT-20 & 2457301.62192 & 0.00086 & $-$291 & \citet{kelt20} & \citet{kelt20} \\
 & 2457544.81092 & 0.00051 & $-$221 & \citet{kelt20} & \citet{kelt20} \\
 & 2457551.75691 & 0.00064 & $-$219 & \citet{kelt20} & \citet{kelt20} \\
 & 2457697.67192 & 0.00072 & $-$177 & \citet{kelt20} & \citet{kelt20} \\
 & 2457881.79960 & 0.00056 & $-$124 & \citet{kelt20} & \citet{kelt20} \\
 & 2457881.79690 & 0.00064 & $-$124 & \citet{kelt20} & \citet{kelt20} \\
 & 2457881.79568 & 0.00087 & $-$124 & \citet{kelt20} & \citet{kelt20} \\
 & 2457888.74555 & 0.00059 & $-$122 & \citet{kelt20} & \citet{kelt20} \\
 & 2457916.53750 & 0.00058 & $-$114 & \citet{kelt20} & \citet{kelt20} \\
 & 2458312.58566 & 0.00022 & 0 & \citet{casasayas2019} & \citet{casasayas2019} \\
 & 2458698.21073 & 0.00014 & 111 & this work & this work \\
\hline
KELT-23A & 2458144.89840 & 0.00046 & $-$277 & \citet{kelt23} & \citet{kelt23} \\
 & 2458144.89724 & 0.00044 & $-$277 & \citet{kelt23} & \citet{kelt23} \\
 & 2458153.91793 & 0.00059 & $-$273 & \citet{kelt23} & \citet{kelt23} \\
 & 2458167.44840 & 0.00110 & $-$267 & \citet{kelt23} & \citet{kelt23} \\
 & 2458187.74583 & 0.00064 & $-$258 & \citet{kelt23} & \citet{kelt23} \\
 & 2458196.77035 & 0.00110 & $-$254 & \citet{kelt23} & \citet{kelt23} \\
 & 2458196.77106 & 0.00063 & $-$254 & \citet{kelt23} & \citet{kelt23} \\
 & 2458196.77350 & 0.00119 & $-$254 & \citet{kelt23} & \citet{kelt23} \\
 & 2458273.45249 & 0.00098 & $-$220 & \citet{kelt23} & \citet{kelt23} \\
 & 2458769.61228 & 0.00006 & 0 & this work & this work \\
\hline
Qatar-1 & 2455640.53380 & 0.00160 & $-$576 & \citet{vonessen2013} & \citet{m15} \\
 & 2455647.63267 & 0.00058 & $-$571 & \citet{vonessen2013} & \citet{m15} \\
 & 2455704.43426 & 0.00059 & $-$531 & \citet{vonessen2013} & \citet{m15} \\
 & 2455711.53450 & 0.00035 & $-$526 & \citet{vonessen2013}, & \citet{m15} \\
 & & & & \citet{covino2013} & \\
 & 2455742.77475 & 0.00022 & $-$504 & \citet{collins2017} & \citet{collins2017} \\
 & 2455752.71499 & 0.00024 & $-$497 & \citet{collins2017} & \citet{collins2017} \\
 & 2455775.43517 & 0.00046 & $-$481 & \citet{vonessen2013}, & \citet{m15} \\
 & & & & \citet{m15} & \\
 & 2455789.63540 & 0.00025 & $-$471 & \citet{collins2017} & \citet{collins2017} \\
 & 2455796.73583 & 0.00021 & $-$466 & \citet{collins2017} & \citet{collins2017} \\
 & 2455799.57580 & 0.00020 & $-$464 & \citet{mislis2015} & \citet{mislis2015} \\
 & 2455799.57550 & 0.00010 & $-$464 & \citet{mislis2015} & \citet{mislis2015} \\
 & 2455799.57590 & 0.00020 & $-$464 & \citet{mislis2015} & \citet{mislis2015} \\
 & 2455799.57560 & 0.00010 & $-$464 & \citet{mislis2015} & \citet{mislis2015} \\
 & 2455799.57630 & 0.00032 & $-$464 & \citet{covino2013} & \citet{m15} \\
 & 2455826.55618 & 0.00022 & $-$445 & \citet{collins2017} & \citet{collins2017} \\
 & 2455836.49672 & 0.00041 & $-$438 & \citet{vonessen2013} & \citet{m15} \\
 & 2455843.59664 & 0.00022 & $-$433 & \citet{collins2017} & \citet{collins2017} \\
 & 2455850.69628 & 0.00019 & $-$428 & \citet{sada2012} & \citet{sada2012} \\
 & 2455897.55768 & 0.00024 & $-$395 & \citet{collins2017} & \citet{collins2017} \\
 & 2455985.60050 & 0.00110 & $-$333 & \citet{vonessen2013} & \citet{m15} \\
 & 2456039.56043 & 0.00075 & $-$295 & \citet{vonessen2013} & \citet{m15} \\
 & 2456049.50010 & 0.00046 & $-$288 & \citet{vonessen2013} & \citet{m15} \\
 & 2456059.43930 & 0.00170 & $-$281 & \citet{vonessen2013} & \citet{m15} \\
 & 2456076.47920 & 0.00100 & $-$269 & \citet{vonessen2013} & \citet{m15} \\
 & 2456097.78070 & 0.00023 & $-$254 & \citet{collins2017} & \citet{collins2017} \\
 & 2456107.72152 & 0.00021 & $-$247 & \citet{collins2017} & \citet{collins2017} \\
 & 2456113.40040 & 0.00110 & $-$243 & \citet{vonessen2013} & \citet{m15} \\
 & 2456130.44153 & 0.00035 & $-$231 & \citet{covino2013} & \citet{m15} \\
 & 2456140.38094 & 0.00055 & $-$224 & \citet{vonessen2013} & \citet{m15} \\
 & 2456141.80175 & 0.00029 & $-$223 & \citet{collins2017} & \citet{collins2017} \\
 & 2456151.74131 & 0.00028 & $-$216 & \citet{collins2017} & \citet{collins2017} \\
 & 2456157.42152 & 0.00044 & $-$212 & \citet{vonessen2013} & \citet{m15} \\
 & 2456161.68167 & 0.00027 & $-$209 & \citet{collins2017} & \citet{collins2017} \\
 & 2456164.52187 & 0.00042 & $-$207 & \citet{covino2013} & \citet{m15} \\
 & 2456181.56264 & 0.00023 & $-$195 & \citet{covino2013} & \citet{m15} \\
 & 2456201.44229 & 0.00066 & $-$181 & \citet{vonessen2013} & \citet{m15} \\
 & 2456225.58336 & 0.00027 & $-$164 & \citet{collins2017} & \citet{collins2017} \\
 & 2456231.26318 & 0.00051 & $-$160 & \citet{vonessen2013} & \citet{m15} \\
 & 2456275.28493 & 0.00043 & $-$129 & \citet{m15} & \citet{m15} \\
 & 2456458.46650 & 0.00020 & 0 & \citet{mislis2015} & \citet{mislis2015} \\
 & 2456489.70762 & 0.00020 & 22 & \citet{collins2017} & \citet{collins2017} \\
 & 2456539.40816 & 0.00065 & 57 & \citet{m15} & \citet{m15} \\
 & 2456566.38885 & 0.00057 & 76 & \citet{m15} & \citet{m15} \\
 & 2456623.18947 & 0.00060 & 116 & \citet{m15} & \citet{m15} \\
 & 2456742.47200 & 0.00110 & 200 & \citet{m15} & \citet{m15} \\
 & 2456749.57192 & 0.00038 & 205 & \citet{m15} & \citet{m15} \\
 & 2456766.61200 & 0.00040 & 217 & \citet{mislis2015} & \citet{mislis2015} \\
 & 2456793.59237 & 0.00053 & 236 & \citet{m15} & \citet{m15} \\
 & 2456803.53269 & 0.00028 & 243 & \citet{m15} & \citet{m15} \\
 & 2456813.47200 & 0.00030 & 250 & \citet{mislis2015} & \citet{mislis2015} \\
 & 2456813.47310 & 0.00020 & 250 & \citet{mislis2015} & \citet{mislis2015} \\
 & 2456823.41500 & 0.00044 & 257 & \citet{puskullu2017} & \citet{puskullu2017} \\
 & 2456830.51327 & 0.00033 & 262 & \citet{m15} & \citet{m15} \\
 & 2456840.45424 & 0.00058 & 269 & \citet{m15} & \citet{m15} \\
 & 2456840.45327 & 0.00042 & 269 & \citet{puskullu2017} & \citet{puskullu2017} \\
 & 2456854.65325 & 0.00031 & 279 & \citet{collins2017} & \citet{collins2017} \\
 & 2456861.75417 & 0.00032 & 284 & \citet{collins2017} & \citet{collins2017} \\
 & 2456867.43480 & 0.00042 & 288 & \citet{puskullu2017} & \citet{puskullu2017} \\
 & 2456888.73423 & 0.00030 & 303 & \citet{collins2017} & \citet{collins2017} \\
 & 2456894.41548 & 0.00047 & 307 & \citet{puskullu2017} & \citet{puskullu2017} \\
 & 2456908.61490 & 0.00020 & 317 & \citet{mislis2015} & \citet{mislis2015} \\
 & 2456911.45488 & 0.00029 & 319 & \citet{m15} & \citet{m15} \\
 & 2456918.55468 & 0.00056 & 324 & \citet{m15} & \citet{m15} \\
 & 2456921.39618 & 0.00048 & 326 & \citet{puskullu2017} & \citet{puskullu2017} \\
 & 2456925.65523 & 0.00025 & 329 & \citet{collins2017} & \citet{collins2017} \\
 & 2456928.49510 & 0.00032 & 331 & \citet{m15} & \citet{m15} \\
 & 2456931.33528 & 0.00053 & 333 & \citet{puskullu2017} & \citet{puskullu2017} \\
 & 2456958.31430 & 0.00120 & 352 & \citet{m15} & \citet{m15} \\
 & 2457124.45935 & 0.00059 & 469 & \citet{puskullu2017} & \citet{puskullu2017} \\
 & 2457168.48014 & 0.00058 & 500 & \citet{puskullu2017} & \citet{puskullu2017} \\
 & 2457330.36216 & 0.00065 & 614 & \citet{puskullu2017} & \citet{puskullu2017} \\
 & 2457340.30315 & 0.00064 & 621 & \citet{puskullu2017} & \citet{puskullu2017} \\
 & 2457347.40134 & 0.00085 & 626 & \citet{puskullu2017} & \citet{puskullu2017} \\
 & 2457570.34628 & 0.00050 & 783 & \citet{thakur2018} & \citet{thakur2018} \\
 & 2457580.28552 & 0.00022 & 790 & \citet{thakur2018} & \citet{thakur2018} \\
 & 2457634.24666 & 0.00035 & 828 & \citet{thakur2018} & \citet{thakur2018} \\
 & 2458959.12981 & 0.00014 & 1761 & this work & this work \\
\hline
TrES-3 & 2454185.91110 & 0.00020 & $-$1076 & \citet{tres3} & \citet{jiang2013} \\
 & 2454198.97359 & 0.00066 & $-$1066 & \citet{tres3} & \citet{jiang2013} \\
 & 2454214.64695 & 0.00036 & $-$1054 & \citet{sozzetti2009} & \citet{jiang2013} \\
 & 2454215.95288 & 0.00033 & $-$1053 & \citet{sozzetti2009} & \citet{jiang2013} \\
 & 2454532.04939 & 0.00033 & $-$811 & \citet{christiansen2011} & \citet{christiansen2011} \\
 & 2454533.35515 & 0.00035 & $-$810 & \citet{christiansen2011} & \citet{christiansen2011} \\
 & 2454534.66317 & 0.00019 & $-$809 & \citet{gibson2009} & \citet{jiang2013} \\
 & 2454535.96903 & 0.00039 & $-$808 & \citet{sozzetti2009} & \citet{jiang2013} \\
 & 2454538.58126 & 0.00035 & $-$806 & \citet{christiansen2011} & \citet{christiansen2011} \\
 & 2454539.88703 & 0.00040 & $-$805 & \citet{christiansen2011} & \citet{christiansen2011} \\
 & 2454541.19261 & 0.00035 & $-$804 & \citet{christiansen2011} & \citet{christiansen2011} \\
 & 2454542.49930 & 0.00041 & $-$803 & \citet{christiansen2011} & \citet{christiansen2011} \\
 & 2454552.94962 & 0.00022 & $-$795 & \citet{sozzetti2009} & \citet{jiang2013} \\
 & 2454569.92982 & 0.00040 & $-$782 & \citet{sozzetti2009} & \citet{jiang2013} \\
 & 2454594.74682 & 0.00037 & $-$763 & \citet{sozzetti2009} & \citet{jiang2013} \\
 & 2454615.64621 & 0.00021 & $-$747 & \citet{gibson2009} & \citet{jiang2013} \\
 & 2454632.62690 & 0.00020 & $-$734 & \citet{gibson2009} & \citet{jiang2013} \\
 & 2454649.60712 & 0.00019 & $-$721 & \citet{gibson2009} & \citet{jiang2013} \\
 & 2454653.52661 & 0.00092 & $-$718 & \citet{gibson2009} & \citet{jiang2013} \\
 & 2454662.66984 & 0.00060 & $-$711 & \citet{gibson2009} & \citet{jiang2013} \\
 & 2454670.50709 & 0.00034 & $-$705 & \citet{gibson2009} & \citet{jiang2013} \\
 & 2454674.42521 & 0.00028 & $-$702 & \citet{gibson2009} & \citet{jiang2013} \\
 & 2454683.56812 & 0.00042 & $-$695 & \citet{gibson2009} & \citet{jiang2013} \\
 & 2454957.86698 & 0.00048 & $-$485 & \citet{sada2012} & \citet{sada2012} \\
 & 2454959.17120 & 0.00110 & $-$484 & \citet{lee2011} & \citet{mannaday2020} \\
 & 2454964.40014 & 0.00095 & $-$480 & \citet{vanko2013} & \citet{mannaday2020} \\
 & 2454965.70470 & 0.00023 & $-$479 & \citet{kundurthy2013} & \citet{mannaday2020} \\
 & 2454977.46000 & 0.00150 & $-$470 & \citet{vanko2013} & \citet{mannaday2020} \\
 & 2454995.74737 & 0.00044 & $-$456 & \citet{turner2013} & \citet{mannaday2020} \\
 & 2454995.74657 & 0.00017 & $-$456 & \citet{kundurthy2013} & \citet{mannaday2020} \\
 & 2455004.88970 & 0.00018 & $-$449 & \citet{turner2013} & \citet{mannaday2020} \\
 & 2455017.95161 & 0.00033 & $-$439 & \citet{turner2013} & \citet{mannaday2020} \\
 & 2455045.38085 & 0.00063 & $-$418 & \citet{vanko2013} & \citet{mannaday2020} \\
 & 2455049.29850 & 0.00150 & $-$415 & \citet{m13c} & \citet{m13c} \\
 & 2455054.52523 & 0.00018 & $-$411 & \citet{colon2010} & \citet{jiang2013} \\
 & 2455058.44480 & 0.00100 & $-$408 & \citet{vanko2013} & \citet{mannaday2020} \\
 & 2455277.88206 & 0.00038 & $-$240 & \citet{kundurthy2013} & \citet{mannaday2020} \\
 & 2455294.86465 & 0.00039 & $-$227 & \citet{lee2011} & \citet{mannaday2020} \\
 & 2455297.47780 & 0.00080 & $-$225 & \citet{m13c} & \citet{m13c} \\
 & 2455314.45500 & 0.00072 & $-$212 & \citet{vanko2013} & \citet{mannaday2020} \\
 & 2455327.51720 & 0.00080 & $-$202 & \citet{m13c} & \citet{m13c} \\
 & 2455332.74259 & 0.00031 & $-$198 & \citet{kundurthy2013} & \citet{mannaday2020} \\
 & 2455341.88380 & 0.00110 & $-$191 & \citet{jiang2013} & \citet{jiang2013} \\
 & 2455358.86606 & 0.00076 & $-$178 & \citet{jiang2013} & \citet{jiang2013} \\
 & 2455358.86723 & 0.00070 & $-$178 & \citet{lee2011} & \citet{mannaday2020} \\
 & 2455362.78568 & 0.00057 & $-$175 & \citet{lee2011} & \citet{mannaday2020} \\
 & 2455362.78470 & 0.00110 & $-$175 & \citet{jiang2013} & \citet{jiang2013} \\
 & 2455365.39650 & 0.00120 & $-$173 & \citet{m13c} & \citet{m13c} \\
 & 2455366.70215 & 0.00080 & $-$172 & \citet{jiang2013} & \citet{jiang2013} \\
 & 2455375.84617 & 0.00090 & $-$165 & \citet{jiang2013} & \citet{jiang2013} \\
 & 2455378.45955 & 0.00090 & $-$163 & \citet{vanko2013} & \citet{mannaday2020} \\
 & 2455416.33972 & 0.00056 & $-$134 & \citet{vanko2013} & \citet{mannaday2020} \\
 & 2455429.39997 & 0.00046 & $-$124 & \citet{vanko2013} & \citet{mannaday2020} \\
 & 2455446.38075 & 0.00021 & $-$111 & \citet{vanko2013} & \citet{mannaday2020} \\
 & 2455479.03425 & 0.00094 & $-$86 & \citet{sun2018} & \citet{sun2018} \\
 & 2455481.64795 & 0.00018 & $-$84 & \citet{kundurthy2013} & \citet{mannaday2020} \\
 & 2455591.36690 & 0.00150 & 0 & \citet{sun2018} & \citet{sun2018} \\
 & 2455643.61454 & 0.00034 & 40 & \citet{vanko2013} & \citet{mannaday2020} \\
 & 2455644.92122 & 0.00019 & 41 & \citet{kundurthy2013} & \citet{mannaday2020} \\
 & 2455678.88252 & 0.00032 & 67 & \citet{kundurthy2013} & \citet{mannaday2020} \\
 & 2455695.86223 & 0.00072 & 80 & \citet{kundurthy2013} & \citet{mannaday2020} \\
 & 2455733.74164 & 0.00035 & 109 & \citet{kundurthy2013} & \citet{mannaday2020} \\
 & 2455797.74568 & 0.00032 & 158 & \citet{kundurthy2013} & \citet{mannaday2020} \\
 & 2455817.33688 & 0.00041 & 173 & \citet{vanko2013} & \citet{mannaday2020} \\
 & 2456011.95934 & 0.00073 & 322 & \citet{turner2013} & \citet{mannaday2020} \\
 & 2456014.57219 & 0.00070 & 324 & \citet{turner2013} & \citet{mannaday2020} \\
 & 2456014.57248 & 0.00065 & 324 & \citet{turner2013} & \citet{mannaday2020} \\
 & 2456028.93996 & 0.00049 & 335 & \citet{turner2013} & \citet{mannaday2020} \\
 & 2456077.27003 & 0.00037 & 372 & \citet{mannaday2020} & \citet{mannaday2020} \\
 & 2456082.49260 & 0.00120 & 376 & \citet{puskullu2017} & \citet{mannaday2020} \\
 & 2456086.41120 & 0.00100 & 379 & \citet{puskullu2017} & \citet{mannaday2020} \\
 & 2456099.47337 & 0.00160 & 389 & \citet{mannaday2020} & \citet{mannaday2020} \\
 & 2456368.54705 & 0.00063 & 595 & \citet{m13c} & \citet{m13c} \\
 & 2456393.36471 & 0.00050 & 614 & \citet{mannaday2020} & \citet{mannaday2020} \\
 & 2456423.40717 & 0.00079 & 637 & \citet{mannaday2020} & \citet{mannaday2020} \\
 & 2456431.24526 & 0.00032 & 643 & \citet{mannaday2020} & \citet{mannaday2020} \\
 & 2456440.38874 & 0.00063 & 650 & \citet{puskullu2017} & \citet{mannaday2020} \\
 & 2456487.41277 & 0.00075 & 686 & \citet{puskullu2017} & \citet{mannaday2020} \\
 & 2456747.34245 & 0.00019 & 885 & \citet{mannaday2020} & \citet{mannaday2020} \\
 & 2456841.38981 & 0.00079 & 957 & \citet{puskullu2017} & \citet{mannaday2020} \\
 & 2457140.50425 & 0.00041 & 1186 & \citet{puskullu2017} & \citet{mannaday2020} \\
 & 2457204.50652 & 0.00089 & 1235 & \citet{puskullu2017} & \citet{mannaday2020} \\
 & 2457212.34484 & 0.00052 & 1241 & \citet{puskullu2017} & \citet{mannaday2020} \\
 & 2457221.48724 & 0.00063 & 1248 & \citet{ricci2017} & \citet{mannaday2020} \\
 & 2457225.40653 & 0.00036 & 1251 & \citet{puskullu2017} & \citet{mannaday2020} \\
 & 2457238.46724 & 0.00066 & 1261 & \citet{ricci2017} & \citet{mannaday2020} \\
 & 2457242.38679 & 0.00025 & 1264 & \citet{puskullu2017} & \citet{mannaday2020} \\
 & 2457256.75464 & 0.00074 & 1275 & \citet{ricci2017} & \citet{mannaday2020} \\
 & 2457259.36698 & 0.00058 & 1277 & \citet{puskullu2017} & \citet{mannaday2020} \\
 & 2457491.86832 & 0.00041 & 1455 & \citet{ricci2017} & \citet{mannaday2020} \\
 & 2457542.80916 & 0.00026 & 1494 & \citet{ricci2017} & \citet{mannaday2020} \\
 & 2458185.45419 & 0.00064 & 1986 & \citet{mannaday2020} & \citet{mannaday2020} \\
 & 2458189.37246 & 0.00068 & 1989 & \citet{mannaday2020} & \citet{mannaday2020} \\
 & 2458202.43178 & 0.00150 & 1999 & \citet{mannaday2020} & \citet{mannaday2020} \\
 & 2458206.35282 & 0.00076 & 2002 & \citet{mannaday2020} & \citet{mannaday2020} \\
 & 2458219.41292 & 0.00058 & 2012 & \citet{mannaday2020} & \citet{mannaday2020} \\
 & 2458223.33346 & 0.00053 & 2015 & \citet{mannaday2020} & \citet{mannaday2020} \\
 & 2458347.42160 & 0.00020 & 2110 & \citet{vonessen2019} & \citet{vonessen2019} \\
 & 2459008.35101 & 0.00022 & 2616 & this work & this work \\
\hline
WASP-3 & 2454143.85104 & 0.00040 & $-$660 & \citet{wasp3} & \citet{wasp3} \\
 & 2454601.86671 & 0.00026 & $-$412 & \citet{tripathi2010} & \citet{nascimbeni2013} \\
 & 2454605.56042 & 0.00030 & $-$410 & \citet{gibson2008} & \citet{nascimbeni2013} \\
 & 2454638.80399 & 0.00034 & $-$392 & \citet{tripathi2010} & \citet{nascimbeni2013} \\
 & 2454660.96479 & 0.00015 & $-$380 & \citet{tripathi2010} & \citet{nascimbeni2013} \\
 & 2454679.43318 & 0.00042 & $-$370 & \citet{christiansen2011} & \citet{nascimbeni2013} \\
 & 2454681.27967 & 0.00034 & $-$369 & \citet{christiansen2011} & \citet{nascimbeni2013} \\
 & 2454683.12798 & 0.00049 & $-$368 & \citet{christiansen2011} & \citet{nascimbeni2013} \\
 & 2454684.97524 & 0.00040 & $-$367 & \citet{christiansen2011} & \citet{nascimbeni2013} \\
 & 2454692.36168 & 0.00056 & $-$363 & \citet{christiansen2011} & \citet{nascimbeni2013} \\
 & 2454694.20776 & 0.00083 & $-$362 & \citet{christiansen2011} & \citet{nascimbeni2013} \\
 & 2454712.67641 & 0.00064 & $-$352 & \citet{nascimbeni2013} & \citet{nascimbeni2013} \\
 & 2454714.52368 & 0.00041 & $-$351 & \citet{gibson2008} & \citet{nascimbeni2013} \\
 & 2454963.84450 & 0.00081 & $-$216 & \citet{tripathi2010} & \citet{nascimbeni2013} \\
 & 2454963.84527 & 0.00118 & $-$216 & \citet{sada2012} & \citet{nascimbeni2013} \\
 & 2454967.53651 & 0.00085 & $-$214 & \citet{montalto2012} & \citet{montalto2012} \\
 & 2454976.77284 & 0.00030 & $-$209 & \citet{tripathi2010} & \citet{nascimbeni2013} \\
 & 2454987.85256 & 0.00093 & $-$203 & \citet{nascimbeni2013} & \citet{nascimbeni2013} \\
 & 2455037.71878 & 0.00086 & $-$176 & \citet{nascimbeni2013} & \citet{nascimbeni2013} \\
 & 2455041.41172 & 0.00035 & $-$174 & \citet{damasso2010} & \citet{nascimbeni2013} \\
 & 2455041.41255 & 0.00058 & $-$174 & \citet{m10} & \citet{nascimbeni2013} \\
 & 2455065.42023 & 0.00036 & $-$161 & \citet{m10} & \citet{nascimbeni2013} \\
 & 2455078.34809 & 0.00114 & $-$154 & \citet{m10} & \citet{nascimbeni2013} \\
 & 2455098.66406 & 0.00044 & $-$143 & \citet{m13b} & \citet{m13b} \\
 & 2455102.36030 & 0.00084 & $-$141 & \citet{m10} & \citet{nascimbeni2013} \\
 & 2455139.29753 & 0.00073 & $-$121 & \citet{m10} & \citet{nascimbeni2013} \\
 & 2455305.51117 & 0.00056 & $-$31 & \citet{m10} & \citet{nascimbeni2013} \\
 & 2455342.44700 & 0.00120 & $-$11 & \citet{m13b} & \citet{m13b} \\
 & 2455349.83390 & 0.00069 & $-$7 & \citet{sada2012} & \citet{nascimbeni2013} \\
 & 2455349.83306 & 0.00090 & $-$7 & \citet{sada2012} & \citet{nascimbeni2013} \\
 & 2455355.37419 & 0.00053 & $-$4 & \citet{m13b} & \citet{m13b} \\
 & 2455362.76233 & 0.00040 & 0 & \citet{m13b} & \citet{m13b} \\
 & 2455366.45610 & 0.00100 & 2 & \citet{m13b} & \citet{m13b} \\
 & 2455401.54564 & 0.00036 & 21 & \citet{m13b} & \citet{m13b} \\
 & 2455423.70889 & 0.00048 & 33 & \citet{nascimbeni2013} & \citet{nascimbeni2013} \\
 & 2455436.63530 & 0.00100 & 40 & \citet{m13b} & \citet{m13b} \\
 & 2455438.48270 & 0.00060 & 41 & \citet{m13b} & \citet{m13b} \\
 & 2455451.41010 & 0.00040 & 48 & \citet{m13b} & \citet{m13b} \\
 & 2455654.56180 & 0.00140 & 158 & \citet{m13b} & \citet{m13b} \\
 & 2455665.64627 & 0.00069 & 164 & \citet{montalto2012} & \citet{montalto2012} \\
 & 2455678.57065 & 0.00106 & 171 & \citet{montalto2012} & \citet{montalto2012} \\
 & 2455689.65263 & 0.00015 & 177 & \citet{nascimbeni2013} & \citet{nascimbeni2013} \\
 & 2455691.49938 & 0.00086 & 178 & \citet{m13b} & \citet{m13b} \\
 & 2455698.88641 & 0.00027 & 182 & \citet{m13b} & \citet{m13b} \\
 & 2455698.88476 & 0.00160 & 182 & \citet{sada2012} & \citet{nascimbeni2013} \\
 & 2455702.58052 & 0.00028 & 184 & \citet{nascimbeni2013} & \citet{nascimbeni2013} \\
 & 2455715.50824 & 0.00072 & 191 & \citet{m13b} & \citet{m13b} \\
 & 2455715.50608 & 0.00074 & 191 & \citet{montalto2012} & \citet{montalto2012} \\
 & 2455728.43608 & 0.00052 & 198 & \citet{m13b} & \citet{m13b} \\
 & 2455739.51620 & 0.00017 & 204 & \citet{nascimbeni2013} & \citet{nascimbeni2013} \\
 & 2455739.51735 & 0.00064 & 204 & \citet{m13b} & \citet{m13b} \\
 & 2455748.75070 & 0.00110 & 209 & \citet{m13b} & \citet{m13b} \\
 & 2455763.52552 & 0.00070 & 217 & \citet{montalto2012} & \citet{montalto2012} \\
 & 2455763.52511 & 0.00031 & 217 & \citet{nascimbeni2013} & \citet{nascimbeni2013} \\
 & 2455765.37180 & 0.00140 & 218 & \citet{m13b} & \citet{m13b} \\
 & 2455776.45452 & 0.00092 & 224 & \citet{m13b} & \citet{m13b} \\
 & 2455787.53583 & 0.00053 & 230 & \citet{m13b} & \citet{m13b} \\
 & 2455787.53379 & 0.00080 & 230 & \citet{montalto2012} & \citet{montalto2012} \\
 & 2455800.46112 & 0.00170 & 237 & \citet{montalto2012} & \citet{montalto2012} \\
 & 2455800.46137 & 0.00055 & 237 & \citet{m13b} & \citet{m13b} \\
 & 2455813.38910 & 0.00150 & 244 & \citet{m13b} & \citet{m13b} \\
 & 2455813.38792 & 0.00098 & 244 & \citet{montalto2012} & \citet{montalto2012} \\
 & 2455837.39876 & 0.00044 & 257 & \citet{m13b} & \citet{m13b} \\
 & 2455850.32740 & 0.00100 & 264 & \citet{m13b} & \citet{m13b} \\
 & 2458223.50975 & 0.00078 & 1549 & \citet{m18b} & \citet{m18b} \\
 & 2459023.18947 & 0.00029 & 1982 & this work & this work \\
 \hline
WASP-12 & 2454515.52496 & 0.00043 & $-$2022 & \citet{wasp12} & \citet{m13a} \\
 & 2454836.40340 & 0.00028 & $-$1728 & \citet{copperwheat2013} & \citet{m16} \\
 & 2454840.76893 & 0.00062 & $-$1724 & \citet{chan2011} & \citet{m16} \\
 & 2455140.90981 & 0.00042 & $-$1449 & \citet{collins2017} & \citet{collins2017} \\
 & 2455147.45861 & 0.00043 & $-$1443 & \citet{m13a} & \citet{m16} \\
 & 2455163.83061 & 0.00032 & $-$1428 & \citet{collins2017} & \citet{collins2017} \\
 & 2455172.56138 & 0.00036 & $-$1420 & \citet{chan2011} & \citet{m16} \\
 & 2455209.66895 & 0.00046 & $-$1386 & \citet{collins2017} & \citet{collins2017} \\
 & 2455210.76151 & 0.00041 & $-$1385 & \citet{collins2017} & \citet{collins2017} \\
 & 2455230.40653 & 0.00024 & $-$1367 & \citet{m11} & \citet{m16} \\
 & 2455254.41761 & 0.00043 & $-$1345 & \citet{m11} & \citet{m16} \\
 & 2455494.52999 & 0.00074 & $-$1125 & \citet{m13a} & \citet{m16} \\
 & 2455498.89590 & 0.00079 & $-$1121 & \citet{sada2012} & \citet{sada2012} \\
 & 2455509.80971 & 0.00037 & $-$1111 & \citet{collins2017} & \citet{collins2017} \\
 & 2455510.90218 & 0.00031 & $-$1110 & \citet{collins2017} & \citet{collins2017} \\
 & 2455518.54147 & 0.00040 & $-$1103 & \citet{cowan2012} & \citet{m16} \\
 & 2455542.55210 & 0.00040 & $-$1081 & \citet{cowan2012} & \citet{cowan2012} \\
 & 2455542.55273 & 0.00029 & $-$1081 & \citet{m13a} & \citet{m16} \\
 & 2455590.57561 & 0.00071 & $-$1037 & \citet{m13a} & \citet{m16} \\
 & 2455598.21552 & 0.00035 & $-$1030 & \citet{m13a} & \citet{m16} \\
 & 2455600.39800 & 0.00030 & $-$1028 & \citet{m13a} & \citet{m16} \\
 & 2455601.49010 & 0.00024 & $-$1027 & \citet{m13a} & \citet{m16} \\
 & 2455603.67261 & 0.00029 & $-$1025 & \citet{collins2017} & \citet{collins2017} \\
 & 2455623.31829 & 0.00039 & $-$1007 & \citet{m13a} & \citet{m16} \\
 & 2455876.52786 & 0.00027 & $-$775 & \citet{m13a} & \citet{m16} \\
 & 2455887.44198 & 0.00021 & $-$765 & \citet{m13a} & \citet{m16} \\
 & 2455888.53340 & 0.00027 & $-$764 & \citet{m13a} & \citet{m16} \\
 & 2455890.71635 & 0.00024 & $-$762 & \citet{m13a} & \citet{m16} \\
 & 2455903.81357 & 0.00032 & $-$750 & \citet{collins2017} & \citet{collins2017} \\
 & 2455920.18422 & 0.00031 & $-$735 & \citet{m13a} & \citet{m16} \\
 & 2455923.45850 & 0.00022 & $-$732 & \citet{m13a} & \citet{m16} \\
 & 2455946.37823 & 0.00018 & $-$711 & \citet{m13a} & \citet{m16} \\
 & 2455947.47015 & 0.00017 & $-$710 & \citet{m13a} & \citet{m16} \\
 & 2455948.56112 & 0.00034 & $-$709 & \citet{m13a} & \citet{m16} \\
 & 2455951.83536 & 0.00011 & $-$706 & \citet{stevenson2014} & \citet{m16} \\
 & 2455952.92708 & 0.00013 & $-$705 & \citet{stevenson2014} & \citet{m16} \\
 & 2455959.47543 & 0.00017 & $-$699 & \citet{m13a} & \citet{m16} \\
 & 2455970.38941 & 0.00040 & $-$689 & \citet{m13a} & \citet{m16} \\
 & 2455971.48111 & 0.00035 & $-$688 & \citet{m13a} & \citet{m16} \\
 & 2455982.39509 & 0.00034 & $-$678 & \citet{m13a} & \citet{m16} \\
 & 2455983.48695 & 0.00035 & $-$677 & \citet{m13a} & \citet{m16} \\
 & 2455984.57797 & 0.00032 & $-$676 & \citet{collins2017} & \citet{collins2017} \\
 & 2455985.66975 & 0.00042 & $-$675 & \citet{collins2017} & \citet{collins2017} \\
 & 2455996.58378 & 0.00037 & $-$665 & \citet{collins2017} & \citet{collins2017} \\
 & 2456005.31533 & 0.00037 & $-$657 & \citet{m13a} & \citet{m16} \\
 & 2456006.40637 & 0.00033 & $-$656 & \citet{m13a} & \citet{m16} \\
 & 2456245.42729 & 0.00033 & $-$437 & \citet{m16} & \citet{m16} \\
 & 2456249.79404 & 0.00039 & $-$433 & \citet{collins2017} & \citet{collins2017} \\
 & 2456273.80514 & 0.00030 & $-$411 & \citet{collins2017} & \citet{collins2017} \\
 & 2456282.53584 & 0.00030 & $-$403 & \citet{m16} & \citet{m16} \\
 & 2456284.71857 & 0.00030 & $-$401 & \citet{collins2017} & \citet{collins2017} \\
 & 2456297.81605 & 0.00030 & $-$389 & \citet{collins2017} & \citet{collins2017} \\
 & 2456302.18179 & 0.00046 & $-$385 & \citet{m16} & \citet{m16} \\
 & 2456305.45536 & 0.00026 & $-$382 & \citet{m16} & \citet{m16} \\
 & 2456319.64424 & 0.00038 & $-$369 & \citet{collins2017} & \citet{collins2017} \\
 & 2456328.37556 & 0.00027 & $-$361 & \citet{m16} & \citet{m16} \\
 & 2456329.46733 & 0.00029 & $-$360 & \citet{m16} & \citet{m16} \\
 & 2456604.50489 & 0.00021 & $-$108 & \citet{m16} & \citet{m16} \\
 & 2456605.59624 & 0.00030 & $-$107 & \citet{m16} & \citet{m16} \\
 & 2456606.68760 & 0.00034 & $-$106 & \citet{m16} & \citet{m16} \\
 & 2456607.77938 & 0.00071 & $-$105 & \citet{collins2017} & \citet{collins2017} \\
 & 2456629.60726 & 0.00019 & $-$85 & \citet{m16} & \citet{m16} \\
 & 2456630.69917 & 0.00043 & $-$84 & \citet{m16} & \citet{m16} \\
 & 2456654.71047 & 0.00034 & $-$62 & \citet{collins2017} & \citet{collins2017} \\
 & 2456662.35014 & 0.00019 & $-$55 & \citet{m16} & \citet{m16} \\
 & 2456663.44136 & 0.00019 & $-$54 & \citet{m16} & \citet{m16} \\
 & 2456664.53256 & 0.00032 & $-$53 & \citet{m16} & \citet{m16} \\
 & 2456677.63039 & 0.00032 & $-$41 & \citet{collins2017} & \citet{collins2017} \\
 & 2456688.54384 & 0.00041 & $-$31 & \citet{m16} & \citet{m16} \\
 & 2456711.46415 & 0.00026 & $-$10 & \citet{m16} & \citet{m16} \\
 & 2456722.37807 & 0.00047 & 0 & \citet{m16} & \citet{m16} \\
 & 2456986.50195 & 0.00043 & 242 & \citet{m16} & \citet{m16} \\
 & 2457010.51298 & 0.00039 & 264 & \citet{m16} & \citet{m16} \\
 & 2457012.69617 & 0.00049 & 266 & \citet{collins2017} & \citet{collins2017} \\
 & 2457045.43831 & 0.00049 & 296 & \citet{m16} & \citet{m16} \\
 & 2457046.53019 & 0.00049 & 297 & \citet{m16} & \citet{m16} \\
 & 2457059.62713 & 0.00035 & 309 & \citet{collins2017} & \citet{collins2017} \\
 & 2457060.71839 & 0.00036 & 310 & \citet{collins2017} & \citet{collins2017} \\
 & 2457067.26715 & 0.00023 & 316 & \citet{m16} & \citet{m16} \\
 & 2457068.35834 & 0.00021 & 317 & \citet{m16} & \citet{m16} \\
 & 2457103.28423 & 0.00031 & 349 & \citet{m16} & \citet{m16} \\
 & 2457345.57867 & 0.00042 & 571 & \citet{m16} & \citet{m16} \\
 & 2457390.32708 & 0.00034 & 612 & \citet{m16} & \citet{m16} \\
 & 2457391.41818 & 0.00033 & 613 & \citet{m16} & \citet{m16} \\
 & 2457426.34324 & 0.00055 & 645 & \citet{m16} & \citet{m16} \\
 & 2457427.43496 & 0.00023 & 646 & \citet{m16} & \citet{m16} \\
 & 2457451.44617 & 0.00021 & 668 & \citet{m18a} & \citet{m18a} \\
 & 2457671.91324 & 0.00035 & 870 & \citet{patra2017} & \citet{patra2017} \\
 & 2457691.55888 & 0.00025 & 888 & \citet{m18a} & \citet{m18a} \\
 & 2457703.56388 & 0.00034 & 899 & \citet{m18a} & \citet{m18a} \\
 & 2457706.83791 & 0.00037 & 902 & \citet{patra2017} & \citet{patra2017} \\
 & 2457726.48400 & 0.00028 & 920 & \citet{m18a} & \citet{m18a} \\
 & 2457727.57547 & 0.00023 & 921 & \citet{m18a} & \citet{m18a} \\
 & 2457765.77515 & 0.00028 & 956 & \citet{patra2017} & \citet{patra2017} \\
 & 2457766.86633 & 0.00039 & 957 & \citet{patra2017} & \citet{patra2017} \\
 & 2457772.32407 & 0.00024 & 962 & \citet{m18a} & \citet{m18a} \\
 & 2457773.41517 & 0.00022 & 963 & \citet{m18a} & \citet{m18a} \\
 & 2457776.68869 & 0.00029 & 966 & \citet{patra2017} & \citet{patra2017} \\
 & 2457781.05566 & 0.00036 & 970 & \citet{patra2020} & \citet{patra2020} \\
 & 2457781.05418 & 0.00043 & 970 & \citet{patra2020} & \citet{patra2020} \\
 & 2457786.51210 & 0.00026 & 975 & \citet{m18a} & \citet{m18a} \\
 & 2457788.69464 & 0.00048 & 977 & \citet{patra2017} & \citet{patra2017} \\
 & 2457800.69978 & 0.00032 & 988 & \citet{patra2017} & \citet{patra2017} \\
 & 2457808.34020 & 0.00040 & 995 & \citet{ozturk2019} & \citet{ozturk2019} \\
 & 2457809.43190 & 0.00018 & 996 & \citet{m18a} & \citet{m18a} \\
 & 2457810.52327 & 0.00021 & 997 & \citet{m18a} & \citet{m18a} \\
 & 2458026.62368 & 0.00056 & 1195 & \citet{m18a} & \citet{m18a} \\
 & 2458050.63519 & 0.00023 & 1217 & \citet{m18a} & \citet{m18a} \\
 & 2458060.45870 & 0.00030 & 1226 & \citet{ozturk2019} & \citet{ozturk2019} \\
 & 2458073.55509 & 0.00022 & 1238 & \citet{m18a} & \citet{m18a} \\
 & 2458074.64651 & 0.00034 & 1239 & \citet{m18a} & \citet{m18a} \\
 & 2458077.92107 & 0.00028 & 1242 & \citet{yee2020} & \citet{yee2020} \\
 & 2458123.76011 & 0.00027 & 1284 & \citet{yee2020} & \citet{yee2020} \\
 & 2458124.85183 & 0.00035 & 1285 & \citet{yee2020} & \citet{yee2020} \\
 & 2458132.49121 & 0.00031 & 1292 & \citet{m18a} & \citet{m18a} \\
 & 2458134.67471 & 0.00032 & 1294 & \citet{yee2020} & \citet{yee2020} \\
 & 2458136.85760 & 0.00033 & 1296 & \citet{yee2020} & \citet{yee2020} \\
 & 2458155.41040 & 0.00050 & 1313 & \citet{ozturk2019} & \citet{ozturk2019} \\
 & 2458155.41152 & 0.00031 & 1313 & \citet{m18a} & \citet{m18a} \\
 & 2458156.50267 & 0.00032 & 1314 & \citet{m18a} & \citet{m18a} \\
 & 2458159.77773 & 0.00091 & 1317 & \citet{yee2020} & \citet{yee2020} \\
 & 2458161.95991 & 0.00035 & 1319 & \citet{patra2020} & \citet{patra2020} \\
 & 2458161.95964 & 0.00026 & 1319 & \citet{patra2020} & \citet{patra2020} \\
 & 2458163.05125 & 0.00021 & 1320 & \citet{patra2020} & \citet{patra2020} \\
 & 2458163.05089 & 0.00034 & 1320 & \citet{patra2020} & \citet{patra2020} \\
 & 2458166.32575 & 0.00034 & 1323 & \citet{m18a} & \citet{m18a} \\
 & 2458178.33104 & 0.00038 & 1334 & \citet{m18a} & \citet{m18a} \\
 & 2458411.89495 & 0.00040 & 1548 & \citet{yee2020} & \citet{yee2020} \\
 & 2458471.92257 & 0.00026 & 1603 & \citet{yee2020} & \citet{yee2020} \\
 & 2458494.84270 & 0.00030 & 1624 & \citet{yee2020} & \citet{yee2020} \\
 & 2458506.84758 & 0.00044 & 1635 & \citet{yee2020} & \citet{yee2020} \\
 & 2458853.91918 & 0.00021 & 1953 & this work & this work \\
 \hline
WASP-92 & 2456381.28418 & 0.00028 & 0 & \citet{wasp9293} & \citet{wasp9293} \\
 & 2458971.32011 & 0.00058 & 1191 & this work & this work \\
\hline
WASP-93 & 2456079.56495 & 0.00046 & 0 & \citet{wasp9293} & \citet{wasp9293} \\
 & 2458779.31199 & 0.00062 & 988 & this work & this work \\
\hline
WASP-135 & 2455230.99020 & 0.00090 & $-$2154 & \citet{wasp135} & \citet{wasp135} \\
 & 2457924.43994 & 0.00022 & $-$232 & \citet{ozturk2021} & \citet{ozturk2021} \\
 & 2458249.56016 & 0.00018 & 0 & \citet{ozturk2021} & \citet{ozturk2021} \\
 & 2458280.38743 & 0.00064 & 22 & \citet{ozturk2021} & \citet{ozturk2021} \\
 & 2458301.41018 & 0.00028 & 37 & \citet{ozturk2021} & \citet{ozturk2021} \\
 & 2458336.44162 & 0.00044 & 62 & \citet{ozturk2021} & \citet{ozturk2021} \\
 & 2458381.28867 & 0.00043 & 94 & \citet{ozturk2021} & \citet{ozturk2021} \\
 & 2458388.29083 & 0.00052 & 99 & \citet{ozturk2021} & \citet{ozturk2021} \\
 & 2459021.71876 & 0.00072 & 551 & this work & this work \\
\enddata
\end{deluxetable*}

\section{Updated Transit Ephemeris Fits}\label{sec:ephemplots}
\restartappendixnumbering

\begin{figure*}
    \centering
    \includegraphics[width=0.90\linewidth]{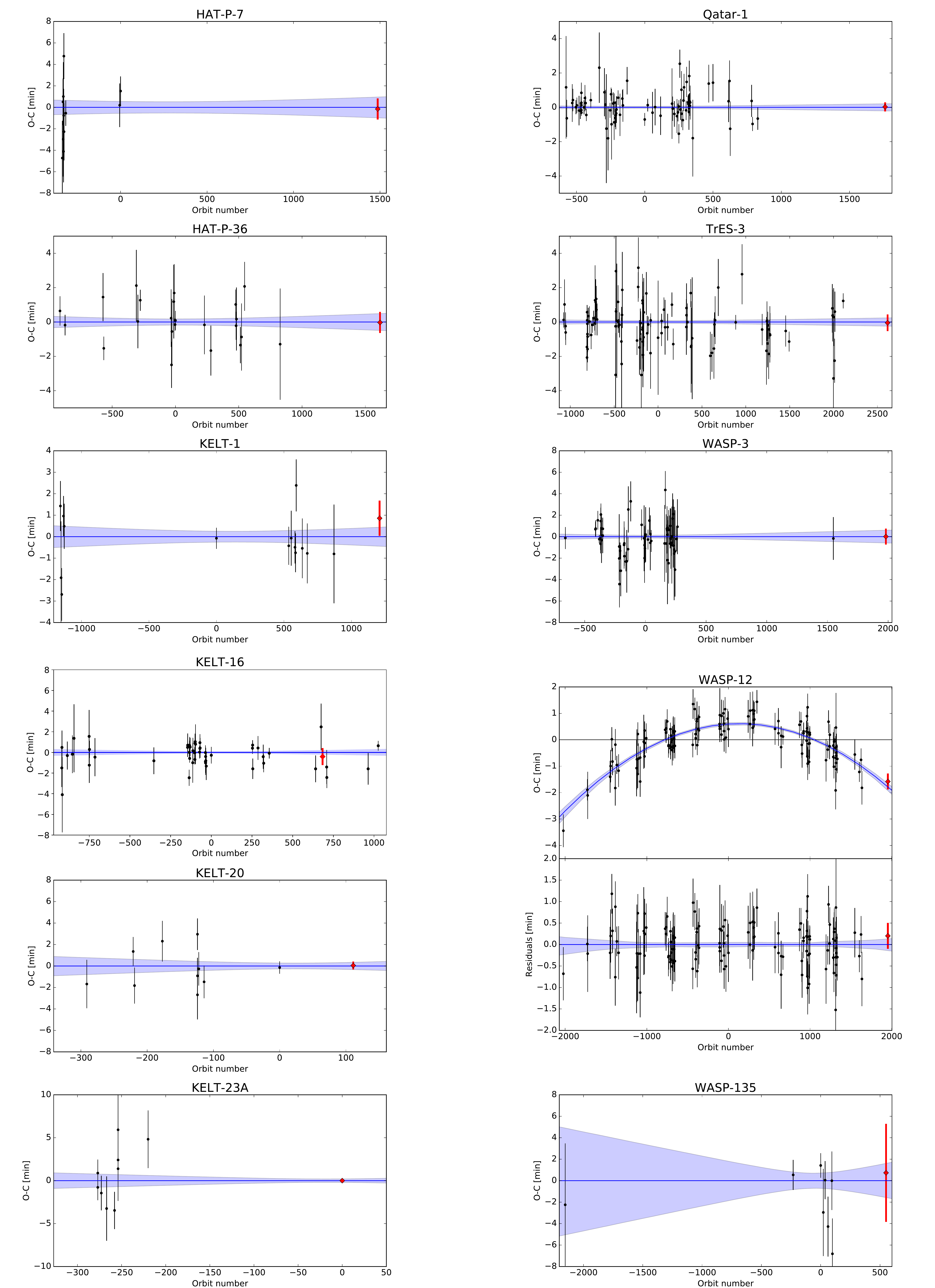}
    \caption{Observed minus calculated ($O-C$) plots for the 11 systems with more than two published transit timings. The TESS measurements are shown in red.}
    \label{fig:ephem}
\end{figure*}

Figure~\ref{fig:ephem} shows the results from our updated transit ephemeris fits for the 11 systems with at least three published transit-timing measurements. The black and red data points correspond to the previously published transit timings and the new TESS-epoch mid-transit times, respectively. The shaded blue region indicates the $1\sigma$ confidence region relative to the best-fit linear transit ephemeris (see Table~\ref{tab:ephem}). The orbit of WASP-12 is decaying, and we have included an additional panel showing the residuals from the best-fit quadratic transit model.

\end{document}